\pdfoutput=1 %% arXiv admin added this line

\documentclass[aps,rmp,a4paper,showpacs,twocolumn,floatfix]{revtex4}
\usepackage[latin1]{inputenc}
\usepackage{amsfonts}
\usepackage{amssymb}
\usepackage{amsmath}
\usepackage{calc}
\usepackage{graphicx}
\usepackage{epsfig}
\usepackage{bm}
\usepackage{ulem}

\begin{document}
%%%%%%%%%%%%%%%%%%%%%%%%%%%%%%%%%%%%%%%%%%%%%%%%
%%%%%%%%%%%%%%%%%%%%%%%%%%%%%%%%%%%%%%%%%%%%%%%%

\title{Quantum repeaters based on atomic ensembles and linear optics}

\date{\today}
\author{Nicolas Sangouard$^{1,2}$}
\author{Christoph Simon$^{1}$}
\author{Hugues de Riedmatten$^{1}$}
\author{Nicolas Gisin$^{1}$}
\affiliation{%
$^{1}$Group of Applied Physics, University of Geneva,
CH-1211 Geneva 4, Switzerland}
\affiliation{$^{2}$Laboratoire Mat\'{e}riaux et
Ph\'{e}nom\`{e}nes Quantiques, UMR CNRS 7162, Universit\'e
Paris 7, 75013 Paris, France}
\begin{abstract}
The distribution of quantum states over long distances is
limited by photon loss. Straightforward amplification as in
classical telecommunications is not an option in quantum
communication because of the no-cloning theorem. This
problem could be overcome by implementing quantum repeater
protocols, which create long-distance entanglement from
shorter-distance entanglement via entanglement swapping.
Such protocols require the capacity to create entanglement
in a heralded fashion, to store it in quantum memories, and
to swap it. One attractive general strategy for realizing
quantum repeaters is based on the use of atomic ensembles
as quantum memories, in combination with linear optical
techniques and photon counting to perform all required
operations. Here we review the theoretical and experimental
status quo of this very active field. We compare the
potential of different approaches quantitatively, with a
focus on the most immediate goal of outperforming the
direct transmission of photons.
\end{abstract}

\maketitle

\tableofcontents

\section{Introduction}
\label{Introduction}

The distribution of quantum states over long distances is
essential for potential future applications of quantum
technology such as long-distance quantum cryptography
\cite{Bennett1984,Gisin2002} and quantum networks
\cite{Kimble2008,Nielsen2000}. In practice quantum channels
such as optical fibers or free-space transmission are
affected by loss and decoherence. This limits the distance
over which quantum information can be transmitted directly
by sending individual quantum systems (typically photons).
In practice the most immediate problem is photon loss. For
example, typical telecommunication optical fibers have
losses of 0.2 dB/km in the optimal wavelength range around
1.5 $\mu$m. In a sense these losses are impressively low.
For example, a piece of fiber that is 1 km long has a
transmission of 95 percent. However, they become
nevertheless very significant once one envisions distances
of hundreds of kilometers or more. Even for very
high-repetition rate sources (say 10 GHz, which is a very
ambitious value for a source of quantum states), the rate
of transmitted photons becomes exponentially low for such
distances. For example, for 500 km one would have a rate of
1 Hz. The rate drops to 0.01 Hz for 600 km, and to
$10^{-10}$ Hz for 1000 km. The latter rate corresponds to 1
photon every 300 years.

In classical telecommunications this problem is overcome
through the use of amplifiers (``repeaters'').
Unfortunately straightforward amplification is not an
option in quantum communication because of the no-cloning
theorem \cite{Wootters1982,Dieks1982}, which shows that
noiseless amplification is impossible unless one restricts
oneself to sets of orthogonal states, whereas the quantum
nature (and thus the advantage) of protocols such as
quantum key distribution arises precisely from the
existence of non-orthogonal states. However, it turns out
that the problem can be overcome using a more sophisticated
method based on entanglement, which is known as the
``quantum repeater'' approach \cite{Briegel1998}.

Entanglement is one of the most counter-intuitive,
non-classical features of quantum physics. Bell's famous
theorem \cite{Bell1964,Mermin1993} states that entangled
states cannot be simulated by local hidden variables, thus
showing that entanglement lies at the heart of quantum
non-locality. A very remarkable feature of entanglement is
that it can be ``swapped'' \cite{Zukowski1993}. Given an
entangled state between two systems $A$ and $B$ and another
entangled state between systems $C$ and $D$, it is possible
to create an entangled state between systems $A$ and $D$ by
performing a joint measurement of systems $B$ and $C$ in a
basis of entangled states, followed by classical
communication of the result to the location of system $A$
and/or $D$. Entanglement between the two latter systems can
be created in this way even though they may never have
interacted.

Entanglement swapping can be seen as the generalization of
quantum teleportation \cite{Bennett1993} to entangled input
states. Conversely, teleportation can be seen as a
consequence of entanglement swapping. Performing a
measurement on system $A$ before the swapping (when $A$ and
$B$ are entangled) projects system $B$ into an un-entangled
quantum state. Performing the same measurement after the
swapping (when $A$ and $D$ are entangled) creates the same
state in system $D$. But the described swapping procedure
commutes with measurements on system $A$. This implies that
the entanglement swapping procedure transfers the quantum
state of system $B$ to system $D$, without any quantum
system physically moving from one location to the other.

Consider a great distance $L$, such that the overall
transmission of the channel is forbiddingly small. As we
have just seen, if one has an entangled state of two
particles separated by the distance $L$ at one's
disposition, one can use this entangled state to teleport
quantum states over this distance. One can also use it to
perform entanglement-based quantum key distribution
\cite{Ekert1991} directly. The creation of entanglement
over some great distance $L$ can also be seen as a
fundamental goal in itself, allowing one to extend tests of
quantum non-locality to a new distance scale.

\begin{figure}
{\includegraphics[scale=0.31]{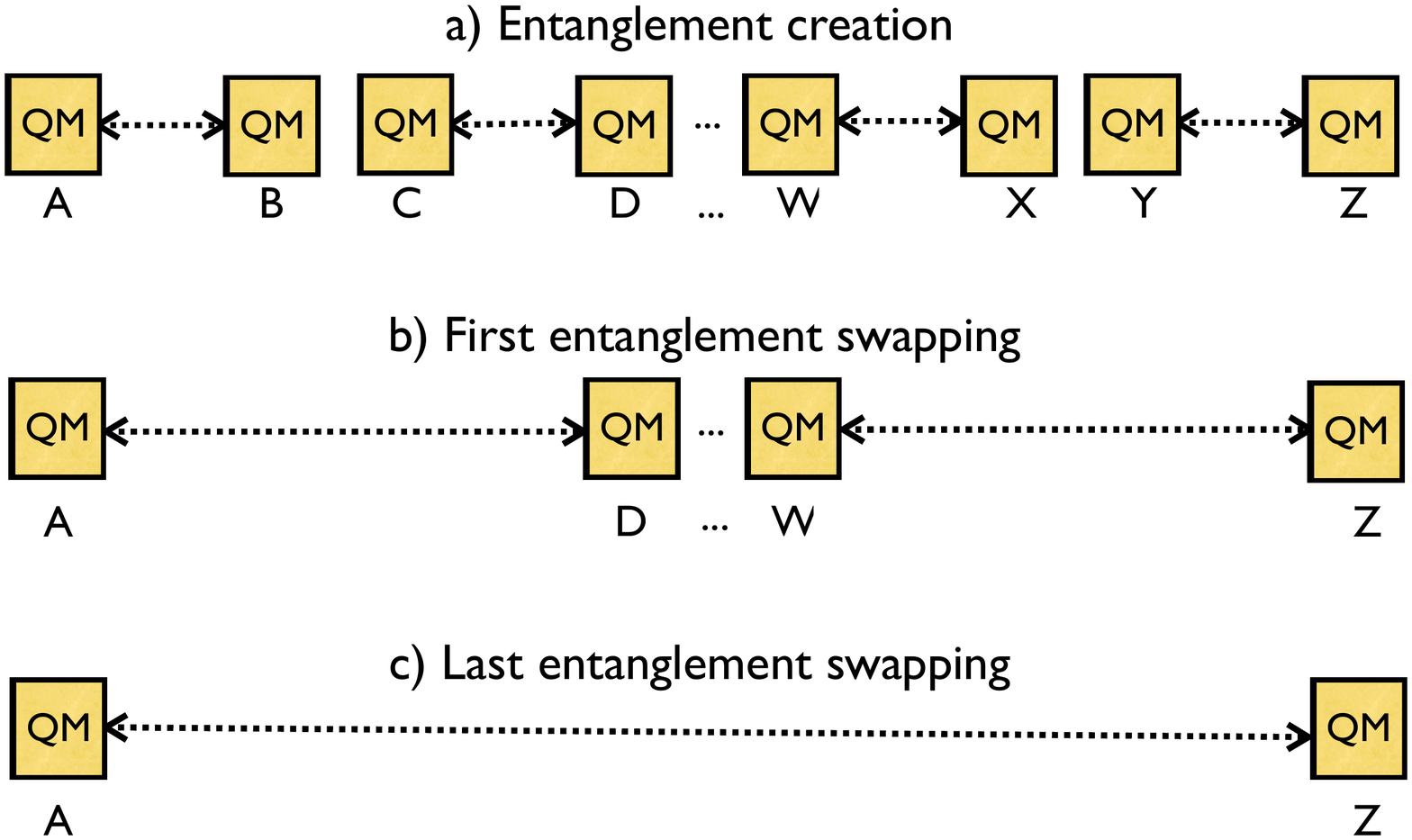}} \caption{(Color
online) Principle of quantum repeaters. In order to
distribute entanglement over long distances, say between
locations A and Z, one proceeds step by step : a)
Entanglement is first created independently within short
elementary links, say between the locations A and B, C and
D, $\hdots,$ W and X, Y and Z. b) Entanglement is then
swapped between neighboring links such that the locations A
and D, $\hdots$ W and Z share entanglement. c) Entanglement
swapping operations are performed successively in a
hierarchical fashion until entanglement is distributed over
the desired distance separating the locations A and Z.
Yellow squares represents quantum memories. The dotted
arrows connecting two remote memories indicate that they
are entangled.} \label{repeater}
\end{figure}

The key idea of the quantum repeater approach
\cite{Briegel1998} is that entanglement over the distance
$L$ can be created by entanglement swapping starting from
two entangled pairs, each of which covers only half the
distance, $\frac{L}{2}$. Moreover, these entangled states
can themselves be created starting from states covering a
distance $\frac{L}{4}$ and so on. If one has a way of
independently establishing entanglement for $N=2^n$
adjacent elementary links each covering a distance
$L_0=\frac{L}{N}$, one can then efficiently create
entanglement over a distance $L$ with $n$ levels of
entanglement swapping operations, cf. Figure \ref{repeater}
($n$ is called the nesting level). For long distances the
described protocol scales much better than direct
transmission.

One essential requirement for the described approach is
thus to be able to establish entanglement for the
elementary links in a ``heralded'' way, i.e. one has to
know when the entanglement has been successfully
established. At first sight, the most direct approach would
be to create entanglement between two systems locally and
then send one of the two systems (e.g. a photon) to the
distant location. However, the elementary links will still
be quite long for realistic protocols, typically of order
100 km, corresponding to a transmission of order $10^{-2}$.
Implementing heralding in such an approach would require
being able to measure that the photon has arrived without
destroying the entanglement, which is very difficult in
practice. A better approach is to create the entanglement
``at a distance''. For example, entanglement between one
atom in $A$ and another atom in $B$ can be created via the
detection of a photon that could have been emitted by
either atom, provided that the measurement of the photon is
performed in such a way that all ``which-way'' information
is erased \cite{Bose1999,Cabrillo1999}. (This can be seen
as another application of the principle of entanglement
swapping, cf. below.) The detection of the photon then
serves as the heralding event for the creation of the
entanglement between the two atoms. If the photon is lost
in transmission, there is no detection and one knows that
one has to try again.

Another essential requirement for the quantum repeater
protocol is that one has to be able to store the created
elementary entanglement until entanglement has been
established in the neighboring link as well, in order to
then be able to perform the required entanglement swapping
operation. The resulting higher-level entanglement again
needs to be stored until the neighboring higher-level link
has been established and so on. Thus quantum repeaters
require the existence of ``quantum memories''
\cite{Hammerer2008,Lukin2003,Tittel2008}. If such memories
are not available, the only solution is to create
entanglement in all links simultaneously. Such memory-less
repeaters, also called ``quantum relays'', do not help to
overcome the problem of photon loss, but can still be
useful to alleviate other problems such as detector dark
counts \cite{Jacobs2002,Collins2005}.

Finally one has to be able to perform the required
entanglement swapping operations between the quantum
memories, i.e. to perform local joint measurements
projecting onto entangled states between two memories. Such
measurements are certainly possible if one has a way of
performing general quantum gates (e.g. CNOT gates) between
neighboring memories. However, this is generally a
difficult task and it is thus of interest to consider
dedicated, simpler solutions, e.g. entangling measurements
that work only with a certain probability, cf. below.

The original quantum repeater protocol of
\cite{Briegel1998} furthermore contains ``entanglement
purification'' \cite{Bennett1996} steps that allow one in
principle to purify the effects of any kind of decoherence.
However, the implementation of such general entanglement
purification requires the preparation of at least two
initial pairs for every purified pair at any given nesting
level for which purification is implemented, leading to
significant overheads and thus to lower rates. This makes
it advantageous to forgo full entanglement purification for
simple architectures of just a few links, where it is not
necessary for small, but realistic error probabilities per
operation. In the present review our focus will be on such
simple architectures, because they offer the most realistic
chance in the short and medium term of achieving the most
immediate goal of a quantum repeater, namely to outperform
the quantum state distribution rate achievable by direct
transmission.

A highly influential proposal for realizing quantum
repeaters was made by \cite{Duan2001}. It is widely known
as the DLCZ protocol (for Duan, Lukin, Cirac and Zoller).
The authors showed how to meet all the above requirements
using atomic ensembles as quantum memories, and linear
optical techniques in combination with photon counting to
perform all the required operations. The use of atomic
ensembles as opposed to single quantum systems such as
individual atoms as memories was motivated by the fact that
collective effects related to the large number of atoms in
the ensemble make it much easier to achieve a strong and
controllable coupling between the memory and the photons
that serve as long-distance quantum information carriers.

The basic process at the heart of the DLCZ protocol is the
spontaneous Raman emission of a photon, which
simultaneously creates a spin excitation in the atomic
ensemble. This correlation between emitted photons and
atomic excitations in each ensemble forms the basis for the
generation of entanglement between distant ensembles (for
each elementary link), which is done via a single photon
detection that erases all ``which-way'' information,
following the principle outlined above for the case of
individual atoms. The spin excitations can be efficiently
reconverted into photons thanks to a collective
interference effect. This forms the basis for the
entanglement swapping operations, which are again done by
detecting single photons while erasing which-way
information.

The DLCZ proposal inspired a large number of highly
successful experiments, for example
\cite{Kuzmich2003,vanderWal2003,Matsukevich2004,Chou2005,Chou2007,
Yuan2008}, showing that the approach of using atomic
ensembles, linear optics and photon counting is indeed very
attractive from a practical point of view. Motivated both
by the impressive experimental progress and by the growing
realization that, while pioneering, the DLCZ protocol does
not yet allow one to outperform the direct transmission of
photons in practice, several papers have proposed
significant improvements to the protocol, while using the
same or very similar experimental ingredients. These
proposals have in turn spurred new experimental
investigations. Here we review this area of research.

We begin with a review of the theoretical proposals,
starting with the DLCZ proposal in section \ref{DLCZ},
including a discussion of its practical limitations,
followed by a review of the most important proposed
improvements in section \ref{Improvements}. In the DLCZ
protocol, both entanglement generation and swapping are
based on one single-photon detection each. Subsection
\ref{twophotonswap} describes a protocol where entanglement
is swapped based on two photon detections, leading to an
improvement in the overall rate. Subsection
\ref{twophotongen} describes protocols where entanglement
is generated based on two photon detections, leading to
enhanced robustness with respect to phase fluctuations in
the channel. Subsections \ref{P2M3} and \ref{Collins} are
devoted to multiplexing. Subsection \ref{P2M3} reviews the
idea of using memories that can store multiple temporal
modes. Their use in the present context is made possible by
the realization that a DLCZ-type atomic ensemble can be
emulated by combining a photon pair source and an
``absorptive'' quantum memory (i.e. a memory that can
absorb and emit photons). This approach promises a great
enhancement in the entanglement generation rate. Subsection
\ref{Collins} reviews work on spatial multiplexing, which
would moreover significantly reduce the requirements on the
memory time. Subsection \ref{SPS} discusses a protocol
based on single-photon sources, which can be effectively
implemented with atomic ensembles, and which yields a
significantly enhanced rate compared to the DLCZ protocol.
Subsection \ref{IPP} describes protocols that are based on
effectively approximating ideal photon pair sources with
atomic ensembles, leading both to enhanced rates and
greatly enhanced robustness.

In section \ref{Comparison} we compare the performance of
different protocols quantitatively. Subsection \ref{Time}
is devoted to the entanglement distribution rates, whereas
subsection \ref{Robustness} discusses the robustness of the
protocols with respect to storage time limitations, phase
errors, and memory and detection inefficiencies. Subsection
\ref{Complexity} discusses complexity issues.

In section \ref{Implementations} we review the experimental
status quo from the point of view of the different
protocols described beforehand. In particular, subsection
V.A is devoted to experiments that realize elements of the
DLCZ protocol, subsection V.B to experiments that are
directed towards the creation and swapping of entanglement
via two-photon detections. Subsection V.C discusses the
implementation of quantum light sources compatible with
ensemble-based quantum memories, while subsection V.D
discusses the realization of the (absorptive) quantum
memories themselves. Subsection V.E is devoted to photon
detectors. Subsection V.F discusses different realizations
of quantum channels. Finally subsection V.G discusses the
issue of coupling losses.

In section \ref{OtherApproaches} we briefly review
approaches to quantum repeaters using other ingredients
besides or instead of atomic ensembles and linear optics,
for example the use of single trapped ions or NV centers as
quantum memories.

In section \ref{Conclusions} we give our conclusions and
look towards the future.

%%%%%%%%%%%%%%%%%%%%%%%%%%%%%%%%
\section{The DLCZ Protocol}
\label{DLCZ}
%%%%%%%%%%%%%%%%%%%%%%%%%%%%%%%%
In this section we review the DLCZ protocol for quantum
repeaters \cite{Duan2001}. We explain the basic physics
underlying the protocol, followed by a description of its
individual steps. We then evaluate the required time for
long-distance entanglement distribution. Finally we discuss
its limitations.

\subsection{Basic Physics}
\label{DLCZ-Basic}

\begin{figure}
    \center
  \includegraphics[angle=0,width=0.8 \columnwidth]{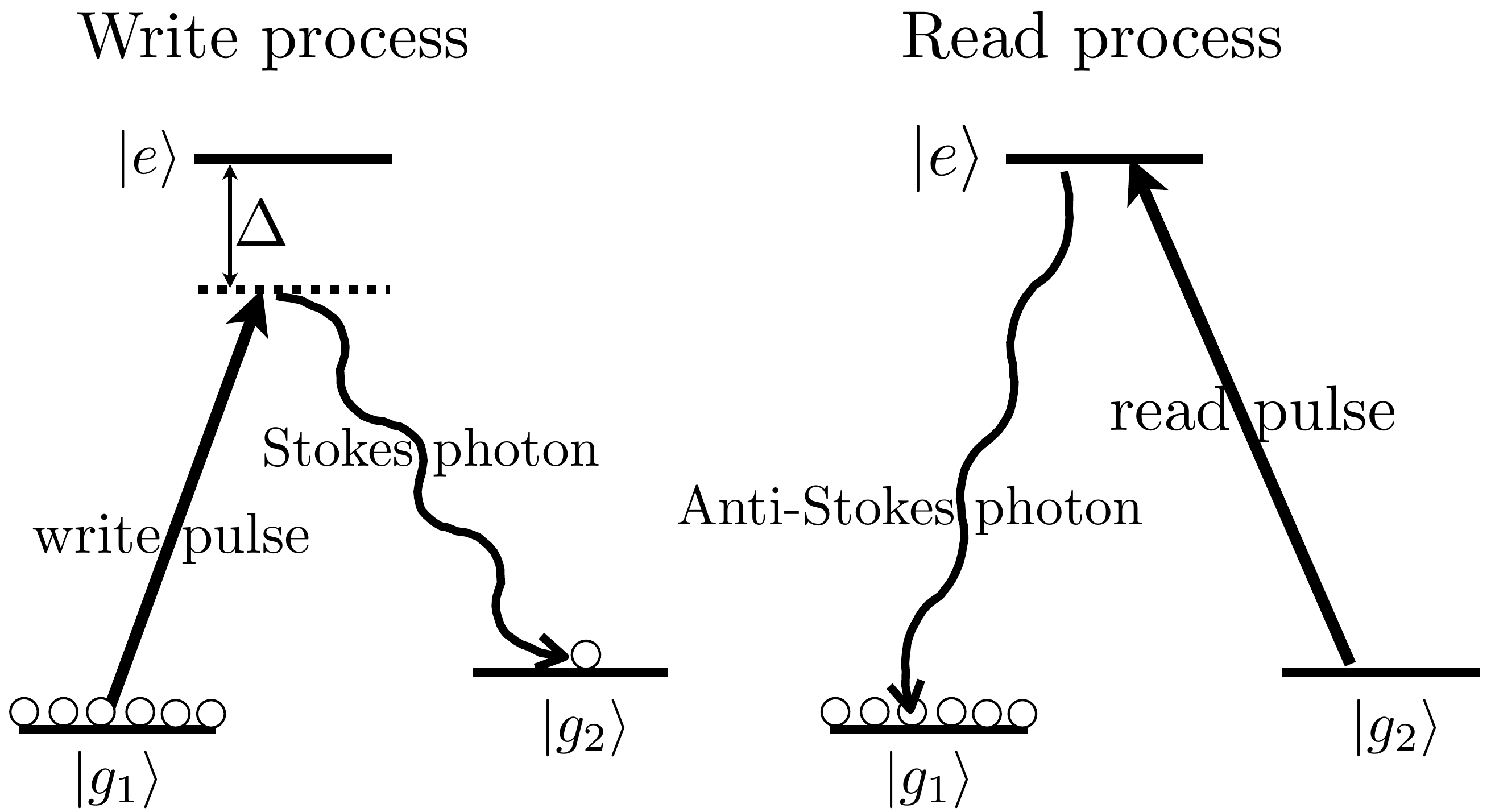}
  \caption{Basic level scheme for the creation of collective atomic
  excitations in atomic ensembles via spontaneous Raman emission (write process)
  and for their readout (read process), as proposed in the DLCZ protocol. Write process: All atoms start
  out in $g_1$. A laser
  pulse off-resonantly excites the $g_1-e$ transition, making it possible for
  a photon to be emitted on the $e-g_2$ transition (with small probability). Read process:
  a resonant laser is applied on the $g_2-e$ transition,
  promoting the single atomic excitation from $g_2$ back to
  $e$, followed by collective emission on the $e-g_1$ transition of a Stokes photon in a well-defined
direction.} \label{levels}
\end{figure}

The DLCZ protocol uses atomic ensembles that can emit
single photons while creating a single atomic excitation
which is stored in the ensemble. The photons can be used to
entangle two distant ensembles. The atomic excitation can
be efficiently converted into a photon thanks to collective
interference, which is used for entanglement swapping and
final use of the entanglement. Here we briefly describe the
underlying physics, the next section explains the protocol.

The basic (idealized) scheme is as follows, cf. Fig.
\ref{levels}. In an ensemble of three-level systems with
two ground states $g_1$ and $g_2$ and an excited state $e$
all $N_A$ atoms are initially in the state $g_1$. An
off-resonant laser pulse on the $g_1-e$ transition (the
{\it write} pulse) leads to the spontaneous emission of a
Raman photon on the $e-g_2$ transition. We will denote this
photon as the {\it Stokes} photon, which corresponds to the
usual Raman scattering terminology, provided that the
energy of $g_2$ is slightly higher than that of $g_1$. We
will adopt this convention throughout this review.
Detection of the Stokes photon in the far field, such that
no information is revealed about which atom it came from,
creates an atomic state that is a {\it coherent}
superposition of all the possible terms with $N_A-1$ atoms
in $g_1$ and one atom in $g_2$, namely
\begin{equation}
\frac{1}{\sqrt{N_A}} \sum \limits_{k=1}^{N_A} e^{i({\bf
k}_w-{\bf k}_S) {\bf x}_k} |g_1\rangle_1
|g_1\rangle_2...|g_2\rangle_k...|g_1\rangle_{N_A},
\label{ideal}
\end{equation}
where ${\bf k}_w$ is the ${\bf k}$ vector of the write
laser, ${\bf k}_S$ is the ${\bf k}$ vector of the detected
Stokes photon, and ${\bf x}_k$ is the position of the
$k$-th atom. In practice the amplitudes of the different
terms may vary, depending on the laser profile and the
shape of the atomic ensemble.

A remarkable feature of such collective excitations that are
of great interest for practical applications is that they
can be {\it read out} very efficiently by converting them
into single photons that propagate in a well-defined
direction, thanks to collective interference
\cite{Duan2001,Laurat2006,Simon2007a}. Resonant laser
excitation of such a state on the $g_2-e$ transition (the
{\it read} laser pulse) leads to an analogous state with
$N_A-1$ atoms in $g_1$ and one delocalized excitation in
$e$, but with supplementary phases $e^{i{\bf k}_r {\bf
x}'_k}$, where ${\bf k}_r$ is the ${\bf k}$ vector of the
read laser and ${\bf x}'_k$ is the position of the $k$-th
atom at the time of the readout (which may be different
from its initial position ${\bf x}_k$ if the atoms are
moving).

All the terms in this state can decay to the initial state
$|g_1\rangle^{\otimes {N_A}}$ while emitting a photon on
the $e-g_1$ transition (the {\it Anti-Stokes} photon). The
total amplitude for this process is then proportional to
\begin{equation}
\sum \limits_{k=1}^{N_A} e^{i({\bf k}_w-{\bf k}_S) {\bf
x}_k} e^{i({\bf k}_r-{\bf k}_{AS}) {\bf x}'_k}.
\end{equation}
The conditions for constructive interference of the $N_A$
terms in this sum depend on whether the atoms are moving
during the storage. If they are at rest (${\bf x}_k={\bf
x}'_k$ for all $k$), then there is constructive
interference whenever the phase matching condition ${\bf
k}_S+{\bf k}_{AS}={\bf k}_w+{\bf k}_r$ is fulfilled,
leading to a very large probability amplitude for emission
of the Anti-Stokes photon in the direction given by ${\bf
k}_w+{\bf k}_r-{\bf k}_S$. For atomic ensembles that
contain sufficiently many atoms, emission in this one
direction can completely dominate all other directions.
This allows a very efficient collection of the Anti-Stokes
photon \cite{Laurat2006,Simon2007a}. If the atoms are
moving, there can still be constructive interference,
provided that ${\bf k}_S={\bf k}_w$ and ${\bf k}_{AS}={\bf
k}_r$. We will come back to this point in section
\ref{Implementations}.A.

Note that there is no collective interference effect for
the emission of the Stokes photon, since its emission by
different atoms corresponds to orthogonal final states,
e.g. the state $|g_2\rangle_1
|g_1\rangle_2...|g_1\rangle_{N_A}$ if the Stokes photon was
emitted by the first atom etc. Full ``which-way''
information about the origin of the photon is thus stored
in the atomic ensemble, making interference impossible
\cite{Scully2003}. As a consequence the total emission
probability for the Stokes photon is simply given by the
sum of the emission probabilities for each atom, and there
is no preferred direction of emission.

We have focused on the emission of a single Stokes photon
into the mode of interest. However, since there is an
ensemble of atoms, there are also amplitudes for the
emission of two or more Stokes photons, accompanied by the
creation of the same number of atomic excitations in $g_2$.
This dynamics can be described by the following
Hamiltonian,
\begin{equation}
H=\chi (a^{\dagger} s^{\dagger}+a s), \label{pairs}
\end{equation}
where $\chi$ is a coupling constant that depends on the
laser intensity, the number of atoms, the detuning and the
transition strengths for the $g_1-e$ and $e-g_2$
transitions, $a^{\dagger}$ is the creation operator for a
Stokes photon and $s^{\dagger}$ is the creation operator
for an atomic excitation in $g_2$. The vacuum state
$|0\rangle$ for the mode $s$ corresponds to the atomic
state with all atoms in $g_1$, the state
$s^{\dagger}|0\rangle$ with one excitation in $s$
corresponds to a state like in Eq. (\ref{ideal}) with one
atom in $g_2$ etc. Here one focuses on one particular ${\bf
k}$ vector for both the Stokes photon and the atomic
excitation.

This Hamiltonian, whose derivation is discussed in much
more detail in \cite{Hammerer2008}, section II.A, thus
describes the creation (and annihilation) of pairs of
bosonic excitations. Note that it is formally equivalent to
the Hamiltonian for the non-linear optical process of
parametric down-conversion
\cite{Burnham1970,Hong1985,Hong1987,Wu1986}. Using operator
ordering techniques developed by \cite{Collett1988}, one
can show that, starting from an initial vacuum state for
both modes $a$ and $s$, it creates the following two-mode
entangled state
\begin{eqnarray}
e^{-iHt}|0\rangle|0\rangle=\frac{1}{\cosh (\chi t)} e^{-i
\tanh (\chi t) a^{\dagger}
s^{\dagger}}|0\rangle|0\rangle=\nonumber\\
\frac{1}{\cosh (\chi t)}\sum\limits_{m=0}^\infty (-i)^m
\tanh^m (\chi t) |m\rangle |m\rangle. \label{twomodes}
\end{eqnarray}
For small values of $\chi t$ this can be expanded
\begin{equation}
\left(1-\frac{1}{2}(\chi t)^2\right)|0\rangle
|0\rangle-i\chi t|1\rangle|1\rangle-(\chi t)^2
|2\rangle|2\rangle +O((\chi t)^3). \label{smallchit}
\end{equation}
Therefore, if the probability to emit one photon and create
one atomic excitation is $(\chi t)^2$, then there is a
probability $(\chi t)^4$ to emit two photons and create two
excitations etc. This possibility of creating multiple
pairs of excitations, which becomes more significant as the
probability of creating a single excitation is increased,
is an important limiting factor in the quantum repeater
protocols discussed in this review, cf. below.

%%%%%%%%%%%%%%%%%%%%%%%%%%%%
\subsection{Protocol}
\label{DLCZ-Protocol}

In this subsection we review the DLCZ protocol. We discuss
the principle for the entanglement creation between two
remote ensembles belonging to the same elementary link and
the method for entanglement swapping between neighboring
links. Finally we show how the created single-photon
entanglement of the form
$|1\rangle|0\rangle+|0\rangle|1\rangle$ can be used to
post-selectively obtain two-photon entanglement that is
useful for applications.
%%%%%%%%%%%%%%
\subsubsection{Entanglement Creation for two Remote Atomic Ensembles}

\begin{figure}[hr!]
{\includegraphics[width=\columnwidth]{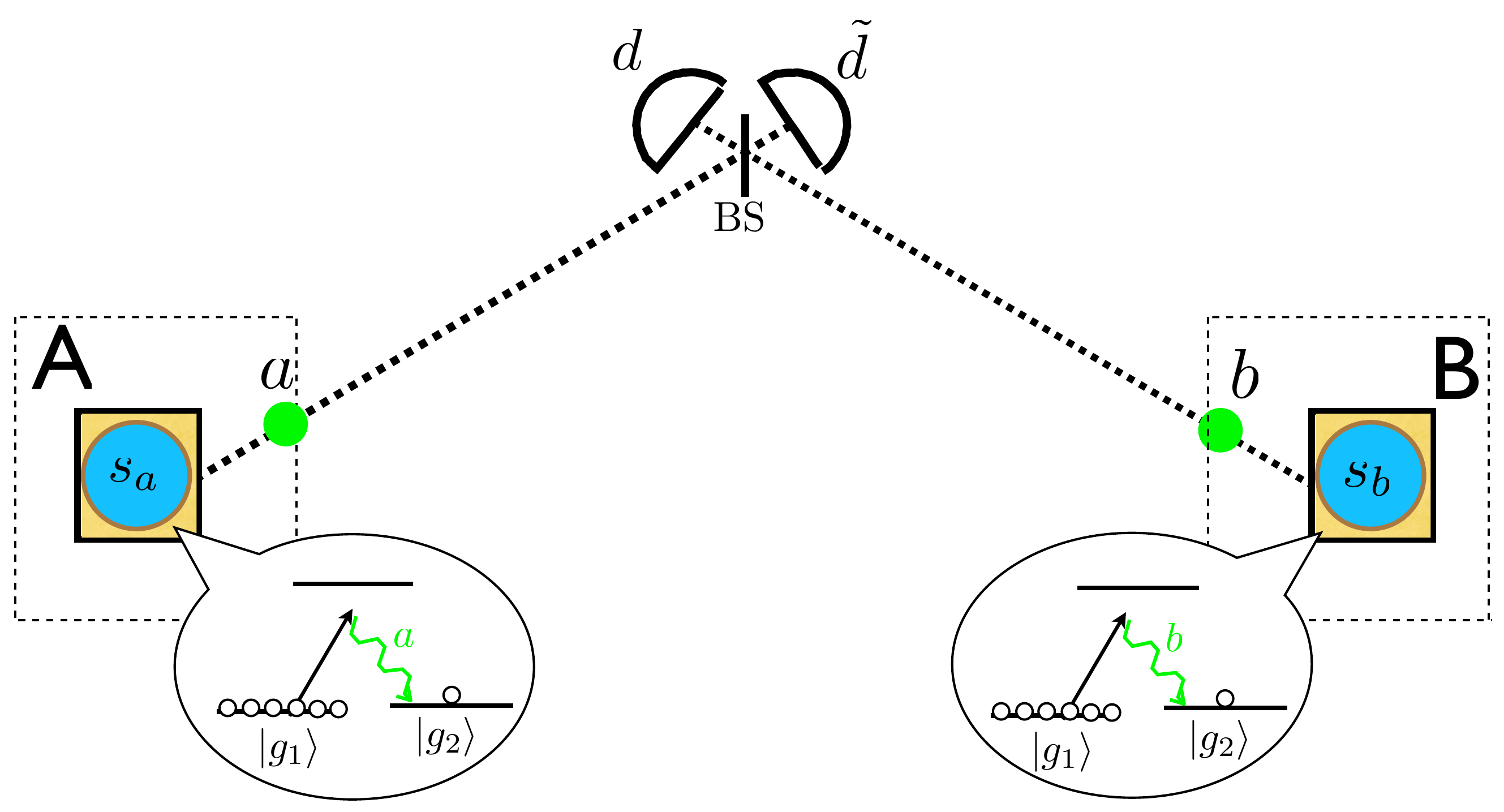}}
\caption{(Color online) Entanglement creation between two
remote ensembles located at $A$ and $B$ within the DLCZ
protocol. The blue circles embedded in yellow squares
represent DLCZ-type atomic ensembles which
probabilistically emit Stokes photons (green dots). These
photons are sent through long optical fibers (dotted line)
to a central station. The detection of a single Stokes
photon at the central station in mode $d$ or $\tilde{d}$,
which could have come either from location $A$ or $B$,
heralds the storage of a single excitation ($s_a$ or $s_b$)
in one of the two ensembles. Half circles represent photon
detectors. The vertical bar represents a beam splitter
(BS).} \label{DLCZcreation}
\end{figure}

The procedure for entanglement creation between two remote
locations $A$ and $B$ requires one ensemble at each
location. The two ensembles are simultaneously excited such
that a single Stokes photon can be emitted, corresponding
to the state
\begin{equation}
\left(1+\sqrt{\frac{p}{2}}\left(s_a^\dagger a^\dagger
e^{i\phi_a}+s_b^\dagger b^\dagger
e^{i\phi_b}\right)+O\left(p\right)\right)|0\rangle.
\label{realstate}
\end{equation}
Here, we assigned bosonic operators $a$ $(b)$ and $s_a$
$(s_b)$ to the Stokes photon and to the atomic excitation
respectively associated to the ensemble $A$ ($B$), $\phi_a$
$(\phi_b)$ is the phase of the pump laser at the location
$A$ $(B)$, and $|0\rangle$ is the vacuum state for all
modes; $O(p)$ represents the multi-photon terms discussed
in the previous subsection.

The Stokes photons are coupled into optical fibers and
combined on a beam splitter at a central station between
$A$ and $B.$ The modes after the beam splitter are
$d=\frac{1}{\sqrt{2}}(ae^{-i\xi_a}+be^{-i\xi_b})$ and
$\tilde{d}=\frac{1}{\sqrt{2}}(ae^{-i\xi_a}-be^{-i\xi_b})$
where $\xi_{a,b}$ stand for the phases acquire by the
photon on their way to the central station. The detection
of a single photon in $d$ for example, projects the state of
the two atomic ensembles in
\begin{equation}
\label{eq2}
|\psi_{ab}\rangle=\frac{1}{\sqrt{2}}\left(s_a^\dagger
e^{i(\phi_a+\xi_a)}+s_b^\dagger
e^{i(\phi_b+\xi_b)}\right)|0\rangle.
\end{equation}
A single atomic excitation is thus delocalized between $A$
and $B.$ This corresponds to an entangled state which can
be rewritten as
\begin{equation}
\label{eq3}
|\psi_{ab}\rangle=\frac{1}{\sqrt{2}}\left(|1_a\rangle|0_b\rangle
+|0_a\rangle|1_b\rangle e^{i\theta_{ab}}\right)
\end{equation}
where $|0_{a(b)}\rangle$ denotes an empty ensemble A (B),
and $|1_{a(b)}\rangle$ denotes the storage of a single
atomic excitation. We have also defined
$\theta_{ab}=\phi_b-\phi_a+\xi_b-\xi_a.$ Taking into
account detections both in $d$ and $\tilde{d}$, the success
probability of the entanglement creation is given by
$P_0=p\eta_d\eta_t$ where $\eta_d$ is the photon detection
efficiency and $\eta_t=\exp(-\frac{L_0}{2 L_{att}})$ is the
transmission efficiency corresponding to a distance of
$\frac{L_0}{2}$, where $L_0$ is the distance between $A$
and $B$ (i.e. the length of the elementary link), and
$L_{att}$ is the fiber attenuation length. (The losses of
0.2 dB/km mentioned previously, which are achievable in the
telecom wavelength range around 1550 nm, correspond to
$L_{att}=22$ km.)

This way of creating entanglement by a single photon
detection was inspired by similar proposals for entangling
two individual quantum systems rather than two ensembles
\cite{Bose1999,Cabrillo1999}. Note that it can be seen as
an implementation of entanglement swapping. One starts with
entangled states between the modes $a$ and $s_a$ as well as
$b$ and $s_b$ as in Eqs. (4) and (\ref{smallchit}). The
single photon detection at the central station projects
onto an entangled state of the photonic modes $a$ and $b$,
creating entanglement between the stored modes $s_a$ and
$s_b$.

%%%%%%%%%%%%%%
\subsubsection{Entanglement Connection between the Elementary Links}
\begin{figure}[hr!]
{\includegraphics[scale=0.32]{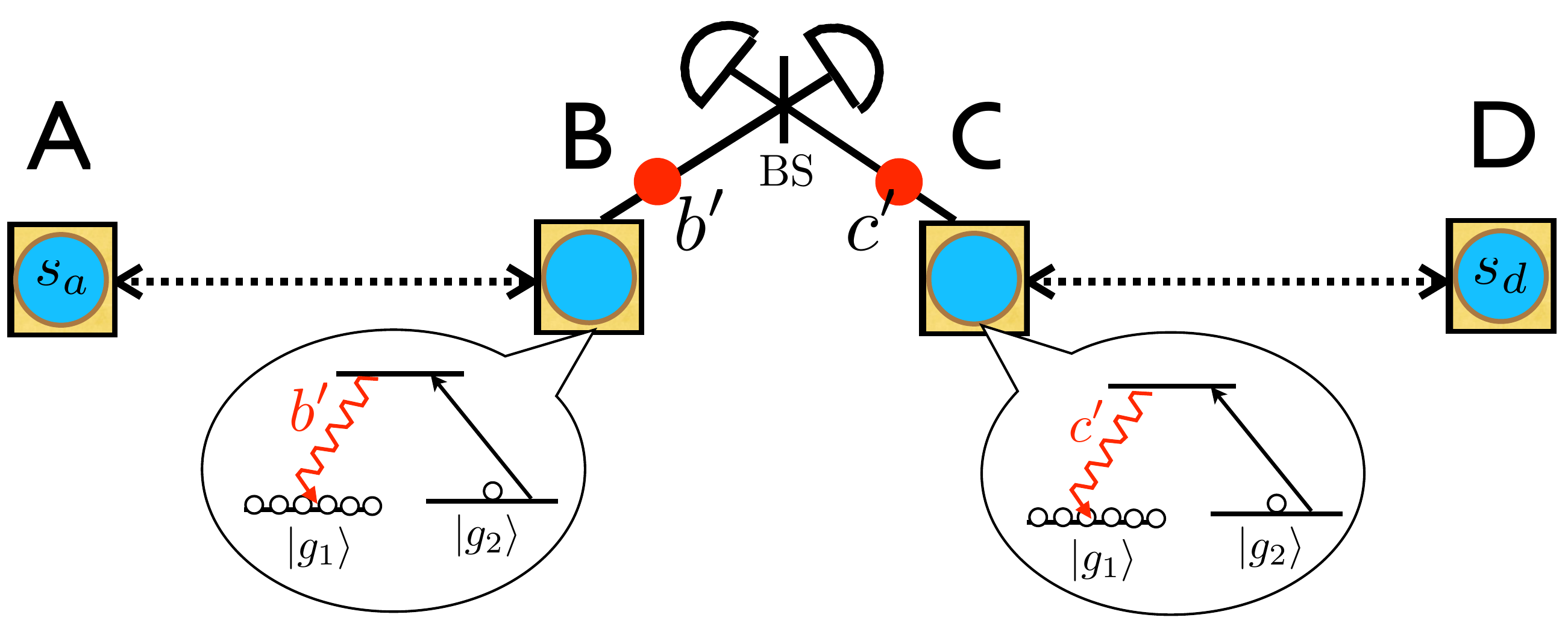}} \caption{(Color
online) Entanglement connection between two links $A$-$B$
and $C$-$D.$ The ensemble $A$ $(C)$ is initially entangled
with $B$ $(D)$ as described by $|\psi_{ab}\rangle$
$(|\psi_{cd}\rangle).$ The memories $B$ and $C$ are
read-out and the resulting anti-Stokes photons are combined
on a beam splitter. The detection of a single photon after
the beam splitter, which could have come either from
location $B$ or $C,$ heralds the storage of a single
excitation ($s_a$ or $s_d$) in the ensembles $A$ and $D$
and projects them into an entangled state
$|\psi_{ad}\rangle.$} \label{DLCZswapping}
\end{figure}

Once entanglement has been heralded within each elementary
link, one wants to connect the links in order to extend the
distance of entanglement. This is done by successive
entanglement swapping between adjacent links with the
procedure shown in Fig. \ref{DLCZswapping}.

Consider two links $AB$ and $CD$ in which the ensembles
$A$-$B$ and $C$-$D$ respectively are entangled by sharing a
single excitation, as before. They are described by the
state $|\psi_{ab}\rangle\otimes|\psi_{cd}\rangle,$ where
$|\psi_{kl}\rangle$ are defined by the equation
(\ref{eq2}). The atomic excitations $s_b$ and $s_c$ that
are probabilistically stored in the ensembles $B$ and $C$
are read-out with a strong, resonant light pulse to be
converted back into ``Anti-Stokes''  photons associated to
the mode $b'$ and $c'$. These two modes are combined on a
beam-splitter and the measurement of a single photon, e.g.
in the mode $\frac{1}{\sqrt{2}}(b'+c')$ will project the
ensembles $A$ and $D$ into the entangled state
\begin{equation}
\label{eq4}
|\psi_{ad}\rangle=\frac{1}{\sqrt{2}}\left(s_a^\dagger
+s_d^\dagger
e^{i(\theta_{ab}+\theta_{cd})}\right)|0\rangle.
\end{equation}
By iterating successive entanglement swapping operations,
it is possible to establish entanglement between more and
more distant ensembles.

We now analyze the effect of nonunit detector efficiency
$\eta_d$ and memory efficiency $\eta_m$ on the entanglement
swapping procedure. The detectors can give the expected
click when two photons are stored in the memories $B$ and
$C$, but only one is detected. In this case, the created
state contains an additional vacuum component
\begin{equation}
\label{eq5} \rho_{ad}=\alpha_1
|\psi_{ad}\rangle\langle\psi_{ad}|+\beta_1|0\rangle\langle
0|
\end{equation}
where $\alpha_1=\frac{1}{2-\eta}$ and
$\beta_1=\frac{1-\eta}{2-\eta}.$ We have defined $\eta$ as
the product of the detector efficiency by the memory
efficiency $\eta=\eta_d\eta_m.$ The success probability for
the first swapping is given by
$P_1=\eta(1-\frac{\eta}{2}).$ Similarly, one can show that
the success probability for entanglement swapping at the
$(i+1)$th level is given by $P_{i+1}=\alpha_i \eta
(1-\frac{\alpha_i\eta}{2})$, where $\alpha_i$ is the weight
of the normalized entangled component in the state
associated to the level $i.$ It is connected to
$\alpha_{i-1}$ by
$\alpha_i=\frac{\alpha_{i-1}}{2-\alpha_{i-1}\eta}$. Using
this last formula, one can easily show that after $n$
nesting levels, the ratio
$\frac{\beta_n}{\alpha_n}=(1-\eta)(2^n -1)$. The relative
weight of the vacuum component thus increases linearly with
the number of elementary links $N=2^n$ composing the
quantum repeater. We will see in what follows that in
schemes where entanglement swapping is performed via
two-photon detections, the vacuum components remain
constant, cf. below.

%%%%%%%%%%%%%%
\subsubsection{Post-Selection of Two-Photon Entanglement}

\begin{figure}[hr!]
{\includegraphics[scale=0.32]{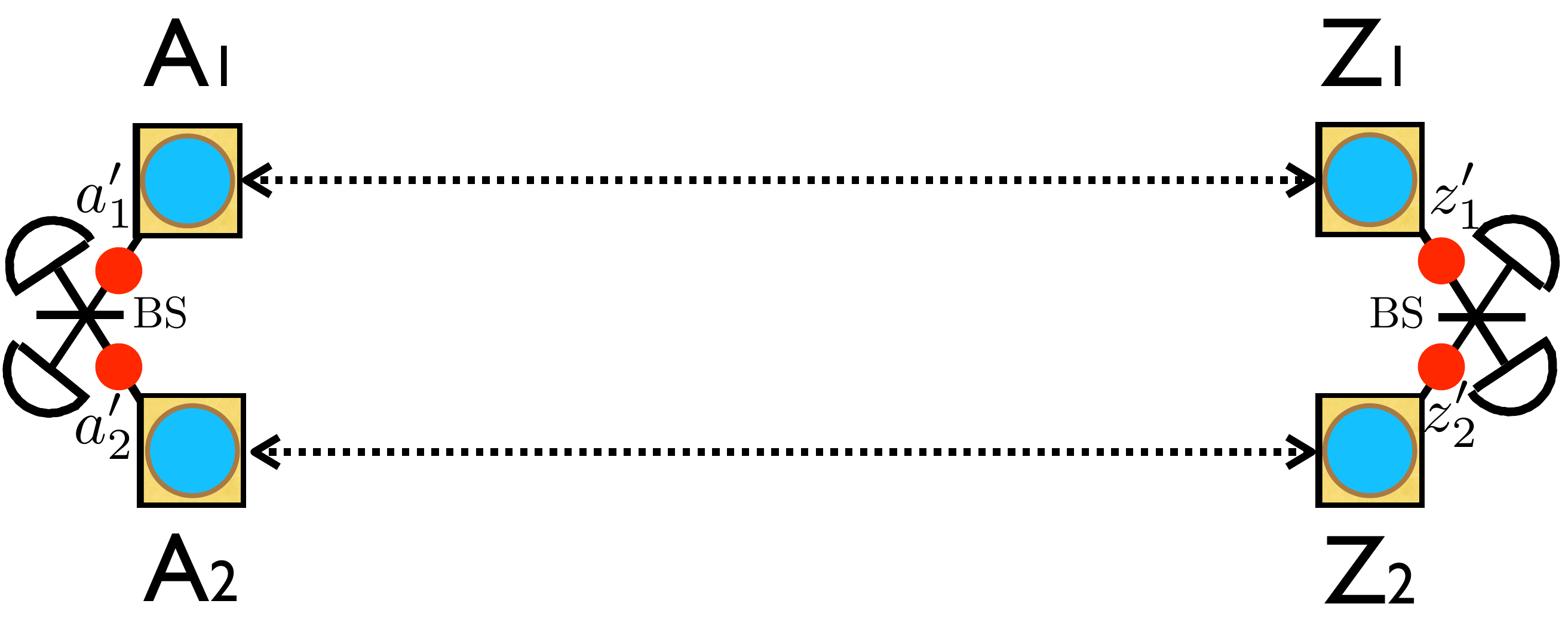}} \caption{(Color
online) Post-selection of two-photon entanglement.
Entanglement has been distributed independently within two
chains (labeled by the subscript 1 or 2) such that the
ensembles $A_1$-$Z_1$ and $A_2$-$Z_2$ share an
entanglement. The atomic excitations at the same location
are read-out and the emitted anti-Stokes photons are
combined into a beam-splitter and then counted.
Measurements in arbitrary basis can be done by changing the
beam-splitter transmission coefficients and phases.}
\label{fig4}
\end{figure}

Suppose that entanglement has been distributed over the
desired distance, say between locations $A$ and $Z.$ The
created entanglement, which consists of $A$ and $Z$ sharing
a single delocalized excitation, is of limited use on its
own, because it is difficult to perform measurements in any
basis other than that of the Fock states $|0\rangle$ and
$|1\rangle$. This is why in the DLCZ protocol the created
single-excitation entanglement is now used as a building
block for more directly useful two-photon entanglement.

One needs two ensembles at each location, labeled $A_1$
$(Z_1)$ and $A_2$ $(Z_2)$ for location $A$ $(Z).$
Entanglement between $A_1$ and $Z_1,$ and between $A_2$ and
$Z_2$ have been established as presented before, such that
we have the state $\frac{1}{2}
(a_1'^\dagger+e^{i\theta_1}z_1'^\dagger)(a_2'^\dagger+e^{i\theta_2}z_2'^\dagger)|0\rangle.$
The projection of this state onto the subspace with one
photon in each location is
\begin{equation}
\label{eq6}
|\Psi_{az}\rangle=\frac{1}{\sqrt{2}}\left(a_1'^\dagger
z_2'^\dagger+e^{i(\theta_2-\theta_1)} a_2'^\dagger
z_1'^\dagger\right)|0\rangle
\end{equation}
which is analogous to conventional polarization or time-bin
entangled states. The required projection can be performed
post-selectively by converting the atomic excitations back
into Anti-Stokes photons and counting the number of photons
in each location. Measurements in arbitrary basis are
possible by combining modes $a_1'$ and $a_2'$ (and also
$z_1'$ and $z_2'$) on beam splitters with appropriate
transmission coefficients and phases. The component
$|\Psi_{az}\rangle$ of the mixed state $\rho_{az}$
distributed after $n$ swapping operations is post-selected
with the probability $P_{ps}=\alpha_n^2\eta^2/2.$
%%%%%%%%%%%%%%%%%%%%%%%%%%%%
\subsection{Performance}

\label{DLCZ-Performance}
%%%%%%%%%%%%%%

%%%%%%%%%%%%%%
\subsubsection{Calculating the Entanglement Distribution Time}

The general formula for calculating the time required for a
successful distribution of an entangled state
$|\Psi_{az}\rangle$ is
\begin{equation}
T_{tot}=\frac{L_0}{c} \frac{f_0 f_1... f_n}{P_0 P_1...P_n
P_{ps}}. \label{Ttot}
\end{equation}
The first factor is the waiting time at the elementary
level, where $L_0=L/2^n$ is the length of the elementary
link, $L$ is the total distance, and $n$ is the ``nesting
level'' of the repeater, as introduced in section
\ref{Introduction}. Entanglement creation attempts for
elementary links only succeed with a probability $P_0$.
After every attempt, one has to wait to find out whether
the attempt has succeeded (whether there was a photon
detection in the central station). If not, the memory has
to be emptied, and one tries again. Assuming that the
repetition rate is not limited by the speed of the write
and read processes themselves, this leads to a basic period
of $\frac{L_0}{c}$, and also explains the factor $1/P_0$.
Furthermore the total time is inversely proportional to the
success probabilities at each level $P_i$, and to the
probability of successful post-selection at the end,
$P_{ps}$. The factors $f_0$ to $f_n$, which all satisfy $1
\leq f_i \leq 2$, take into account the fact that for every
$i$-th level swapping attempt one has to establish {\it
two} neighboring links at level $i-1$. This takes longer
than establishing a single such link by a factor $f_{i-1}$.
The precise values of these factors depend on the ensemble
of success probabilities up to the given level. No analytic
expression for them is known at this point for general $i$.
However, it is easy to show that $f_0=\frac{3}{2}$ for $P_0
\ll 1$ \cite{Collins2007,Brask2008}. Intuitively, if the
waiting time for a single link is $T$, one only has to wait
a time $T/2$ for a success in one of two neighboring links.
Then one still has to wait a time $T$ for the second link.
Moreover numerical evidence shows that setting
$f_i=\frac{3}{2}$ for all $i$ is a good approximation
\cite{Jiang2007,Brask2008}. We will use this approximation
in the following to calculate and compare entanglement
distribution times for different protocols. A more detailed
justification for Eq. (\ref{Ttot}) and a more detailed
discussion of the $f_i$ is given in Appendix
\ref{calculatetime}.

Plugging the expressions for the success probabilities
$P_0$, $P_i$ and $P_{ps}$ into Eq. (\ref{Ttot}), one finds
\begin{equation}
\label{evaluation_time} T_{tot}=3^{n+1}\frac{L_0}{c}
\frac{\prod_{k=1}^{n}\left(2^k-\left(2^k-1\right)\eta\right)}{\eta_d\eta_{t}p
\eta^{n+2}}.
\end{equation}

This still contains the pair emission probability $p$. The
value of $p$ is constrained by the fact that there is a
probability of order $p^2$ for the emission of two Stokes
photons into the desired mode (associated with the creation
of two atomic excitations), as discussed in section
\ref{DLCZ-Basic}. This leads to errors which reduce the
fidelity $F$ of the distributed state. For the present
protocol these errors grow approximately quadratically with
the number of links. To first order in $p$, $F$ is always
of the form $1-A_n p (1-\eta)$, with, for example $A_0=8,
A_1=18, A_2=56, A_3=204, A_4=788$. More details on the
multi-photon error calculation are given in Appendix
\ref{errors}. In the following we will assume that we can
tolerate a fidelity reduction $1-F=0.1$ due to multiphoton
errors. The maximum allowed value of $p$ for a given $n$
can then be determined directly from the above values for
$A_n$.

\subsubsection{Comparison to Direct Transmission}

We will now compare the entanglement distribution time for
the DLCZ protocol to the time for quantum state
distribution using direct transmission. We consider a fiber
attenuation of $0.2$ dB/km ($\eta_t=e^{-L_0/(2L_{att})},$
with $L_{att}=22$km) corresponding to typical telecom
fibers and telecom wavelength photons. We furthermore take
into account the reduced photon velocity within the fiber,
$c=2\times10^8$m/s. We assume equal memory and
photon-resolving detector efficiencies $\eta_m=\eta_d=0.9$.
This is certainly a demanding choice, however it is far
from the performance levels typically required for
fault-tolerant quantum computing, for example. We will
describe and discuss the experimental status quo in some
detail in section \ref{Implementations}. We optimize the
nesting level $n$ and thus the number of links $2^n$ for
each distance $L$.

For direct transmission we assume a single-photon source
with a repetition rate of 10 GHz, as we did in section
\ref{Introduction}. This will be our reference all through
this paper. This is certainly an ambitious value. It might
one day be achieved for single-photon sources based on
quantum dots in high-finesse semiconductor micro-cavities
\cite{Moreau2001,Santori2002}, for example, where lifetimes
can be of order 100 ps. However, for now these sources are
not very efficient. Moreover, the source would have to
operate at telecom wavelengths (i.e. around 1.5 $\mu$m)
\cite{Ward2005,Hostein2009}. Of course, any such choice is
somewhat arbitrary. However, it will become clear later on
that our conclusions don't depend very strongly on the
exact choice of rate for the reference source, essentially
because the scaling with distance is very different for
direct transmission and for quantum repeaters. As a
consequence, the curve corresponding to direct transmission
in our main comparison figure, Fig. \ref{comparison} in
section \ref{Comparison}, has a much steeper slope than the
curves corresponding to all the considered repeater
protocols, so that it intersects them all in the distance
range $L=500$ to 650 km. Changing the reference rate would
change the cross-over distances only slightly. The most
important question in the short and medium term is what
distribution rates a given repeater protocol can achieve in
that distance range.

For the DLCZ protocol, we find a cross-over point of
$L=630$ km, with an entanglement distribution time
$T_{tot}=340$ seconds at that distance, for $n=2$ (4
links). The corresponding value of $p=0.01$. Note that,
thanks to the very different scaling, for longer distances,
the repeater is much faster than direct transmission, for
example for 1000 km $T_{tot}=4100$ seconds, compared to
$10^{10}$ seconds for direct transmission. Nevertheless,
this result is somewhat disappointing. On the one hand, a
single entangled pair every 340 seconds is of course a very
low rate. Even more importantly, for the repeater to work,
the memory storage time has to be comparable to the
mentioned 340 seconds. In particular, it has to be long
enough for the final post-selection to be possible, i.e.
long enough to create two independent single-photon
entangled states over the whole distance. This is extremely
challenging. Briefly anticipating the detailed discussion
in section \ref{Implementations}, the best current results
for quantum memory times in DLCZ-type experiments with
atomic gases are in the few ms range
\cite{Zhao2009a,Zhao2009}. A storage time of order 1 s was
achieved in a solid state system for a memory protocol
based on electromagnetically induced transparency
\cite{Longdell2005}, though not yet at the quantum level.
Decoherence times as long as 30 seconds have been
demonstrated for the same kind of solid state system
(Pr:YSO) in \cite{Fraval2005}. Going even further and
really implementing the whole protocol at such timescales
is likely to be extremely challenging. There is thus strong
motivation to try to invent protocols that allow faster
generation of long-distance entanglement. Section
\ref{Improvements} is dedicated to various such proposals.

%%%%%%%%%%%%%%%%%%%%%%%%%%%%
\subsection{Discussion - Limitations}
\label{DLCZ-Limitations}

Motivated by the results in the last subsection, we now
describe several limitations of the DLCZ protocol, which
have become starting points for further developments that
will be described in section \ref{Implementations}.

(i) There is a trade-off between high fidelity of the
distributed state and high distribution rate. We have seen
that the errors due to multiple emissions from individual
ensembles grow quadratically with the number of elementary
links $N$. In order to suppress these errors, one then has
to work with very low emission probability $p$, which
limits the achievable rate. This quadratic growth of the
multi-photon errors is related to the fact that the vacuum
component in the created single-photon entangled state
grows linearly with $N$, cf. the discussion at the end of
Appendix \ref{errors}. In sections \ref{twophotonswap},
\ref{twophotongen} and \ref{IPP} we will see schemes where
the vacuum component remains constant, and, as a
consequence, the multi-photon errors grow only linearly
with $N$, thanks to the use of entanglement swapping
operations based on two photon detections instead of a
single photon detection. In section \ref{SPS} we review a
scheme where multi-photon errors are greatly reduced
through the use of single-photon sources, which can be
effectively realized with atomic ensembles.

(ii) The entanglement creation between two remote ensembles
requires interferometric stability over long distances. To
illustrate the challenge this represents, let us consider
an elementary link with $L_0=125$km. The entanglement in
equation (\ref{eq6}) depends on the phase
$\theta_2-\theta_1,$ the contribution to which from the
given elementary link can be rewritten as
$\big[\theta_{B_2}(t_2)-\theta_{B_1}(t_1)\big]-\big[\theta_{A_2}(t_2)-\theta_{A_1}(t_1)\big].$
We have defined $t_1$ ($t_2$) as the moment where the first
(second) single-photon entangled state Eq. (\ref{eq3}) was
created in the elementary link. The phase thus has to
remain stable over the time scale given by the mean value
of $t_2-t_1$ which is $L_0/(cP_0).$ For the considered
example, this gives $\langle t_2-t_1 \rangle=4.5$s. Over
such long time scales, both the phases of the pump lasers
and the fiber lengths are expected to fluctuate
significantly. This problem has to be addressed in any
practical implementation of the protocol, either through
active stabilization of the fiber lengths, or possibly
through the use of self-compensating Sagnac-type
configurations, cf. section V.F. The described problem
stems from the fact that in the DLCZ protocol long-distance
entanglement is generated via single photon detections. In
sections \ref{twophotongen} and \ref{IPP} we review schemes
where entanglement is instead generated via two-photon
detections, greatly reducing the stability requirements for
the channels.

(iii) We have argued before that in the DLCZ repeater
protocol one is a priori limited to a single entanglement
generation attempt per elementary link per time interval
$L_0/c.$ In sections \ref{P2M3} and \ref{Collins} we will
describe how this limitation can be overcome using memories
that can store a large number of distinguishable modes, cf.
below.

(iv) For long communication distances to be realistic, the
wavelength of the Stokes photons has to be in the optimal
range for telecom fibers (around 1.5 $\mu$m). This either
severely restricts the choice of atomic species or forces
one to use wavelength conversion techniques
\cite{Tanzilli2005}, which for now are not very efficient
at the single photon level, mostly due to coupling losses.
In section \ref{P2M3} we describe how this requirement can
be overcome by separating entanglement generation and
storage.

%%%%%%%%%%%%%%%%%%%%%%%%%%%%%%%%
\section{Improvements}
\label{Improvements}
%%%%%%%%%%%%%%%%%%%%%%%%%%%%%%%%

In this section we review various improvements over the
DLCZ protocol that have been proposed over the last few
years. We only discuss architectures that use essentially
the same ingredients, i.e. atomic ensembles, linear optics,
and photon counting, but that use them in different ways in
order to achieve improved performance. We have seen that in
the DLCZ protocol both entanglement generation and swapping
are based on a single photon detection. Subsection
\ref{twophotonswap} describes a protocol where entanglement
is swapped based on two photon detections, which leads to a
constant (rather than growing) vacuum component in the
created state, resulting in an improvement in the overall
entanglement distribution rate. Subsection
\ref{twophotongen} describes protocols where entanglement
is moreover generated based on two photon detections,
leading to enhanced robustness with respect to phase
fluctuations in the channel. Subsections \ref{P2M3} and
\ref{Collins} are devoted to multiplexing. Subsection
\ref{P2M3} reviews the idea of using memories that can
store multiple temporal modes. Such memories can be
realized using inhomogeneously broadened atomic ensembles
in certain solid-state systems. Their use in the present
context is made possible by the realization that a
DLCZ-type atomic ensemble can be emulated by combining a
photon pair source with a memory that can absorb and emit
photons. This approach promises a great enhancement in the
entanglement generation rate. Subsection \ref{Collins}
reviews work on spatial multiplexing, which would be even
more powerful than the temporal variety. Subsection
\ref{SPS} discusses a protocol based on single-photon
sources, which can be effectively implemented with atomic
ensembles, and which yields a significantly enhanced rate
compared to the DLCZ protocol. Subsection \ref{IPP}
describes protocols that are based on effectively
approximating ideal photon pair sources with atomic
ensembles, leading both to enhanced rates and greatly
enhanced robustness.
 %%%%%%%%%%%%%%%%%%%%%%%%%%%%%%%%
\subsection{Entanglement Swapping via Two-Photon Detections}
\label{twophotonswap}

In the previous chapter, it has been pointed out that
entanglement swapping based on single-photon detections
leads to the growth (linear with the number of links) of
vacuum components in the generated state, and to the rapid
growth (quadratic with the number of links) of errors due
to multiple emissions from individual ensembles. The vacuum
components are detected at the post-selection level and
thus reduce the achievable rate. The multi-photon errors
reduce the fidelity of the distributed states. In order to
suppress these errors, one has to work with very low
emission probabilities, further reducing the distribution
rate. One possible way of addressing this problem is the
use of entanglement swapping operations that are based on
two photon detections instead of a single detection
\cite{Jiang2007,Chen2007,Zhao2007}. It turns out that in
this case the vacuum component remains stationary under
entanglement swapping, and the multi-photon related errors
grow only linearly with the number of links. In the present
subsection we review the proposal of \cite{Jiang2007},
where the elementary entanglement is generated by
single-photon detections as in the DLCZ protocol, but the
entanglement swapping is based on two-photon detections. In
the next subsection we discuss the proposals of
\cite{Chen2007,Zhao2007} where the entanglement is
generated by two photon detections as well.

%%%%%%%%%%%%%%
\subsubsection{First-Level Entanglement Swapping}

\begin{figure}[hr!]
{\includegraphics[scale=0.32]{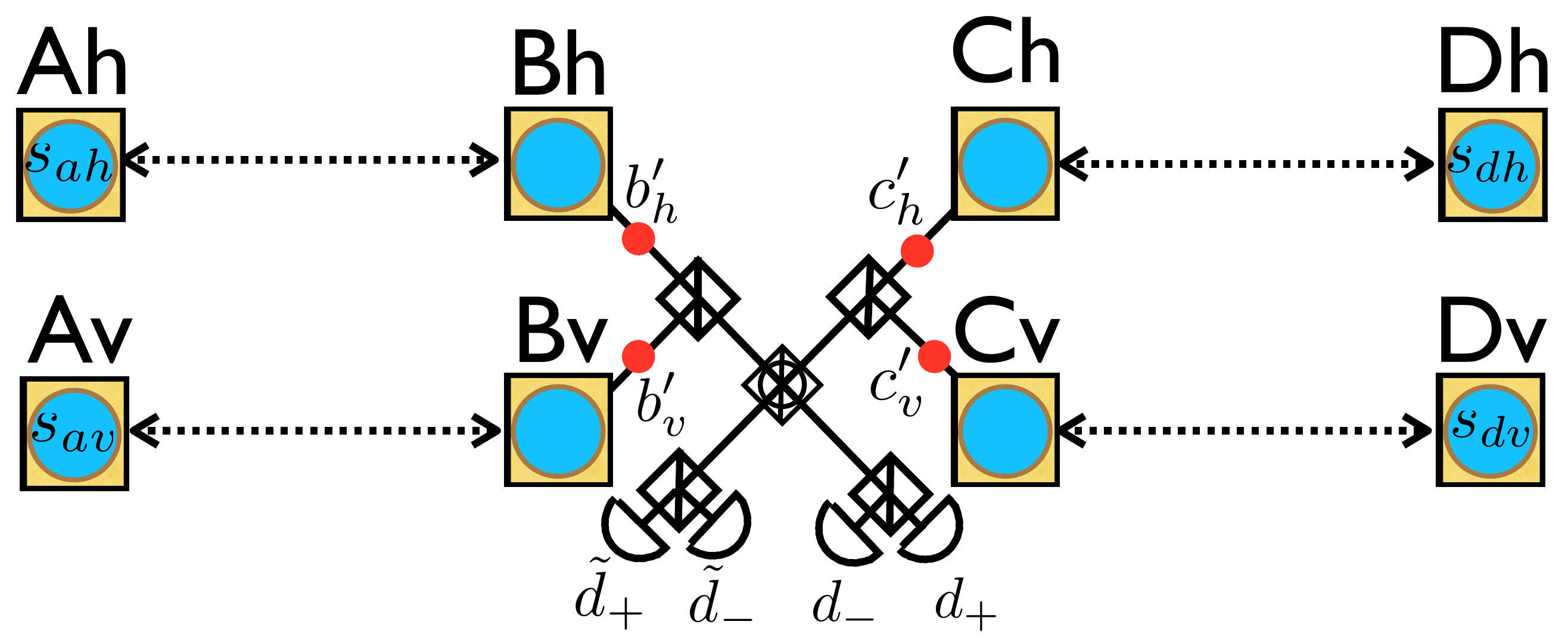}} \caption{(Color
online) First-level entanglement swapping based on a
two-photon detection. Long-distance entanglement between
the ensembles located at $A$ and $B$ ($C$ and $D$) is
created via single Stokes photon detections following the
DLCZ protocol, see Fig. \ref{DLCZcreation}. The subscript
$h$ or $v$ refers to horizontal or vertical polarization.
The spin-wave excitation stored in atomic ensembles $B_h,$
$B_v,$ $C_h,$ $C_v$ are read out and the corresponding
anti-Stokes photon in modes $b'_h,$ $b'_v,$ $c'_h,$ $c'_v$
are combined at a central station using the setup shown.
Vertical bar within squares label polarizing beam spitters
(PBSs) that transmit (reflect) $h$ $(v)$-polarized photons.
The central PBS with a circle performs the same action in
the $\pm45°$ basis. The coincident detection between the
modes $d_+$ and $\tilde{d}_+$ heralds the storage of two
excitations ($s_{ah}$-$s_{dh}$ or $s_{av}$-$s_{dv}$),
either in the ensembles $A_h$ and $D_h$ or in the ensembles
$A_v$ and $D_v$, cf. Eq. (\ref{eq11}).} \label{jiang}
\end{figure}

The elementary links in the protocol of \cite{Jiang2007}
have the same form as in the DLCZ protocol. Depending on
the nesting level, two distinct swapping operations are
performed. At the first swapping level, the principle of
the entanglement connection is shown in Fig. \ref{jiang}.
This requires two ensembles at each locations emitting
photons with well defined polarization : the horizontally
(vertically) polarized modes are produced from upper
(lower) atomic ensembles $A_h$ and $B_h$ ($A_v$ and $B_v$).
Suppose that the ensembles $A_h$ and $B_h$ ($A_v$ and
$B_v$) are entangled as in the DLCZ protocol, i.e. based on
the detection of a single Stokes photon at a central
station which could have been emitted by either of the two
ensembles. Further suppose that entanglement between $C_h$
and $D_h$ ($C_v$ and $D_v$) has also been heralded in the
same way. The average time for the entanglement creation of
these four links is $T_0=\frac{25}{12} \frac{1}{P_0}
\frac{L_0}{c}$, where $P_0=p \eta_d \eta_t$. The prefactor
$\frac{25}{12}$ can be obtained using the same methods as
in Appendix A for four variables instead of two. In order
to swap the entanglement toward the ensembles $A$ and $D,$
the spin-wave stored in the memories $B_h$-$B_v$ and
$C_h$-$C_v$ are readout and the emitted anti-Stokes modes,
labelled $b'_h$-$b'_v$ and $c'_h$-$c'_v,$ are combined at a
central station where they are detected in modes
$d_\pm=b'_h+b'_v\pm c'_h\mp c'_v$ and $\tilde{d}_\pm=\pm
b'_h \mp b'_v+ c'_h+ c'_v$ using the setup shown in Fig.
\ref{jiang}. In the ideal case, a twofold coincident
detection between $d_+$ and $\tilde{d}_+$ projects the
state of the two remaining spin-wave modes nondestructively
into
\begin{equation}
\label{eq11}
|\Psi_{ad}\rangle=\frac{1}{\sqrt{2}}\left(s^\dagger_{ah}s^\dagger_{dh}+s^\dagger_{av}s^\dagger_{dh}\right)|0\rangle.
\end{equation}
This operation thus allows one to exchange single spin-wave entanglement of the form (\ref{eq2}) with more standard two-particle entanglement of the form (\ref{eq11}). However, only four out of the 16 terms in the Schmidt decomposition of $|\psi_{ahbh}\rangle \otimes |\psi_{avbv}\rangle \otimes |\psi_{chdh}\rangle \otimes |\psi_{cvdv}\rangle$ have a contribution to the output state, the remainders being eliminated by projective measurement, reducing the success probability for entanglement swapping to $1/8.$\\
Taking into account nonunit detector efficiency and memory
recall, one can get the expected coincident detection when
more than two spin-waves are stored in the memories B and C
but only two are detected. In this case, the created state
contains additional terms including single spin-wave modes
and a vacuum component
\begin{eqnarray}
\label{eq12}
\rho_{ad}^1&=&\nonumber c_2^1|\Psi_{ad}\rangle\langle\Psi_{ad}|+ \\
& &\nonumber c_1^1\left(|s_{ah}\rangle\langle
s_{ah}|+|s_{av}\rangle\langle s_{av}|+|s_{dh}\rangle\langle
s_{dh}|+
|s_{dv}\rangle\langle s_{dv}|\right)+\\
& &c_0^1|0\rangle\langle0|,
\end{eqnarray}
where $c_2^1=\frac{\eta^2}{8P_1},$ $c_1^1=\frac{(1-\eta)\eta^2}{16P_1}$ and $c_0^1=\frac{(1-\eta)^2\eta^2}{8P_1}.$ We have introduced a superscript $1$ to label the level of the entanglement swapping. The probability for the successful preparation of this mixed state is $P_1=\frac{1}{8}\eta^2(2-\eta)^2.$ \\

%%%%%%%%%%%%%%
\subsubsection{Higher-Level Entanglement Swapping}

\begin{figure}[hr!]
{\includegraphics[scale=0.32]{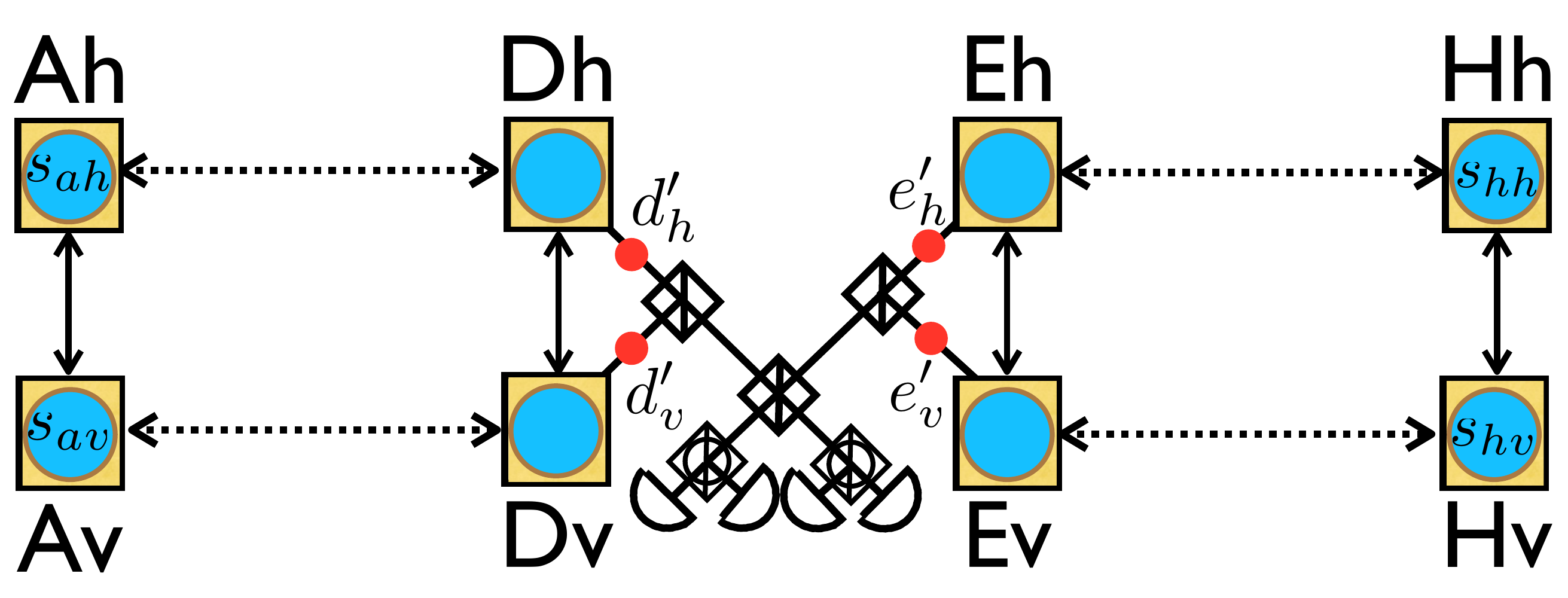}} \caption{(Color
online) Higher-level entanglement swapping based on a
two-photon detection. The ensembles located at $A$ and $D$
($E$ and $H$) are entangled based on the principle shown in
Fig. \ref{jiang} and are described by the state
$|\Psi_{ad}\rangle$ $(|\Psi_{eh}\rangle)$ (see eq.
(\ref{eq11})). The spin-wave stored in ensembles $D_h,$
$D_v,$ $E_h$ and $E_v$ are converted back into anti-Stokes
photons which are combined using the set of linear optics
shown. A twofold coincident detection between $d'_h+e'_v$
and $e'_h+d'_v$ nondestructively projects the ensembles $A$
and $H$ into the state $|\Psi_{ah}\rangle.$}
\label{2phswapping}
\end{figure}

For further distribution over longer distances of the
previous two spin-wave entangled state, one uses the setup
shown in Fig. \ref{2phswapping}. To illustrate the
higher-level swapping operations, suppose that two
spin-wave entangled states of the form (\ref{eq11}) are
distributed between ensembles $A$-$D$ and $E$-$H$ leading
to the state $|\Psi_{ad}\rangle\otimes|\Psi_{eh}\rangle.$
This entanglement can be swapped toward the ensembles
$A$-$H$ by combining two anti-Stokes photons at a central
station, where one photon is released from the $D$
ensembles, the other from the $E$ ensembles, and performing
a projective measurement onto the modes $d'_h\pm e'_v$ and
$e'_h\pm d'_v.$ The twofold coincident detection between
$d'_h+ e'_v$ and $e'_h+d'_v$ for example, collapses the two
remaining full memories into
$|\Psi_{ah}\rangle=1/\sqrt{2}(s^\dagger_{ah}
s^\dagger_{hh}+s^\dagger_{av} s^\dagger_{hv})|0\rangle.$

Using the same set of linear optics and detectors shown in
Fig. \ref{2phswapping}, one can perform successive
entanglement swapping operations, such that the state
$|\Psi_{az}\rangle$ can be distributed over the full
distance after $n$ swapping steps, between the locations
$A$ and $Z.$  Due to imperfect detection and memory
efficiency, the distributed state $\rho_{az}^{n}$ includes
single spin-wave and vacuum modes. One can show that their
weights $c_2^n,$ $c_1^n$ and $c_0^n$ are unchanged compared
to the weights $c_2^1,$ $c_1^1$ and $c_0^1.$ Indeed, the
condition for having a stationary state is
$c_0c_2=4(c_1)^2,$ which is fulfilled by $c_2^1,$ $c_1^1$
and $c_0^1.$ This is in contrast to the DLCZ protocol,
where the vacuum component is amplified (approximately
doubled) through every entanglement swapping operation, cf.
section \ref{DLCZ}. The success probability for the $i$th
entanglement connection is therefore given by
$P_i=2\eta^2(c_2^1/2+c_1^1)^2=\frac{\eta^2}{2(2-\eta)^2},$
for $i>1.$ The entangled component $|\Psi_{az}\rangle$ of
the distributed state $\rho_{az}^n$ can be post-selected
subsequently. The probability for such a successful
post-selection is
$P_{ps}=\eta^2c_2^n=\frac{\eta^2}{(2-\eta)^2}.$
%%%%%%%%%%%%%%

\subsubsection{Performance}

%The average time for a successful distribution of the state
%$|\Psi_{az}\rangle$ is given by
%\begin{equation}
%\label{eq13}
%T_{tot}=\left(\frac{3}{2}\right)^{n+1}\frac{L_0}{c}
%\frac{1}{P_0P_1\cdots P_nP_{ps}}
%\end{equation}
Using the expressions of $T_0,$ $P_1,$ $P_i$ for $n\geq
i>1$ and $P_{ps},$ one can write $T_{tot}$ as
\begin{equation}
\label{eq14} T_{tot}=\frac{50}{3} 3^{n-1}
\frac{L_0}{c}\frac{(2-\eta)^{2(n-1)}}{p\eta_t\eta_d\eta^{2n+2}},
\end{equation}
As in the DLCZ protocol, one has to take into account the
possible errors due to multiple-pair emissions within an
elementary link. The fidelity of the distributed state that
one wants, fixes the value for the success probability $p$
of the Stokes emission that one can use to estimate the
distribution rate based on eq. (\ref{eq14}). It is shown in
Ref. \cite{Jiang2007} that the errors grow only linearly
with the number of elementary links, whereas they grow
quadratically with the number of links when entanglement
connection is based on single-photon detections, as seen
before for the DLCZ protocol. This improved scaling is
related to the fact that the vacuum component is stationary
in the present protocol, since the errors in the final
state arise from the interaction of the vacuum and
multi-photon components, cf. Appendix \ref{errors}.

The protocol of \cite{Jiang2007} begins to outperform
direct transmission (with a 10 GHz single-photon source as
before) for a distance of 610 km, achieving an entanglement
distribution time of 190 seconds, cf. also Fig.
\ref{comparison} in section \ref{Comparison}. Here we
assume the same values of $\eta_m=0.9$ and $\eta_d=0.9$ as
before, and the same desired final fidelity $F=0.9$. The
optimum number of links for this distance is 4. This is
about a factor of 4 faster than the performance of the DLCZ
protocol, however it is clearly still a very long time for
creating a single entangled pair, which moreover is
probably still out of reach for realistic quantum memory
storage times. The advantage compared to the DLCZ protocol
is larger for longer distances, but of course the overall
entanglement distribution times are even longer.

The main reason why the improvement is so relatively modest
is that errors in the elementary link due to multiple
excitations still force one to work with low emission
probability $p$. The optimum value for 4 links is $p=0.037$
\cite{Jiang2007}, compared to $p=0.010$ for the DLCZ
protocol. Multiple excitations  are hard to detect in the
entanglement generation process (which is the same as in
the DLCZ protocol), because the corresponding Stokes
photons have to propagate far and are lost with high
probability. Moreover the entanglement generation based on
a single photon detection also leads to the phase stability
issues discussed at the end of the previous section. It is
then natural to consider changing the elementary link. This
is the topic of the next subsection.

\subsection{Entanglement Generation via Two-Photon Detections}
\label{twophotongen}

Simultaneously with the proposal by \cite{Jiang2007},
several schemes were proposed where not only the
entanglement swapping, but also the elementary entanglement
generation step is done via a two-photon detection
\cite{Chen2007,Zhao2007}. This approach was inspired by
earlier proposals for entanglement creation between distant
atoms or ions based on two-photon detections
\cite{Duan2003,Feng2003,Simon2003}. The main advantage of
generating entanglement in this way is that long-distance
phase stability is no longer required. In the protocols
discussed so far the detection of a single photon which
could have come from either of two distant locations
creates a single delocalized atomic excitation whose
entangled nature depends on the propagation phases of the
photon for the two possible paths. In contrast, in the
present case entanglement between two distant memories is
generated by projecting two photons, one coming from each
location, into an entangled state of their internal degrees
of freedom. This operation, and thus the created
long-distance entanglement, is insensitive to the
propagation phases of the two photons, which only
contribute an irrelevant global phase to the pair wave
function.

In this subsection we will focus on the protocol presented
in section III.B of \cite{Chen2007} because it is very
similar to the protocol of \cite{Jiang2007} and achieves a
better performance than the simpler protocol of
\cite{Zhao2007}, which we will also discuss briefly. The
protocol of section III.C of \cite{Chen2007}, which is
based on the local preparation of entangled pairs, followed
by two-photon entanglement generation and swapping, and its
improved version by \cite{Sangouard2008}, which achieve
significantly better performance while being equally
robust, are discussed separately in section \ref{IPP}.

%%%%%%%%%%%%%%
\subsubsection{Principle}
Interestingly, even though it was proposed simultaneously
and independently, the protocol of Ref. \cite{Chen2007}
section III.B can be presented as a simple variation of the
protocol of \cite{Jiang2007} discussed in the previous
subsection, in which the entanglement generation step of
\cite{Jiang2007} is performed locally and the first
entanglement swapping step of \cite{Jiang2007} is performed
remotely.

\begin{figure}[hr!]
{\includegraphics[width=\columnwidth]{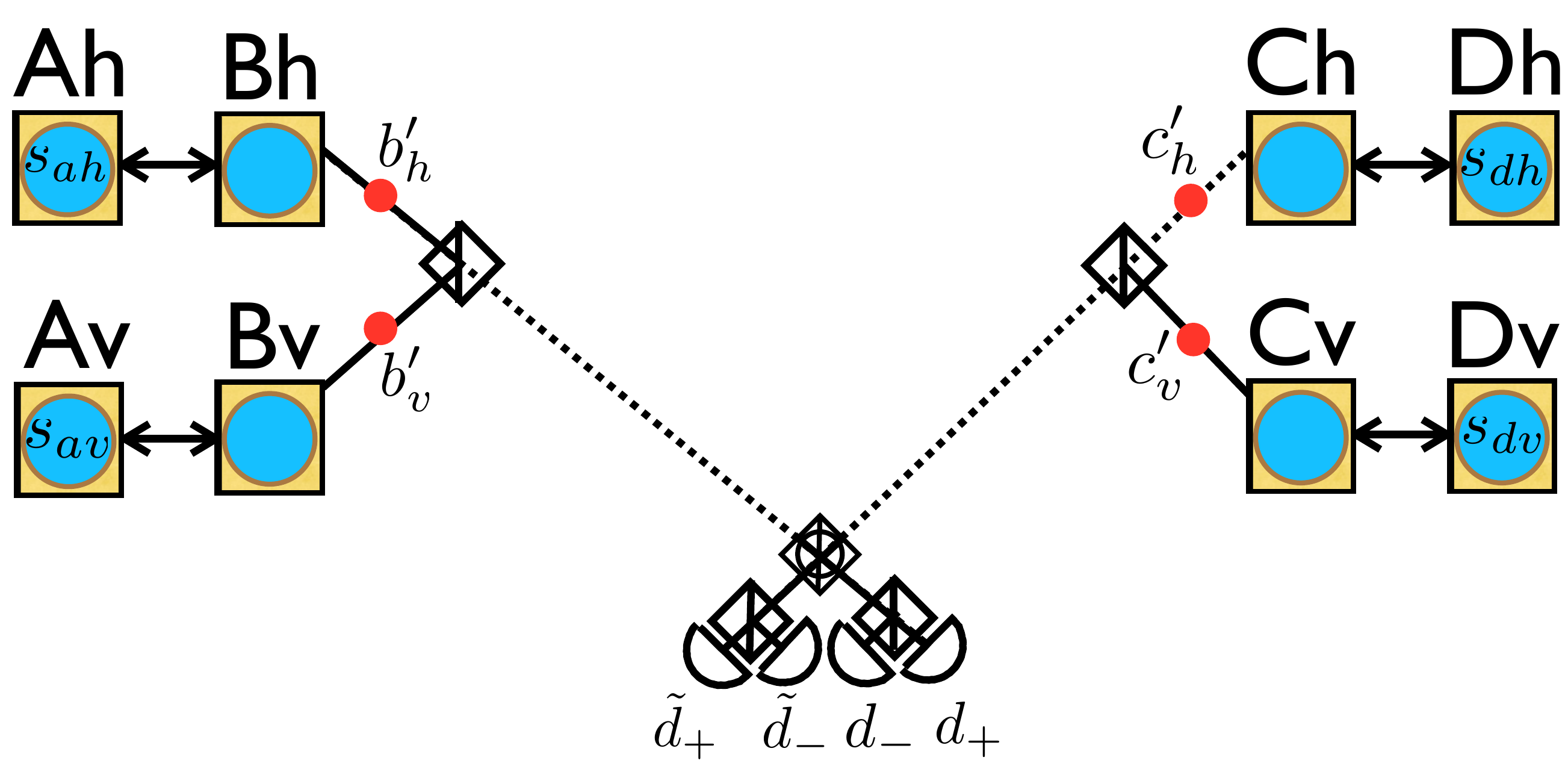}}
\caption{(Color online) Setup for entanglement creation
based on two-photon detection as proposed in Ref.
\cite{Chen2007} section III.B. The ensembles $A_h$ and
$B_h$, as well as $A_v$ and $B_v$, where all four ensembles
are located at the same node $AB$, are entangled two by two
as in the DLCZ protocol, cf. Fig. \ref{DLCZcreation}. The
ensembles $A_h$-$B_h$ ($A_v$-$B_v$) store a single
delocalized photon with horizontal (vertical) polarization.
In a similar way, the ensembles $C_h$ and $D_h$ ($C_v$ and
$D_v$), located at a different node $CD$, have been
entangled independently. The excitations stored in the
ensembles $B_h$, $B_v$, $C_h$, $C_v$ are read out and the
resulting photonic modes are combined at a central station
using the set-up shown. Ideally, the coincident detection
of two photons in ${d}_+$ and ${\tilde{d}}_+$ projects
non-destructively the atomic cells $A$-$D$ into the
entangled state $|\Psi_{\rm ad}\rangle$ of Eq.
(\ref{eq11}). } \label{2phcreation}
\end{figure}

At each node, one needs four ensembles, say $A_h$, $A_v$,
$B_h$ and $B_v$ at location $AB$ and $C_h$, $C_v$, $D_h$
and $D_v$ at location $CD$ as in Fig. \ref{2phswapping},
except that the $A$ and $B$ ensembles are close to each
other, but far from the ensembles $C$-$D.$ At each node,
the ensembles with identical subscripts are entangled by
sharing a single spin-wave excitation leading to the state
$|\psi_{ahbh}\rangle\otimes|\psi_{avbv}\rangle\otimes
|\psi_{chdh}\rangle\otimes |\psi_{cvdv}\rangle.$

The average waiting time for the creation of this product
state is approximatively given by $T_{prep}=\frac{25}{12}
\frac{1}{r P_s}$ with $r$ the repetition rate of the
elementary sources and $P_s=p\eta_d.$ The prefactor
$\frac{25}{12}$ is a very good approximation for $P_s \ll
1$. It can be obtained using the same methods as in
Appendix A for four variables instead of two. Note that
multiple emissions from the same ensemble are detected with
high probability, in contrast to the two previous
protocols, since the corresponding Stokes photon does not
propagate far and thus does not undergo significant losses.

Entanglement over the distance between $AB$ and $CD$ is now
generated using the setup shown in Fig. \ref{2phcreation},
i.e. by converting the atomic excitation stored in the $B$
and $C$ ensembles into anti-Stokes photons followed by the
detection in modes $d\pm$ and $\tilde{d}_\pm.$ Taking into
account imperfections of detectors and memories, the
twofold coincident detection $d_+$-$\tilde{d}_+$ projects
the state of the ensembles $A$ and $D$ into
\begin{eqnarray}
\label{eq16}
\rho_{ad}^0&=&\nonumber c_2^0|\Psi_{ad}\rangle\langle\Psi_{ad}|+ \\
& &\nonumber c_1^0\left(|s_{ah}\rangle\langle
s_{ah}|+|s_{av}\rangle\langle s_{av}|+|s_{dh}\rangle\langle
s_{dh}|+
|s_{dv}\rangle\langle s_{dv}|\right)\\
& &+c_0^0|0\rangle\langle0|;
\end{eqnarray}
where $c_2^0=\frac{1}{(2-\eta\eta_t)^2},$
$c_1^0=\frac{(1-\eta\eta_t)}{2(2-\eta\eta_t)^2}$ and
$c_0^0=\frac{(1-\eta\eta_t)^2}{(2-\eta\eta_t)^2}.$ The
probability for the successful preparation of this mixed
state is $P_0 = \frac{1}{8}\eta^2\eta_t^2(2-\eta\eta_t)^2.$

Figure \ref{2phswapping} shows how, using the same
combination of linear optical elements as for the previous
protocol, one can perform successive entanglement swapping
operations in order to distribute the state $\rho_{az}^n$
after $n$ swapping operations.
In analogy to above, the state $\rho_{az}^n$ includes vacuum and single spin-wave components with unchanged weights with respect to the initial ones, i.e. $c_2^i=c_2^0,$ $c_1^i=c_1^0$ and $c_0^i=c_0^0$ where the superscript $i$ refers to the $i$th swapping. The probability for the swapping operation to succeed is given by $P_i=2\eta^2(\frac{c_2^0}{2}+c_1^0)^2=\frac{\eta^2}{2(2-\eta\eta_t)^2}.$\\
Finally, one can perform a post-selection of the entangled component $|\Psi_{az}\rangle$ of the state $\rho_{az}^n.$ The probability for the successful post-selection is given by $P_{ps}=\eta^2 c_2^n=\frac{\eta^2}{(2-\eta\eta_t)^2}.$\\

%The average time for the distribution of the state
%$\Psi_{az}$ is given by
%\begin{equation}
%\label{eq17}
%T_{tot}=\left(\frac{3}{2}\right)^{n}\frac{L_0}{c}
%\frac{1}{P_0P_1\cdots P_nP_{ps}}.
%\end{equation}

\subsubsection{Performance}

Taking into account the expressions of $P_0,$ $P_i$ with
$n\geq i\geq 1$ and $P_{ps},$ one can write $T_{tot}$ as
\begin{equation}
T_{tot}=8\times3^{n}\times\frac{L_0}{c}\frac{(2-\eta\eta_t)^{2n}}{\eta_t^2\eta^{2n+4}}.
\label{chentime}
\end{equation}
For this formula to be strictly valid, the time $T_{prep}$
required to prepare entanglement between local ensembles
has to be negligible compared to the communication time,
i.e. in our case $T_{prep}=\frac{25}{12rp\eta_d} \ll
\frac{L_0}{c}.$ Otherwise, one has to replace
$\frac{L_0}{c}$ by $\frac{L_0}{c}+T_{prep}.$ For a
realistic source repetition rate of 10 MHz, preparation
time and communication time become comparable for
$p=10^{-3}$. The authors did not quantify the multi-photon
errors in the protocol in detail, making it difficult to
say for which link number this value of $p$ is attained.
Unfortunately the results of \cite{Jiang2007} on the
multi-photon errors cannot be taken over directly because
the strong photon loss corresponding to long-distance
propagation intervenes at different stages in the two
protocols, even though they are otherwise formally
equivalent. For the following estimate we take the simple
formula Eq. (\ref{chentime}), which gives a lower bound for
the entanglement distribution time. Since two photons have
to reach the central station, the square of the
transmission $\eta_t$ intervenes in this formula, making
the distribution time more sensitive to losses in the
elementary link and thus favoring more and shorter links
compared to the DLCZ protocol and the protocol of
\cite{Jiang2007}. We limit the total number of links to 16
in order to stay in a regime where it is reasonably
plausible that entanglement purification may not be
required. (It is worth noting that increasing the link
number improves the rate by less than a factor of 2 in the
distance range that we are focusing on.) We choose the same
detection and memory efficiencies, $\eta_m=\eta_d=0.9$, as
before. With all the mentioned assumptions, the protocol
starts to outperform direct transmission (with a 10 GHz
source, as before) for a distance of 640 km, achieving an
entanglement distribution time of 610 seconds, see also
Fig. \ref{comparison} in section \ref{Comparison}. The
performance in terms of rate is thus comparable to (but
slightly worse than) for the DLCZ protocol.

One important reason for the long time required is that
excess photon emissions (three or four photons) in the
long-distance entanglement generation step typically remain
undetected due to large fiber losses. As a consequence, the
generated state has large vacuum and single-photon
components, which lead to small success probabilities for
the subsequent swapping steps, and thus to a rather low
overall entanglement distribution rate. This makes it very
difficult to really profit in practice from the main
advantage of the protocol, which is its increased
robustness with respect to phase fluctuations in the
fibers. We will see below that the protocols discussed in
section \ref{IPP} have the same advantage in robustness
while achieving much faster entanglement distribution.

\subsubsection{Other Protocol}

As mentioned before, another protocol based exclusively on
two-photon detections was proposed by \cite{Zhao2007}
simultaneously with the work by \cite{Chen2007}. In the
scheme of Ref. \cite{Zhao2007}, entanglement is directly
generated over long distances, without a preceding local
DLCZ-type step. Since only a small excitation probability
can be used for each entanglement generation attempt in
order to avoid multiphoton errors, and since after each
attempt one has to communicate its success or failure over
a long distance, the required entanglement generation time
becomes significantly longer than for the DLCZ protocol. In
fact, since the success probability for every entanglement
generation attempt is proportional to $p^2$ (where $p$ is
the emission probability as before), the entanglement
distribution time for this protocol is about a factor of
$\frac{1}{p^2}$ longer than for the protocol of section
\ref{IPP}. For typical link numbers $p$ has to be smaller
than $0.01$ in order to avoid multi-photon errors,
resulting in a factor of at least $10^4$ between these two
protocols, cf. also Fig. \ref{comparison} in section
\ref{Comparison}.

So far we have seen that using improved protocols compared
to the original DLCZ proposal it is possible to achieve
moderately faster entanglement distribution or to eliminate
the need for interferometric stability. However the
achievable rates are still far too low. The next two
sections are devoted to multiplexing, an approach which
holds great promise for overcoming this key difficulty.

\subsection{Photon Pair Sources and Multimode Memories}
\label{P2M3}

In this subsection we review an approach towards
multiplexing \cite{Simon2007} that starts from the
realization that a DLCZ-type atomic ensemble can be
emulated by the combination of a photon pair source and a
quantum memory that can absorb and re-emit photons. By
itself, this has the advantage of allowing greater
wavelength flexibility for the memory compared to the DLCZ
situation where the Stokes photon has to be emitted at a
telecom wavelength. If the memory furthermore has the
capacity of storing and re-emitting light in a (possibly
large) number of different temporal modes, the approach
described below promises greatly improved entanglement
distribution rates. The implementation of such memories is
within reach in certain solid-state atomic ensembles, as
will be discussed in more detail in section
\ref{Implementations}.D.

\subsubsection{Separation of Entanglement Generation and
Storage}

The basic element of all the protocols discussed so far is
an ensemble of three-level atoms that is coherently excited
in order to generate a Stokes-photon by Raman scattering,
heralding the storage of an atomic spin excitation which
can later be reconverted into an anti-Stokes photon.
Depending of the protocol, either the Stokes photon or the
anti-Stokes photon is used to create entanglement between
remote memories. One of these photon therefore has to
propagate over long distances and we want its wavelength to
match the telecom wavelengths where the fiber attenuation
is small (around 1550 nm). This gives a significant
constraint on the operating wavelength of the memory. None
of the quantum memories that have been demonstrated so far
work at this wavelength. Possible technological solutions
include the use of wavelength conversion, which however so
far is not very efficient at the single photon level
\cite{Tanzilli2005} (primarily due to coupling losses), or
the use of Erbium-doped crystals as quantum memories, where
first experimental \cite{Staudt2006,Staudt2007,Staudt2007a}
and theoretical \cite{Ottaviani2009} investigations have
been performed, but implementing the DLCZ protocol is still
a distant and uncertain prospect.

\begin{figure}
{\includegraphics[width=\columnwidth]{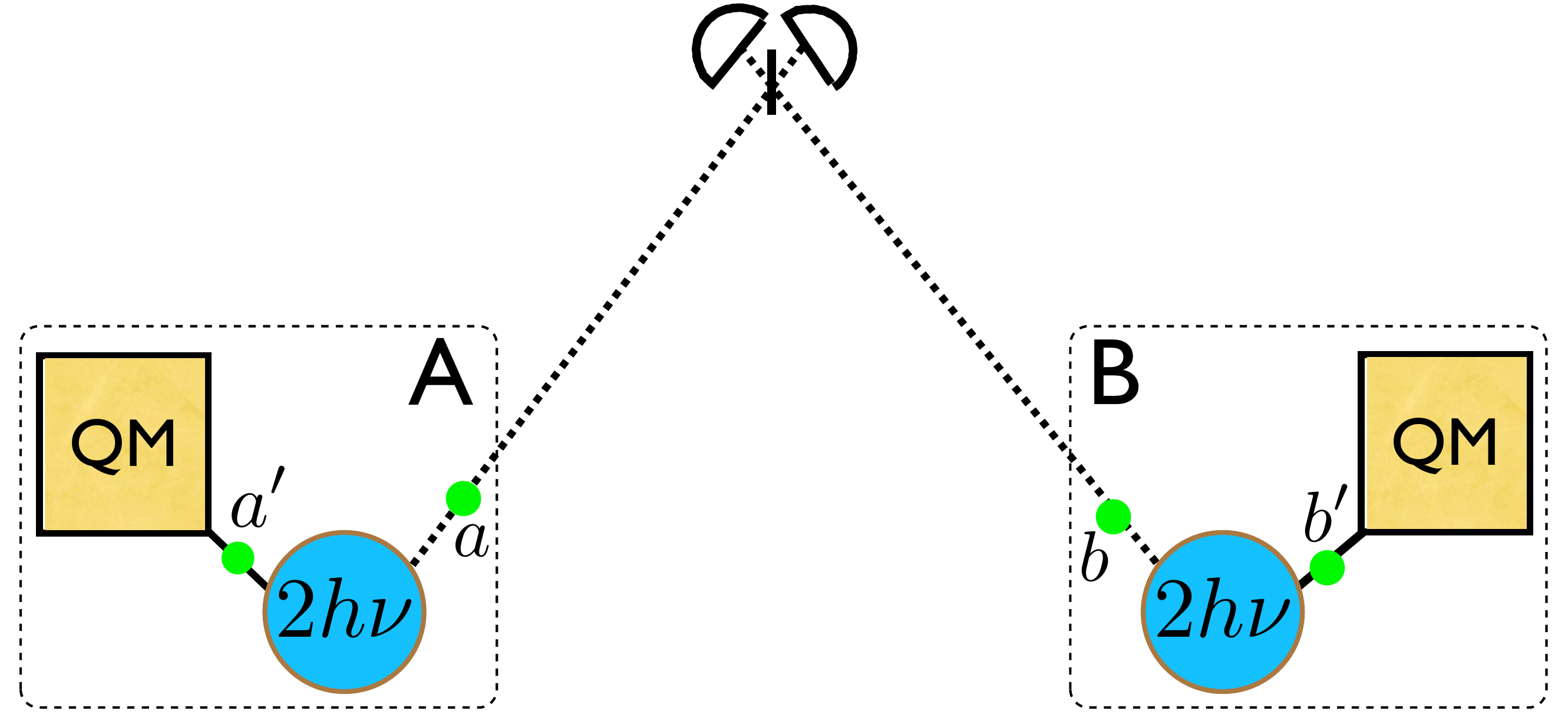}}
\caption{(Color online) Separation between entanglement
creation and storage using photon pair sources and
absorptive quantum memories. Brown circles represent
sources emitting photon-pairs in modes $a$-$a'$ for
location $A$ and in modes $b$-$b'$ for location $B.$ The
prime modes $a'$ and $b'$ are stored in neighboring quantum
memories (yellow squares) whereas the mode $a$ and $b$ are
combined on a beam splitter (vertical bar) at a central
station such that the detection of a single photon in one
of the output modes heralds the entanglement between the
quantum memories in $A$ and $B.$} \label{P2M3creation}
\end{figure}

A different approach for long-distance entanglement
creation was proposed in \cite{Simon2007}. It combines pair
sources and absorptive memories to emulate the DLCZ
protocol, cf. Fig. \ref{P2M3creation}. The basic procedure
for entanglement creation between two remote locations $A$
and $B$ requires one photon-pair source and one memory at
each location. The sources are simultaneously and
coherently excited such that each of them has a small
probability $p/2$ to emit a pair, corresponding to the
state
\begin{equation}
\label{eq18} \big[1+\sqrt{\frac{p}{2}}\left(a^\dagger
a'^\dagger + b^\dagger
b'^\dagger\right)+O(p)\big]|0\rangle.
\end{equation}
here $a$ and $a'$ ($b$ and $b'$) are two modes,
corresponding e.g. to two different directions of emission,
cf. Fig. \ref{P2M3creation}. The $O(p)$ term describes the
possibility of multiple pair emissions. It introduces
errors in the protocol, implying that $p$ has to be kept
small, in analogy with the DLCZ protocol. The modes $a'$
and $b'$ are stored in local memories whereas the modes $a$
and $b$ are combined on a beam splitter at a central
station. The modes $a$ and $b$ should thus be at telecom
wavelength, but there is no such requirement for the modes
$a'$ and $b'$. Similarly to the entanglement creation in
the DLCZ protocol, the detection of a single photon after
the beam splitter heralds the storage of a single photon in
the memories $A$ and $B,$ leading to the state (\ref{eq3}).
Note that we have set the phases to zero for simplicity.
The entanglement can be extended to longer distances by
successive entanglement swapping as in the DLCZ protocol.
The required photon pair sources could be realized with
atomic ensembles. For example, \cite{Chaneliere2006} have
proposed to use a specific atomic cascade in Rb for which
the first photon has a wavelength of 1.53 $\mu$m. There are
also convenient ways of implementing pair sources not based
on atomic ensembles, notably parametric down-conversion
\cite{Burnham1970,Hong1985,Hong1987,Wu1986} in non-linear
optical crystals, which allows a lot of wavelength
flexibility. Pair sources can also be realized based on the
DLCZ protocol, by applying the write and read pulses with a
small time interval or even simultaneously, cf. section
\ref{Implementations}.

%%%%%%%%%%%%%%%%%%%%%%%%%%%%%%%%

\subsubsection{Protocol with Temporal Multimode Memories}

\begin{figure}
{\includegraphics[width=\columnwidth]{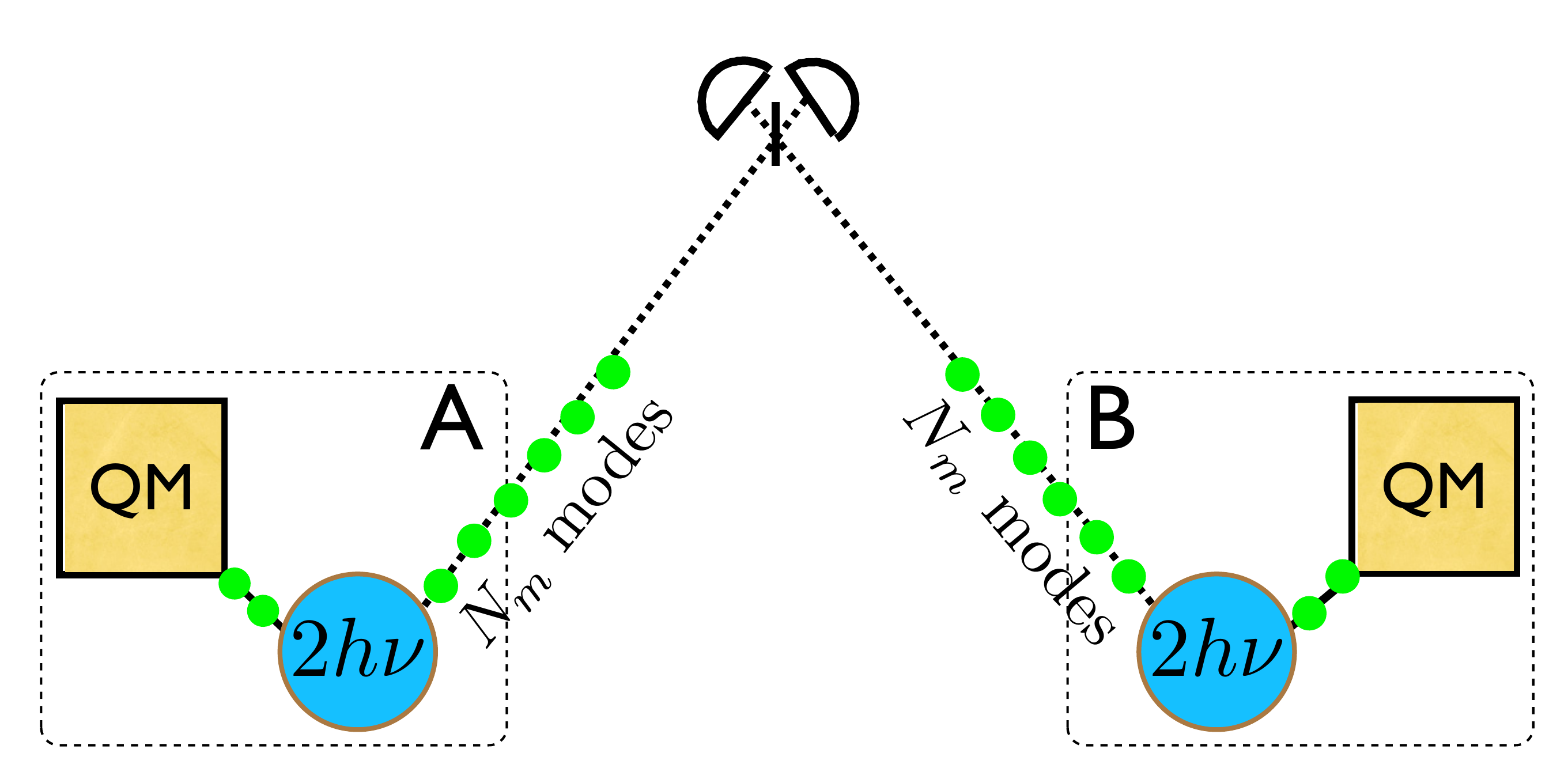}}
\caption{Entanglement creation with temporal multimode
memories. The source can be triggered a large number of
times in every communication time interval $\frac{L_0}{c}$.
One mode from each pair is sent to the central station, the
other one is stored in the multimode memory.}
\label{creationMMM}
\end{figure}

In section \ref{DLCZ-Limitations} we pointed out that in
the DLCZ protocol one is a priori limited to a single
entanglement generation attempt per communication time
interval $\frac{L_0}{c}$, because the memories have to be
emptied after every unsuccessful attempt. The same is true
for the protocols of sections \ref{twophotonswap} and
\ref{twophotongen}. The architecture described in the
present section is particularly well adapted for temporal
multiplexing, which overcomes this limitation, cf. Fig.
\ref{creationMMM}. If the memories can store not only one
mode but a train of pulses, one can trigger the sources
many times per communication time $L_0/c,$ potentially
creating pairs into modes $a_i$ and $a'_i$ ($b_i$ and
$b'_i$), where $i$ ($i=1,...,N_m$) labels the respective
``time-bin'', and $N_m$ is the total number of temporal
modes. All the modes $a'_i$ and $b'_i$ are stored in the
respective memories at $A$ and $B$. Any of the modes $a_i$
or $b_i$ can now give rise to a detection after the central
beam splitter. This leads to an increase of the
entanglement generation probability $P_0$ by a factor of
$N_m$ (for $N_m P_0 \ll 1$), which directly translates into
an increase of the entanglement distribution rate by the
same factor. The speed-up is thus achieved at the most
elementary level, that of entanglement generation. As a
consequence, the same principle could also be applied to
other quantum repeater protocols, although the
technological challenges vary depending on the protocol,
cf. below.

\begin{figure}
{\includegraphics[width=\columnwidth]{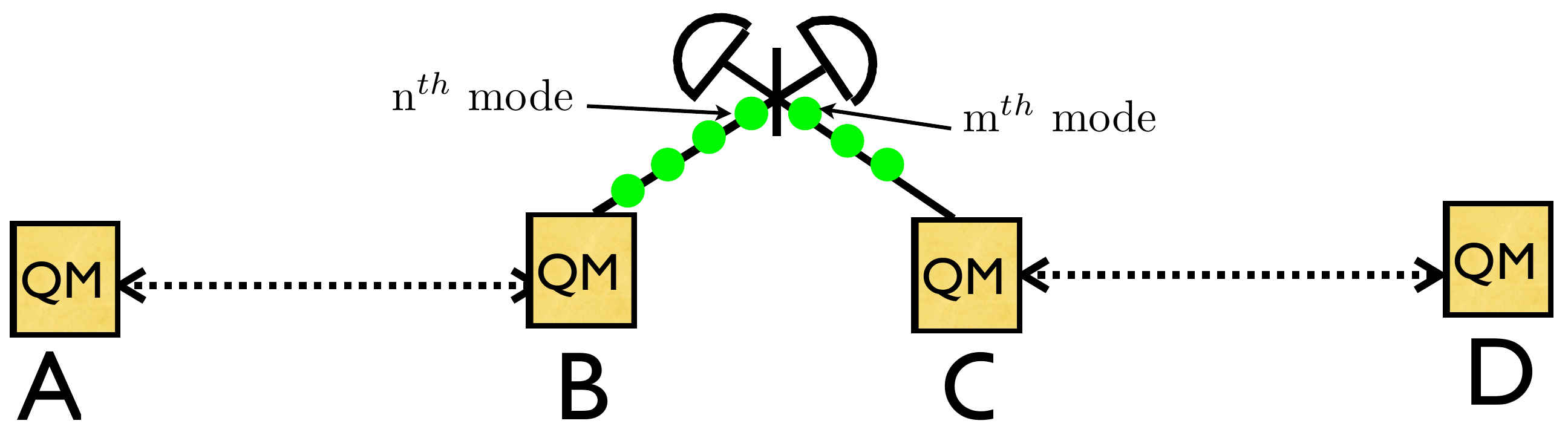}}
\caption{Entanglement swapping with temporal multimode
memories. One has to be capable of recombining at the beam
splitter exactly those modes whose partners have
participated in a successful entanglement generation at the
lower level.} \label{swapMMM}
\end{figure}

In order to do entanglement swapping using multimode memories, one
has to be able to recombine exactly those modes, whose partners
have given rise to a detection, and thus a successful entanglement
generation, in the respective links, cf. Fig. \ref{swapMMM}. If
this is ensured, entanglement swapping can again proceed in
analogy with the DLCZ protocol. Temporal multimode memories with
the required characteristics can be realized for example based on
the photon echo principle in inhomogeneously broadened solid-state
atomic ensembles (notably in rare-earth ion doped crystals). A
particularly promising approach towards the efficient realization
of such memories is based on ``atomic frequency combs''
\cite{Afzelius2009a}. This is discussed in more detail in section
\ref{Implementations}.D. In addition to the requirements on memory
efficiency and storage time discussed previously, an important
characteristic for such a multimode memory is its bandwidth, since
this may limit the number of modes that can be stored in a given
time interval $\frac{L_0}{c}$, even if the memory is in principle
capable of storing more modes. This is not a major limitation for
the present protocol. In \cite{Afzelius2009a} it is argued, for
example, that a memory based on an Eu-doped crystal with a
bandwidth of 12 MHz (limited by hyperfine transition spacings in
Eu) would be capable of storing a train of $N_m=100$ pulses with a
total length of 50 $\mu$s, which is still much shorter than the
communication time for a typical link length $L_0=100$ km, which
is of order 500 $\mu$s, taking the reduced speed of light in the
fiber into account. Note that a first experimental demonstration
of an interface with $N_m=32$ was recently performed in a Nd-doped
crystal \cite{Usmani2009}. Assuming $N_m=100$, and
$\eta_m=\eta_d=F=0.9$ as before, the protocol of \cite{Simon2007}
starts to outperform direct transmission (assuming the usual 10
GHz single-photon source) for a distance of 510 km, achieving an
entanglement distribution time of 1.4 seconds, using a repeater
architecture with 4 elementary links, see also Fig.
\ref{comparison} in section \ref{Comparison}. This is a
significantly improved rate compared to the previous sections.
Moreover this timescale is also much more compatible with
realistically achievable quantum memory times
\cite{Longdell2005,Fraval2005}, as mentioned before and discussed
in more detail in section \ref{Implementations}.

The present protocol closely follows the original DLCZ
protocol, in particular relying on entanglement generation
via a single photon detection, leading to similar phase
stability issues, even though they are somewhat reduced by
the shorter timescale of entanglement distribution. It is
natural to ask whether it is possible to implement
multimode versions of more robust protocols, such as the
ones of section \ref{twophotongen} or the one (still to be
discussed) of section \ref{IPP}. This may well be possible.
It is however more challenging than for the present
protocol, mainly because the fastest robust protocols rely
on the local preparation of stored single-photon (section
\ref{twophotongen}) or two-photon (section \ref{IPP})
entanglement. Such preparation requires a lot of
repetitions, which reduces the time available for temporal
multiplexing per time interval $\frac{L_0}{c}$.

\subsection{Spatially Multiplexed Memories}
\label{Collins}

\begin{figure}
{\includegraphics[width=\columnwidth]{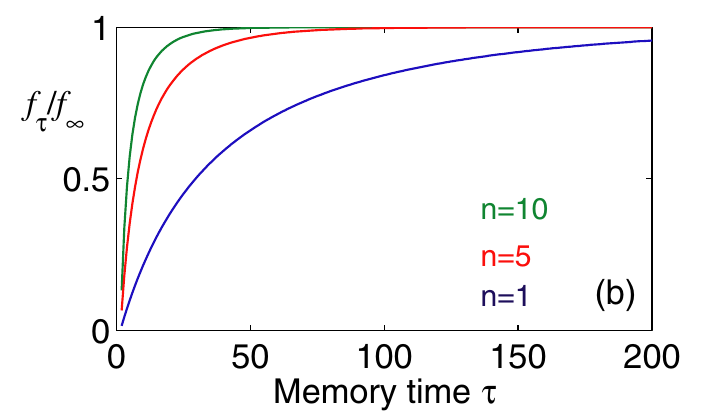}}
\caption{(Color online) Reduced memory time requirements
thanks to spatial multiplexing. Here $n$ is the number of
multiplexed memories per site. Memory time requirements for
strictly parallel operation always correspond to $n=1$.
Figure from Ref. \cite{Collins2007}.} \label{collins}
\end{figure}

For temporal multimode memories as described in the
previous section, the gain in entanglement distribution
time is due to the lowest level of the repeater protocol,
as $P_0$ is enhanced by a factor of $N_m$. All higher
levels of the protocol are unchanged with respect to the
protocols without multiplexing. Ref. \cite{Collins2007},
which was published before Ref. \cite{Simon2007}, studies a
more far-reaching form of multiplexing, which might be
possible in particular in the spatial domain. They envision
a situation where several different sub-ensembles of the
DLCZ type can be addressed completely independently. In
particular, photons can be retrieved independently from
each sub-ensemble and combined at will. Most importantly,
even when some of the sub-ensembles have been ``filled'',
i.e. entangled atomic excitations that involve these
sub-ensembles have been created, one can use others that
are still empty to make new attempts at entanglement
creation. There is no known way of implementing such a
step-by-step ``accumulation'' of stored entanglement in the
temporal multimode case. The temporal multimode memories
discussed in the previous subsection, can be charged only
once (although with a large number of modes), then they
have to be read out before being useful again, cf. section
\ref{Implementations}.

Ref. \cite{Collins2007} compares such strongly multiplexed
repeaters to the case where $N_r$ completely independent
repeater architectures are used in parallel (for which the
rate enhancement is exactly equal to $N_r$, of course). They
find a moderate advantage in terms of rate for the strongly
multiplexed case, without explicitly quantifying the
advantage. Note that \cite{Jiang2007} address the same
question in section V of their paper and find a scaling of
the entanglement distribution rate with $N_r^{1.12}$ instead
of $N_r$, which is consistent with the modest improvement
found by \cite{Collins2007}. However, \cite{Collins2007}
also show that there is a very significant advantage for
the multiplexed approach in terms of the necessary memory
time. Whereas for strictly parallel repeaters the necessary
memory times are determined by the waiting times for each
repeater individually (and thus are extremely long), the
multiplexed architecture leads to greatly reduced
requirements on the storage times, cf. Figure
\ref{collins}. First experimental efforts towards spatial
multiplexing are described in section
\ref{entanglementphotonexcitation}. Further theoretical
work includes \cite{Vasilyev2008,surmacz-2007}. The ideal
in the long run would clearly be to combine temporal and
spatial multiplexing in the same system, in order to
maximize potential quantum repeater rates. Multiplexing of
different completely independent frequency channels is
another attractive possibility, in particular for the
inhomogeneously broadened solid-state ensembles that are
being investigated in the context of temporal multimode
storage.

\subsection{Single-Photon Source based Protocol}
\label{SPS}

In section \ref{DLCZ} and Appendix \ref{errors} it was
shown that multi-photon emission events impose significant
limitations on the performance of the DLCZ protocol.
Motivated by this fact, \cite{Sangouard2007} suggested a
protocol based on single-photon sources, which makes it
possible to eliminate such errors. The protocol was
conceived for ideal single-photon sources. However, a good
approximation of such a source can be implemented with
atomic ensembles. The resulting scheme leads to a
significantly improved entanglement distribution rate
compared to the DLCZ protocol. In the following we first
describe the ideal protocol and its performance, then we
discuss how to implement a single-photon source to good
approximation with atomic ensembles.

%%%%%%%%%%%%%%
\subsubsection{Principle}

\begin{figure}[hr!]
{\includegraphics[scale=0.32]{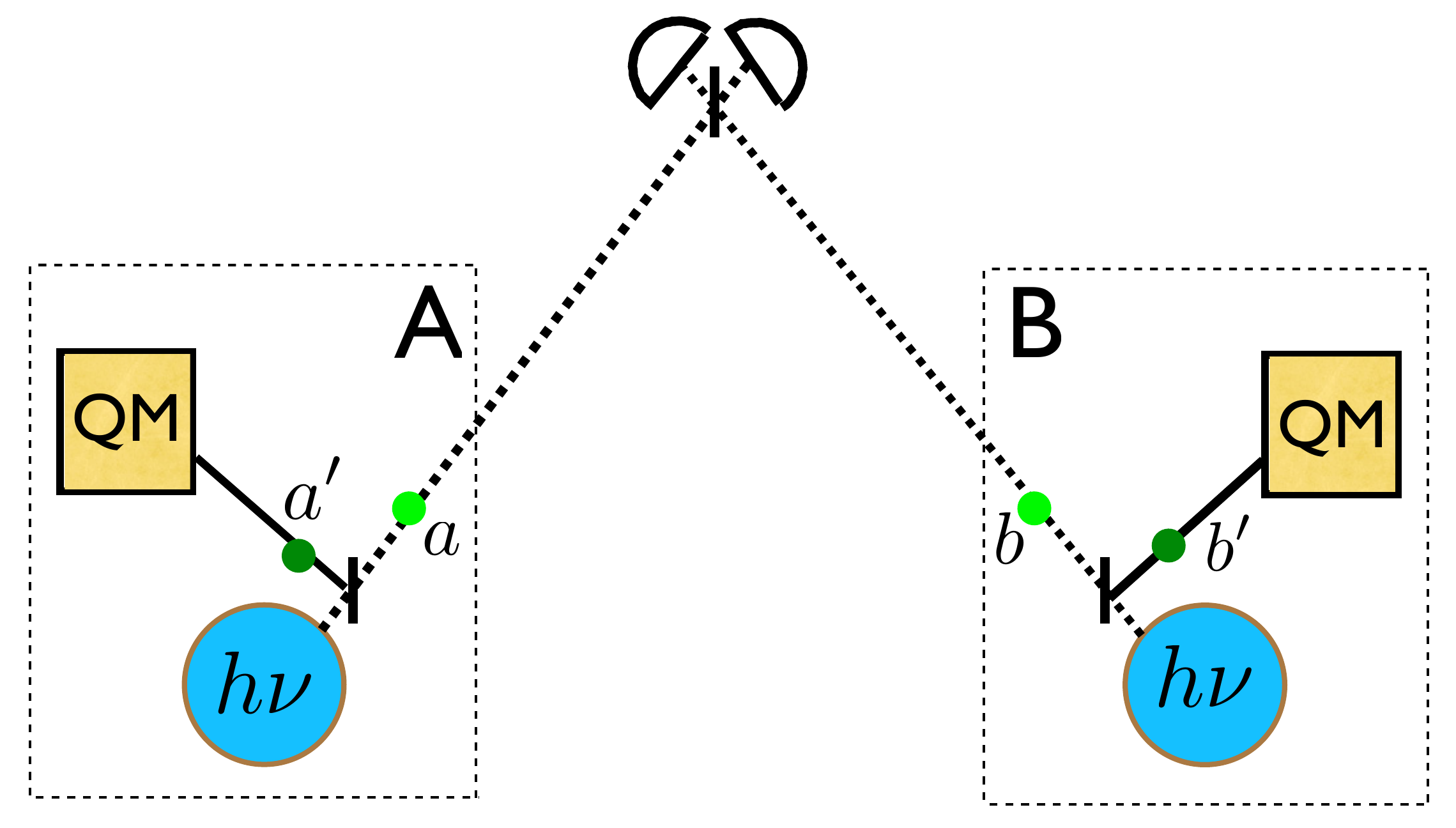}} \caption{(Color
online) Entanglement creation between two remote ensembles
located at $A$ and $B$ based on single-photon sources. Blue
circles represent single-photon sources. A single-photon is
generated at each location and sent through an asymmetric
beam splitter with small transmission and high
reflectivity, leading to a superposition of modes $a$ and
$a'$ ($b$ and $b'$) at location $A$ ($B$). The modes $a'$
and $b'$ are stored in local memories whereas $a$ and $b$
are sent to a central station where they are combined on a
50/50 beam splitter. The detection of a single photon at
the central station heralds the storage of the second one
within the memories with high probability, due to the
asymmetry of the local beam splitters. This creates the
entanglement of the two remote memories which share a
single excitation.} \label{SPS-setup}
\end{figure}

The architecture of the scheme proposed in Ref.
\cite{Sangouard2007} is represented in Fig.
\ref{SPS-setup}. The locations $A$ and $B$ contain each one
single-photon source and one memory. When they are excited,
each of the two sources ideally creates one photon which is
sent through a beam splitter with reflection and
transmission coefficients $\alpha$ and $\beta$ satisfying
$|\alpha|^2+|\beta|^2=1,$ such that after the
beam-splitters, the state of the two photons is $(\alpha
a'^{\dag}+\beta a^{\dag})(\alpha b'^{\dag}+\beta
b^{\dag})|0\rangle$, which can be developed as
\begin{equation}
\label{eq19} \left(\alpha^2 a'^{\dag}b'^{\dag}+\alpha\beta
\left(a^{\dag}b'^{\dag}
+a'^{\dag}b^{\dag}\right)+\beta^2a^{\dag}b^{\dag}\right)|0\rangle.
\end{equation}
The modes $a',$ $b'$ are stored in local memories. The
modes $a$, $b$ are coupled into optical fibers and combined
on a beam splitter at a central station, with the modes
after the
beam-splitter denoted by $d=\frac{1}{\sqrt{2}}(a+b)$ and $\tilde{d}=\frac{1}{\sqrt{2}}(a-b),$ as before. We are interested in the detection of one photon, for example in mode $d$. Let us detail separately the contributions from the three terms in Eq. (\ref{eq19}).\\
The term $a'^{\dag}b'^{\dag}|0\rangle$ which corresponds to
two full memories, cannot generate the expected detection and thus does not contribute to the entanglement creation.\\
The term $(a^{\dag}b'^{\dag}+a'^{\dag}b^{\dag})|0\rangle$
may induce the detection of a single photon in mode $d$
with
probability $\alpha^2\beta^2  \eta_t \eta_d.$ Such detection creates the desired state $\frac{1}{\sqrt{2}}(a'^{\dag}+b'^{\dag})|0\rangle$ associated to entangled memories.\\
Finally the term $a^{\dag}b^{\dag}|0\rangle$ may also
produce a single photon in mode $d$ if one of the two
photons is lost. The probability this term produces the
expected detection  is approximately $\beta^4\eta_t
\eta_d$, since for long distances
$\eta_t \ll 1.$ This detection heralds the vacuum state $|0\rangle$ for the remaining modes $a'$ and $b'.$ \\
Taking into account all these contributions, the state
created by the detection of a single photon in mode $d$ is
thus given by
\begin{equation}
\label{eq20}  \alpha^2
|\psi_{ab}\rangle\langle\psi_{ab}|+\beta^2|0\rangle\langle0|
\end{equation}
where
$|\psi_{ab}\rangle=\frac{1}{\sqrt{2}}(a'^{\dag}+b'^{\dag})|0\rangle.$
The state $|\psi_{ab}\rangle$ describes the entanglement of
the two memories located at $A$ and $B$, while the vacuum
state $|0\rangle$ corresponds to empty memories. We
emphasize that none of the three terms in Eq. (\ref{eq19})
leads to an error of the form $a'^\dag b'^\dag|0\rangle.$
This is a crucial difference compared to the DLCZ protocol,
cf. Appendix B. By considering both detections in mode $d$
and $\tilde{d}$, one can show that the success probability
for entanglement creation in an elementary link is
$P_0=2p_1\beta^2\eta_t \eta_d$ with $p_1$ the probability
that the source emits one photon ($p_1=1$ in the ideal
case).
\\

The further steps are as for the DLCZ protocol: Neighboring
links are connected via entanglement swapping, creating the
entanglement between two distant locations $A$ and $Z$. One
shows that the success probability for entanglement
swapping at the $i$-th level is given by
$P_i=\frac{p_1\alpha^2\eta}{2}
\frac{(2^i-(2^i-1)p_1\alpha^2\eta)}{(2^{i-1}-(2^{i-1}-1)p_1\alpha^2\eta)^2}$
(with $i \geq 1$). Moreover each location contains two
memories, denoted $A_1$ and $A_2$ for location $A$ etc.
Entangled states of the given type are established between
$A_1$ and $Z_1$, and between $A_2$ and $Z_2$. By
post-selecting the case where there is one excitation in
each location, one generates an effective state of the form
\begin{equation}
\label{eq21} \frac{1}{\sqrt{2}}\left(|1_{\rm A1}1_{\rm
Z2}\rangle+|1_{\rm A2}1_{\rm Z1}\rangle\right).
\end{equation}
The probability for a successful projection onto the state
Eq. (\ref{eq21}) is given by
$P_{ps}=\frac{\eta^2}{2}\frac{(p_1
\alpha^2)^2}{(2^i-(2^i-1)p_1\alpha^2\eta)^2}.$ The vacuum
component in Eq. (\ref{eq20}) does not contribute to this
final state, since if one of the two pairs of memories
contains no excitation, it is impossible to detect one
excitation in each location. The vacuum components thus
have no impact on the fidelity of the final state. This is
not the case for components involving two full memories as
in Refs. \cite{Duan2001} and \cite{Simon2007}, which may
induce one excitation in each location and thus decrease
the fidelity. Note that vacuum components, which exist for
the single-photon source protocol already at the level of
the elementary links, occur for the DLCZ protocols as well,
starting after the first entanglement swapping procedure.

%%%%%%%%%%%%%%
\subsubsection{Performance}

As indicated before, the absence of fundamental errors
proportional to the entanglement creation probability leads
to significantly improved entanglement distribution rates
for the single-photon source protocol with respect to the DLCZ protocol.
We now discuss this improvement quantitatively. The weight of the vacuum
component at each nesting level is larger in the
single-photon source protocol, and thus the success
probabilities $P_i$ (with $i \geq 1)$ for entanglement
swapping are somewhat lower. However, the probability $P_0$
can be made much larger than in the photon-pair source
protocols. Overall, this leads to higher entanglement
distribution rates, as we detail now. Taking into account
the expression of $P_0, P_i$ and $P_{ps},$ one can show
that the total time required for entanglement distribution
with the single-photon protocol is
\begin{equation}
\label{eq23} T_{\rm tot}=\frac{3^{n+1}}{2}\frac{L_0}{c}
\frac{\prod_{k=1}^{n}\left(2^k-\left(2^k-1\right)p_1\alpha^2\eta\right)}{\eta_d\eta_{t}p_1^{n+3}\beta^2\alpha^{2n+4}
\eta^{n+2}}.
\end{equation}

Assuming $\eta_m=\eta_d=p_1=0.9$ as before, the present
protocol starts to outperform direct transmission (with a
10 GHz single-photon source) for a distance of 580 km,
achieving an entanglement distribution time of 44 seconds,
with a repeater composed of 4 links and a beam splitter
transmission $\beta^2=0.16$, see also Fig. \ref{comparison}
in section \ref{Comparison}. This is about an order of
magnitude faster than the DLCZ protocol. The significant
improvement compared to the DLCZ protocol can be understood
as due to higher value of $P_0$ ($9.2 \times 10^{-3}$ as
opposed to $3.4 \times 10^{-4}$ for DLCZ). On the other
hand the vacuum component is larger in the present
protocol, reducing the success probability for the swapping
operations, which is why the improvement is not as large as
the difference in $P_0$.

\subsubsection{Implementing the Single Photon Source Protocol with
Atomic Ensembles}

In section \ref{DLCZ} we explained how the emission of a
Stokes photon in a DLCZ-type atomic ensemble creates a
single stored atomic excitation, and how this atomic
excitation can subsequently be reconverted into a photon.
This implies that an ensemble charged with a single
excitation can serve as a single-photon source. The
possibility of multi-photon emissions, which can go
undetected even for photon-number resolving detectors since
they do not have perfect efficiency in practice, means that
this source is not ideal, but that there is a two-photon
contribution with an amplitude equal to $p_2=2 p (1-\eta_d)
\eta_m$, where $p$ is the emission probability for the
Stokes photon, $\eta_m$ is the memory efficiency, and
$\eta_d$ is the efficiency of the detector that detects the
Stokes photon and thus announces that the ensemble is
charged. If one chooses $p$ sufficiently small, one can
therefore realize a very good approximation to an ideal
single-photon source. However, this means that the
excitation triggering the potential Stokes photon emission
has to be repeated many times before the ensemble is
successfully charged. In an implementation one has to check
whether this imposes significant limits on the distribution
rate. Fortunately, this is not the case for realistic
repetition rates for the Stokes emission, say 10 MHz. For a
repeater with 4 links one can show that the maximum value
of $p_2$ compatible with $F=0.9$ is $p_2=0.0011$, giving
$p=0.006$ for the emission probability. With a 10 MHz
repetition rate, the ensemble will thus be charged on
average every 18 $\mu$s. Comparing to a typical
communication time $\frac{L_0}{c}$ of order 750 $\mu$s for
a 150 km link, this even leaves considerable scope for
temporal multiplexing, provided one has appropriate
multimode memories. It should be emphasized in particular
that in the present protocol both the source and the memory
have to be at telecom wavelength. Note that a single-photon
source can also be realized by combining a photon pair
source (which can be ensemble-based, but also e.g. based on
parametric down-conversion) and an absorptive memory in
analogy with the approach described in section \ref{P2M3},
see also section \ref{Implementations}.

\subsubsection{Alternative Implementation via Partial
Readout}

\begin{figure}
{\includegraphics[width=\columnwidth]{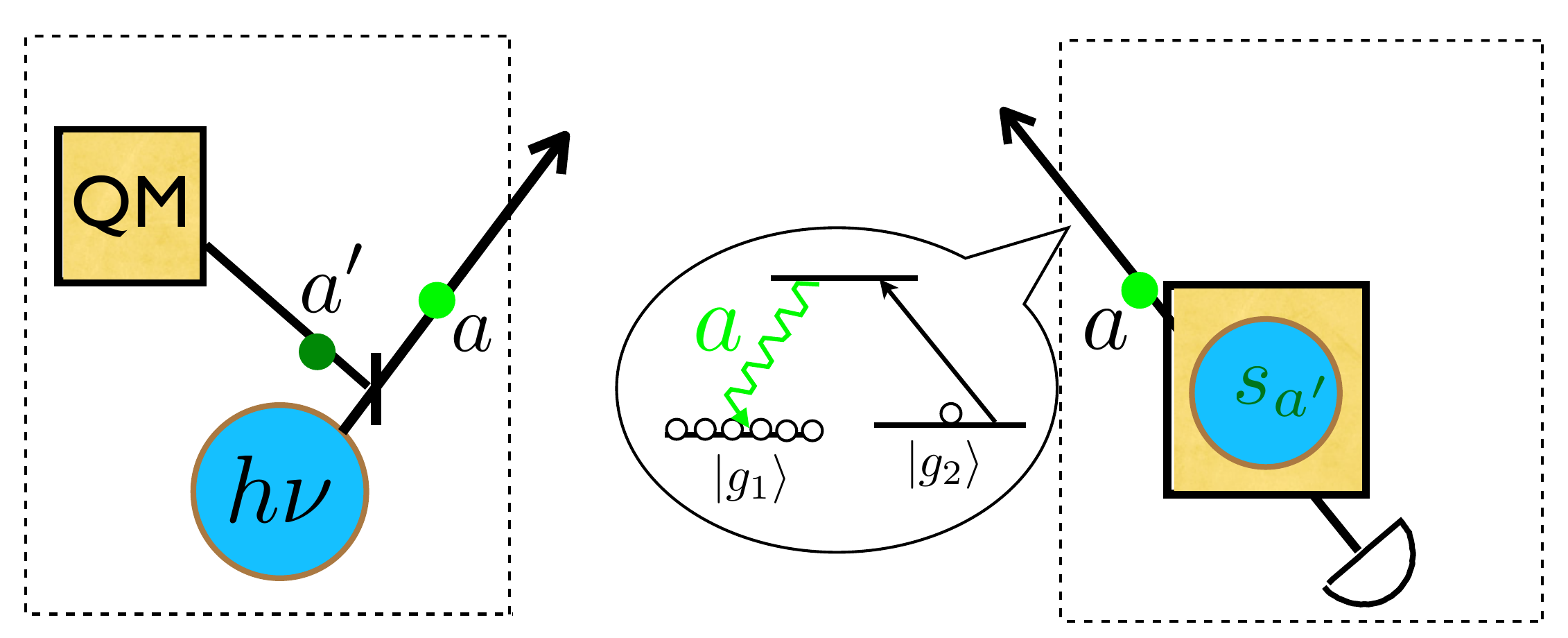}}
\caption{(Color online) A single-photon source whose output
is partially stored in a memory, as required in the
protocol of \cite{Sangouard2007} described in section
\ref{SPS}, can be emulated with a DLCZ-type atomic
ensemble, in which an atomic excitation is first created,
accompanied by a Stokes photon emission, and then partially
read out through the application of a read pluse with an
area that is smaller than $\pi$. The same principle is also
used in the protcol of \cite{Sangouard2008} described in
section \ref{IPP}.} \label{SPS-partial}
\end{figure}

We presented the protocol as consisting of the creation of
single photons, followed by their {\it partial storage},
i.e. the storage of one of the two output modes of a beam
splitter. Alternatively, once the ensemble that serves as
the single-photon source has been charged by the emission
of a Stokes photon as described before, it can be {\it
partially read out}, i.e. the atomic spin waves can be
partially converted back into propagating photons, see Fig.
\ref{SPS-partial}. In principle this could be done using
read pulses whose area is smaller than the standard $\pi$,
chosen to give the same values of $\alpha$ and $\beta$ as
above. There is a subtlety concerning this idea in the
usual DLCZ-type experiments because the Anti-Stokes photon
is typically emitted during the duration of the read pulse,
such that it seems difficult to assign a fixed pulse area
to the read. However, a Rabi oscillation regime was
nevertheless observed by \cite{Balic2005} and theoretically
described by \cite{Kolchin2007} even for ensembles with
large optical depth. It may thus be possible nevertheless
to pick a pulse area that corresponds to the desired values
of $\alpha$ and $\beta$. The described idea certainly works
for other kinds of memories, where readout pulse and
emission are separated in time, such as the memories based
on the photon echo principle discussed in section
\ref{Implementations}.D.2, which is particularly relevant
if the single-photon source is realized by combining a
photon pair source and an absorptive memory as discussed
above. The partial readout approach discussed here is
important for the protocol described in the following
section \ref{IPP} as well, see below.

%%%%%%%%%%%%%%%%%%%%%%%%%%%%%%%%
\subsection{Protocols based on Local Generation of Entangled Pairs
and Two-Photon Entanglement Swapping} \label{IPP}

In the previous section we saw that for quantum repeaters
where the entanglement creation is based on a single photon
detection, it is advantageous to have an ideal
single-photon source, or even a good approximation of such
a source realized with atomic ensembles. Analogously, it is
fairly natural to ask whether it may be possible to achieve
an efficient repeater protocol with entanglement creation
based on two-photon detections if one had an ideal photon
pair source, or a good approximation of such a source
implemented with atomic ensembles. This is indeed a
fruitful approach.

The first such protocol was proposed in section III.C of
\cite{Chen2007}, without evaluating its performance. Ref.
\cite{Sangouard2008} proposed an improved version of the
same approach and showed that this leads to a powerful
quantum repeater protocol, which is both robust under phase
fluctuations, and achieves the best entanglement creation
time of all known non-multiplexed protocols with ensembles
and linear optics, cf. below.

In order to effectively realize a single-pair source,
Reference \cite{Chen2007} proposed to generate entangled
pairs of atomic excitations locally by using four
single-photon sources (which, as we saw in the previous
section, can be realized with DLCZ-type ensembles), linear
optical elements, and two quantum memories based on
Electromagnetically Induced Transparency (EIT), cf. Fig. 11
of Ref. \cite{Chen2007}. Four photons are emitted by the
ensembles serving as sources, two of them are detected, two
are absorbed again by the EIT memories. This double use of
ensembles (emission followed by storage) leads to
relatively large errors (vacuum and single-photon
contributions) in the created state if the memory
efficiencies are smaller than one. These errors then have a
negative impact on the success probabilities of the
entanglement generation and swapping operations, and thus
on the overall time needed for long-distance entanglement
distribution.

The proposal of Ref. \cite{Sangouard2008}, which is based
on partial readout of the memories, allows one to produce
entangled pairs of atomic excitations with higher-fidelity.
Indeed, this scheme does not use any emission followed by
storage. For the same memory and detection efficiency, it
thus leads to higher quality entangled pairs compared to
the method of Ref. \cite{Chen2007}, and as a consequence to
a significantly improved rate for the overall quantum
repeater protocol. In the following we describe the
proposal of Ref. \cite{Sangouard2008} in more detail and
evaluate its performance.

%%%%%%%%%%%%%%
\subsubsection{Local Generation of Entangled Pairs of Atomic Excitations}

\begin{figure}[hr!]
{\includegraphics[scale=0.32]{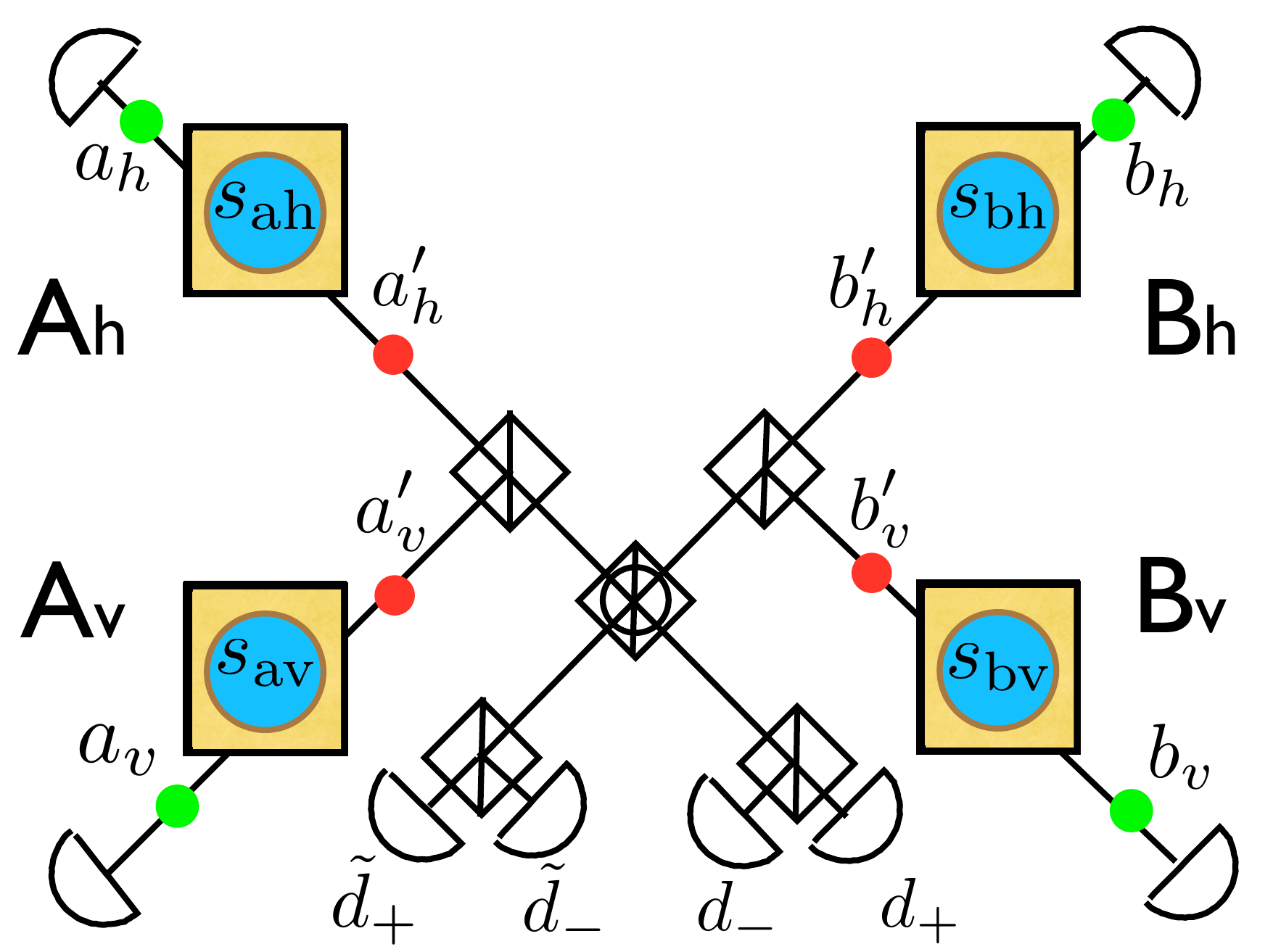}} \caption{(Color
online) Setup for generating high-fidelity entangled pairs
of atomic excitations. Yellow squares containing a blue
circle represent atomic ensembles which probabilistically
emit Stokes photons (green dots). The conditional detection
of a single Stokes photon heralds the storage of one atomic
spin-wave excitation. In this way an atomic excitation is
created and stored independently in each ensemble. Then all
four ensembles are simultaneously read out partially,
creating a probability amplitude to emit an Anti-Stokes
photon (red dots). The coincident detection of two
Anti-Stokes photons in ${d}_+$ and ${\tilde{d}}_+$ projects
non-destructively the atomic cells into the entangled state
$|\Psi_{\rm ab}\rangle$ of Eq. (\ref{eq25}).}
\label{IPPsource}
\end{figure}

The proposed setup for the generation of high-fidelity
entangled pairs requires four atomic ensembles. As before,
the four ensembles are repeatedly excited independently
with a repetition rate $r$ until four Stokes photons are
detected, heralding the storage of an atomic spin-wave in
each ensemble. The Stokes photons have a well defined
polarization : the horizontally (vertically) polarized
modes are labeled by ${a}^{\dagger}_h$ and
${b}^{\dagger}_h,$ $({a}^{\dagger}_v$ and
${b}^{\dagger}_v)$ and are produced from upper (lower)
atomic ensembles $A_h$ and $B_h$ ($A_v$ and $B_v$) as
represented in Fig. \ref{IPPsource}. The associated single
atomic spin excitation are labeled by ${s}^{\dagger}_{\rm
ah}, {s}^{\dagger}_{\rm av}, {s}^{\dagger}_{\rm bh}$ or
${s}^{\dagger}_{\rm bv}$ depending on the location. The
average waiting time for successful charging of all four
ensembles is approximately given by $T=\frac{25}{12 rp}$,
as can be shown by the same methods that are used in
Appendix A. Thanks to the independent creation and storage,
it scales only like $1/p,$ with $p$ the probability for a
Stokes photon to be emitted.

Once all ensembles are charged, the four stored spin-wave
modes are then partially converted back into photonic
excitations, leading to the state $ (\alpha
{a}'^{\dagger}_h+\beta {s}^{\dagger}_{\rm ah})\otimes
(\alpha {a}'^{\dagger}_v+\beta {s}^{\dagger}_{\rm
av})\otimes (\alpha {b}'^{\dagger}_h+\beta
{s}^{\dagger}_{\rm bh})\otimes (\alpha
{b}'^{\dagger}_v+\beta {s}^{\dagger}_{\rm bv})|0\rangle $
with $|\alpha|^2+|\beta|^2=1.$ The primed modes
${a}'^{\dagger}_h, {a}'^{\dagger}_v,$ $({b}'^{\dagger}_h,
{b}'^{\dagger}_v)$ refer to the emitted anti-Stokes photons
from memories located at $A_h$ and $A_v$ ($B_h$ and $B_v$)
respectively. The discussion concerning the implementation
of partial readout at the end of the preceding section also
applies in the present case. The released anti-Stokes
photons are combined at a central station where they are
detected in modes ${d}_{\pm}={a}'_{h}+{a}'_{v}\pm
{b}'_{h}\mp {b}'_{v}$ and ${\tilde{d}}_\pm=\pm {a}'_{h} \mp
{a}'_{v}+{b}'_{h}+{b}'_{v}$, using the setup shown in Fig.
\ref{IPPsource}. In the ideal case, a twofold coincident
detection between ${d}_+$ and ${\tilde{d}}_+$ projects the
state of the two remaining spin-wave modes
non-destructively onto
\begin{equation}
\label{eq25}
|\Psi_{\rm{ab}}\rangle=1/\sqrt{2}({s}_{\rm{ah}}^{\dagger}{s}_{\rm{bh}}^{\dagger}+{s}_{\rm{av}}^{\dagger}{s}_{\rm{bv}}^{\dagger})|0\rangle.
\end{equation}
The stored atomic excitations can be reconverted into
photons as desired. In the proposed quantum repeater
protocol, one excitation (e.g. the one in the $B$
ensembles) is reconverted into a photon right away and used
for entanglement generation. The other excitation is
reconverted later for entanglement swapping or for the
final use of the entanglement. Note that the setup can also
be used as a source of single photon pairs, if both
excitations are converted into photons simultaneously.

Given an initial state where all four memories are charged,
the probability for a coincidence between ${d}_+$ and
${\tilde{d}}_+$ is given by $\frac{1}{2}\alpha^4 \beta^4.$
Since the twofold coincidences ${d}_+$-${\tilde{d}}_-$,
${d}_-$-${\tilde{d}}_+$, ${d}_-$-${\tilde{d}}_-$ combined
with the appropriate one-qubit transformation also collapse
the state of the atomic ensembles into
$|\Psi_{\rm{ab}}\rangle,$ the overall success probability
for the entangled pair preparation is given by
$P_s=2\alpha^4\beta^4.$\\

We now analyze the effect of nonunit detector efficiency
and memory recall efficiency. The waiting time for the
memories to be charged is now $T^\eta=T/\eta_d=\frac{25}{12
rp \eta_d}$. Furthermore, the detectors can now give the
expected coincidences when three or four anti-Stokes
photons are released by the memories, but only two are
detected. In this case, the created state contains
additional terms including single spin-wave modes and a
vacuum component,
\begin{eqnarray}
\label{eq26} {\rho}_{\rm ab}^{s}&=&\nonumber
c_2^{s} |\Psi_{\rm{ab}}\rangle\langle \Psi_{\rm{ab}} | \\
&& \nonumber +c_1^{s} \Big( |{s}_{{\rm ah}}\rangle\langle
{s}_{{\rm ah}}| +|{s}_{{\rm av}}\rangle\langle {s}_{{\rm
av}}| +|{s}_{{\rm bh}}\rangle\langle {s}_{{\rm bh}}|
+|{s}_{{\rm bv}}\rangle\langle {s}_{{\rm bv}}|\Big)\\
&&+c_0^{s} |0\rangle\langle0|;
\end{eqnarray}
where $c_2^{s}=2\alpha^4\beta^4\eta^2/P_s^{\eta},$
$c_1^{s}= \alpha^6 \beta^2\eta^2(1-\eta)/P_s^{\eta}$ and
$c_0^{s}=2\alpha^8(1-\eta)^2\eta^2/P_s^{\eta}$. We have
introduced a superscript $s$ for``source''. The probability
for the successful preparation of this mixed state is
$P_s^{\eta} = 2 \eta^2 \alpha^4(1-\alpha^2\eta)^2.$ The
fidelity of the conditionally prepared state is equal to
the two-photon component
$c_2^s={\beta^4}/{(1-\alpha^2\eta)^2}.$ As can be seen from
the two previous equations, there is a tradeoff on the
readout coefficients $\alpha, \beta$. The creation of an
entangled state with a high fidelity favors $\alpha \approx
0$, whereas a high success probability favors $\alpha
\approx \beta \approx 1/\sqrt{2}.$

\subsubsection{Repeater Protocol using Two-Photon
Detections}

\begin{figure}
{\includegraphics[scale=0.32]{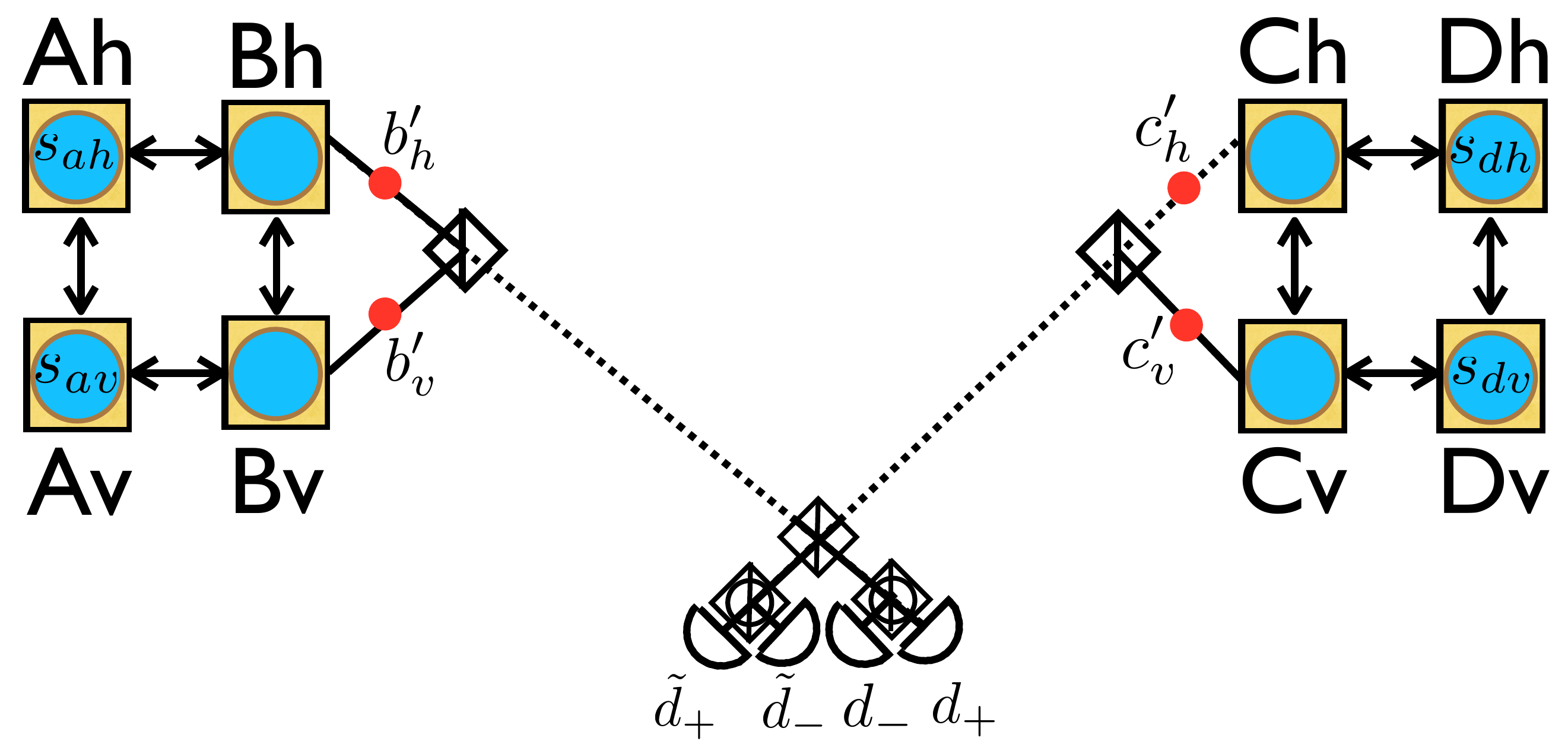}}
\caption{(Color online) Entanglement creation based on
two-photon detections using two locally prepared entangled
pairs. The excitations stored in the
ensembles $B_h$, $B_v$, $C_h$, $C_v$ are read out and the
resulting photonic modes are combined at a central station
using the set-up shown. Ideally, the coincident detection
of two photons in ${d}_+$ and ${\tilde{d}}_+$ projects
non-destructively the atomic cells $A$-$D$ into the
entangled state $|\Psi_{\rm ad}\rangle.$} \label{IPPcreation}
\end{figure}

We now describe how this source of heralded pairs can be
inserted within a quantum repeater protocol. The setup for
entanglement creation between two remote sources involving
the ensembles AB and CD is shown in Fig. \ref{IPPcreation}.
The central station is identical to the one used for the
higher-level entanglement swapping operations in the
protocol of \cite{Jiang2007} see also section \ref{jiang},
Fig. 7, and for all swapping operations in the protocol of
section III.B of \cite{Chen2007}, see also section
\ref{twophotongen} of the present review. Two anti-Stokes
photons are combined at a central station, where one photon
is released from the B ensembles and the other from the C
ensembles and a projective measurement is performed into
the modes ${D}_\pm^{bc}={b}'_{\rm{h}}\pm {c}'_{\rm{v}}$ and
${D}_{\pm}^{cb}={c}'_{\rm{h}}\pm {b}'_{\rm{v}}.$ The
twofold coincident detection ${D}_+^{bc}$-${D}_+^{cb}$
 (${D}_+^{bc}$-${D}_-^{cb}$, ${D}_-^{bc}$-${D}_+^{cb}$, or
 ${D}_-^{bc}$-${D}_-^{cb}$ combined with the appropriate
one-qubit operations) collapses the two remaining full
memories into $|\Psi_{\rm ad}\rangle.$ Due to
imperfections, the distributed state $\rho_{\rm ad}^{0}$
includes vacuum and single spin-wave modes. The weights
$c_2^s, c_1^s, c_0^s$ of the source state $\rho_{\rm ab}^s$
satisfy $c_0^s c_2^s=4 (c_1^s)^2$ such that they are
unchanged after the entanglement creation, as before. We
thus have
 $c_2^0=c_2^s,$ $c_1^0=c_1^s$ and  $c_1^0=c_0^s.$
 The success probability for the entanglement creation is given by
$P_0=2\eta^2\eta_t^2\left(c_2^s/2+c_1^s \right)^2.$

\begin{figure}
{\includegraphics[scale=0.32]{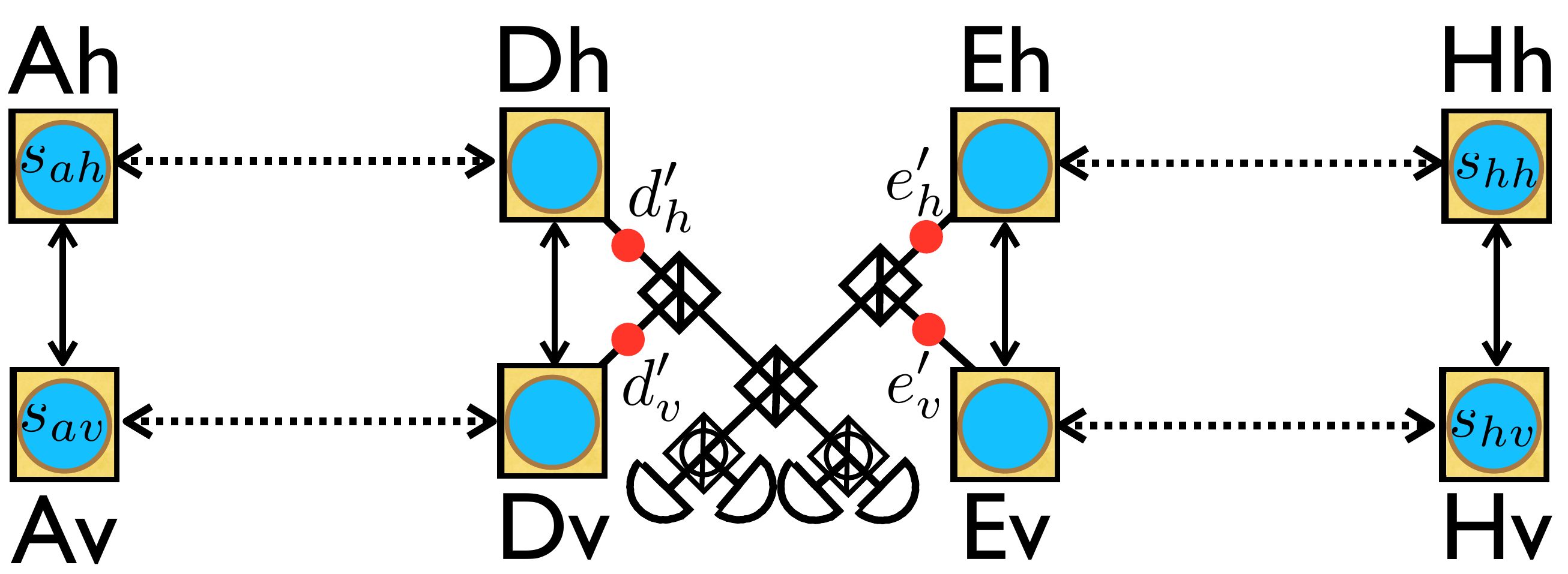}}
\caption{(Color online) Entanglement swapping based on
two-photon detections. The spin-wave stored in ensembles $D_h,$
$D_v,$ $E_h$ and $E_v$ are converted back into anti-Stokes
photons which are combined using the set of linear optics
shown. A twofold coincident detection between $d'_h+e'_v$
and $e'_h+d'_v$ nondestructively projects the ensembles $A$
and $H$ into the state $|\Psi_{ah}\rangle.$} \label{IPPswapping}
\end{figure}

Using the same set of linear optical elements and detectors
(see Fig. \ref{IPPswapping}), one can perform $n$
successive entanglement swapping such that the state
$\rho_{\rm az}^{n}$ is distributed between the distant
locations A and Z. In analogy to above, the distributed
state $\rho_{\rm az}^{n}$ includes vacuum and single
spin-wave components with unchanged weights with respect to
the initial ones, i.e. $c_2^{n}=c_2^{s}, $ $c_1^{n}=c_1^s$
and $c_0^{n}=c_0^s.$ From the expression of $P_0$ and
keeping in mind that the entanglement swapping operations
are performed locally such that there are no transmission
losses, one deduces the success probability for the $i$-th
swapping, $P_i=2\eta^2\left(c_2^s/2+c_1^s \right)^2.$ The
two-spin-wave component of the distributed mixed state
$|\Psi_{\rm az}\rangle$ is finally post-selected
with the probability $P_{ps}=c_2^s\eta^2.$\\

%%%%%%%%%%%%%%
\subsubsection{Performance}

From the expressions of $P_0,$ $P_i$ (with $i \geq 1$) and
$P_{ps}$, one can rewrite $T_{\rm tot}$ as
\begin{equation}
T_{\rm tot}=2\times3^{n} \times (T_s + \frac{L_0}{c})
\frac{(1-\alpha^2\eta)^{2(n+2)}}{\eta_{t}^2\eta^{2(n+2)}\beta^{4(n+2)}}.
\end{equation}
For realistic values of the repetition rate (say $r=10$
MHz) the source preparation time $T_s=\frac{3 T^\eta}{2
P_s^\eta}$ will be comparable to the communication time
$\frac{L_0}{c}$. With the usual assumptions
$\eta_m=\eta_d=F=0.9$, the protocol starts to outperform
direct transmission with a 10 GHz single-photon source for
a distance of 560 km, with an entanglement distribution
time of 15 seconds, for a repeater with 8 links, $p=0.013$
and $\alpha^2=0.26$, see also Fig. \ref{comparison} in
section \ref{Comparison}. Note that for these values
$T_s=380$ $\mu$s and $\frac{L_0}{c}=350$ $\mu$s, with
$c=2\times 10^8$m/s in the fiber as usual. Since the
repetition rate is already limited by the source
preparation time, multiplexing is difficult in the present
case. However, the protocol achieves the best performance
of all known non-multiplexed protocols using only atomic
ensembles and linear optics, and is moreover robust with
respect to channel phase fluctuations thanks to the use of
two-photon detections for entanglement generation.
Multiplexing would become possible if the source
preparation could be accelerated, in particular if an ideal
photon pair source (such as a single-atom cascade) in
combination with an appropriate memory was available.

%%%%%%%%%%%%%%%%%%%%%%%%%%%%%%%%
\section{Comparison and Discussion}
%%%%%%%%%%%%%%%%%%%%%%%%%%%%%%%%

\label{Comparison}

We now compare the performance of the various proposed
protocols in more detail. The first subsection is devoted
to the time needed for entanglement distribution. Then we
review the robustness of the protocols with respect to
several important technological imperfections. Finally we
briefly discuss the complexity of implementing the proposed
protocols.

%%%%%%%%%%%%%%%%%%%%%%%%%%%%%%%%
\subsection{Entanglement Distribution Time}
\label{Time}

\begin{figure}
{\includegraphics[scale=0.32]{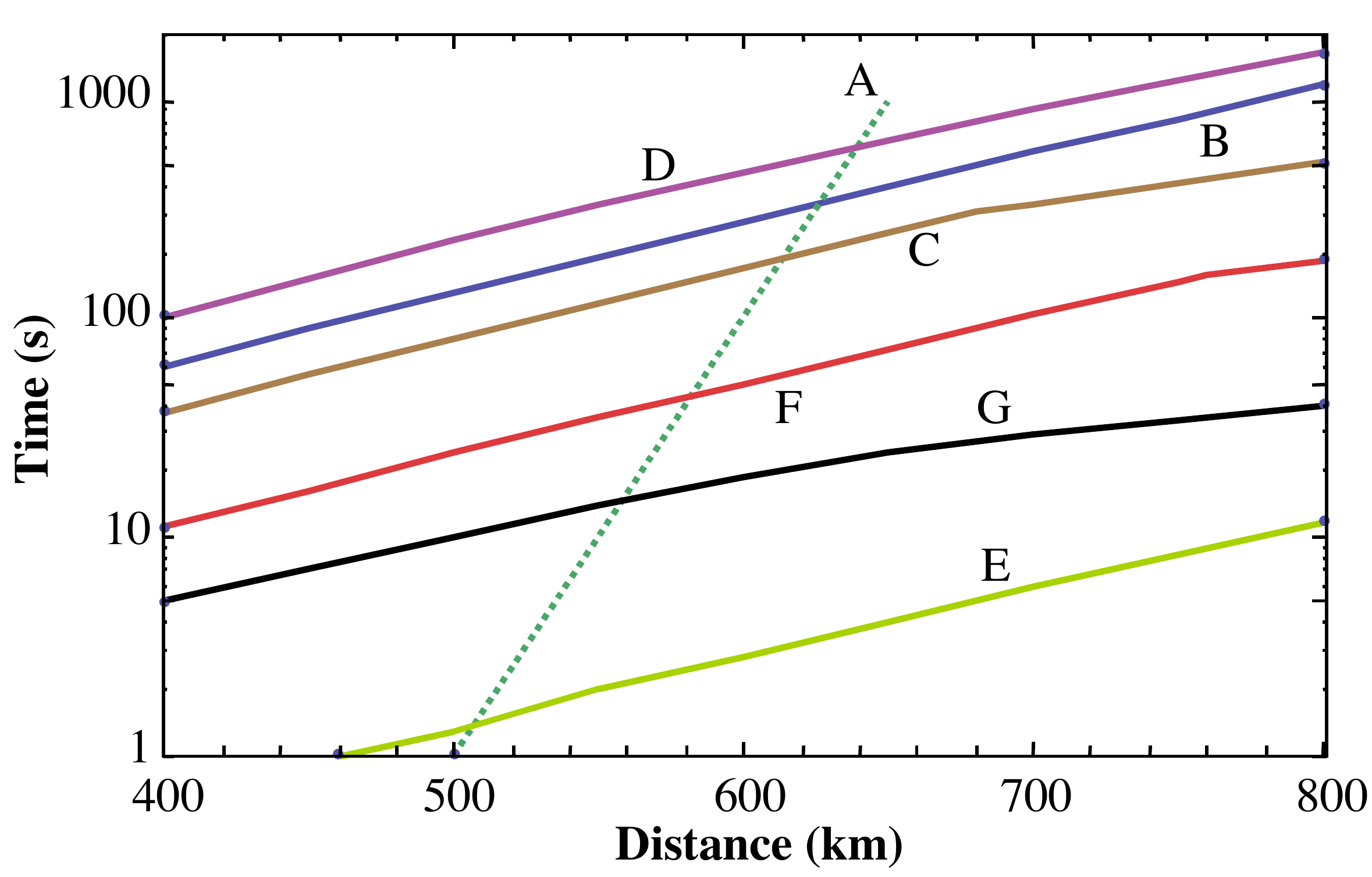}}
\caption{(Color online) Comparison of quantum repeater
protocols based on atomic ensembles and linear optics. The
quantity shown is the average time needed to distribute a
single entangled pair with a final fidelity $F=0.9$ for the
given distance. We assume losses of 0.2 dB/km,
corresponding to telecom fibers at a wavelength of 1.5
$\mu$m. A: as a reference, the time required using direct
transmission of photons through optical fibers with a
single-photon generation rate of 10 GHz. B: original DLCZ
protocol that uses single-photon detections for both
entanglement generation and swapping \cite{Duan2001}. C:
protocol of \cite{Jiang2007} which uses entanglement
swapping based on two-photon detections. D: the protocol of
section III.B of \cite{Chen2007} which first creates
single-photon entanglement locally using single-photon
detections, and then generates long-distance entanglement
using two-photon detections. E: protocol of
\cite{Simon2007} that uses photon pair sources (which can
be realized with ensembles) and multi-mode memories to
implement a temporally multiplexed version of the DLCZ
protocol. We have assumed a memory that can store $N=100$
temporal modes. F: protocol of \cite{Sangouard2007} that
uses quasi-ideal single photon sources (which can be
implemented with atomic ensembles) plus single-photon
detections for generation and swapping. G: protocol of
\cite{Sangouard2008}, which creates high-fidelity entangled
pairs locally and uses two-photon detections for
entanglement generation and swapping, thus following the
approach of section III.C of \cite{Chen2007}, but using an
improved method of generating the local entanglement. We
have assumed a basic source repetition rate of 10 MHz,
which is a limiting factor for this protocol. For all the
curves we have assumed memory and detector efficiencies of
90\%. We imposed a maximum number of 16 links, which is
larger than or equal to the optimal link number for all
protocols apart from curve D, for which the effect is also
less than a factor of two.} \label{comparison}
\end{figure}

Fig. \ref{comparison} shows the time required for
distributing a single entangled pair as a function of
distance for the protocols discussed in detail in the
previous sections. We have again chosen a final target
fidelity $F=0.9$. It should be noted that this takes into
account only errors due to multi-photon emission, which
occur in all the discussed protocols. In practice there are
other sources of errors in addition, such as imperfect mode
overlap and phase fluctuations, that affect different
protocols differently, as we have previously discussed,
requiring e.g. different degrees of fiber length
stabilization depending on whether entanglement is
generated via single-photon or two-photon detections.

Schemes that use two-photon detections for long-distance
entanglement generation are more sensitive to photon losses
than schemes that use single-photon detections for the same
purpose. As a consequence, two-photon protocols favor
larger numbers of elementary links for the same distance
compared to the single-photon schemes. In Fig.
\ref{comparison} we have limited the maximum number of
links to 16, to make the protocols more comparable, and to
have link numbers for which it is plausible that
entanglement purification may not be necessary. In the
shown distance range, this has no effect on the performance
of single-photon entanglement generation based protocols
(curves B, C, E and F), which favor fewer links, nor on the
two-photon detection based protocol of \cite{Sangouard2008}
(curve G). It has a slight effect (of order a factor of 2)
for the protocol of \cite{Chen2007} (curve D).

Fig. \ref{comparison} shows that all protocols start to
outperform direct transmission somewhere in the range 500
to 650 km. It also shows the significant differences in the
required entanglement distribution time that we have
already seen in section \ref{Improvements}. Focusing first
on protocols that create entanglement by single photon
detection, one can see the improvement in going from the
original DLCZ protocol (curve B) to the protocol of Ref.
\cite{Jiang2007} (curve C) and then to the protocol of Ref
\cite{Sangouard2007} (curve F). On the other hand, for
protocols where entanglement is created by two-photon
detection, one sees that the protocol of Ref.
\cite{Chen2007} section III.B (curve D) already achieves a
performance that is fairly similar to the DLCZ protocol
(while significantly improving its robustness), whereas the
protocol of Ref. \cite{Sangouard2008} is significantly
faster (curve G). However, even the fastest protocol (curve
G) still requires very long times for entanglement
distribution, which not only leads to very low rates of
quantum communication, but also is extremely taxing in
terms of quantum memory requirements. Note that the
two-photon detection based protocol of \cite{Zhao2007} is
slower than the protocols shown, cf. section
\ref{twophotongen}.

Curve E, which corresponds to the multi-mode memory based
protocol of Ref. \cite{Simon2007}, emphasizes the advantage
of multiplexing. In this protocol, which is essentially a
multiplexed version of the DLCZ protocol, entanglement is
created by single-photon detections, requiring
long-distance phase stability. Multiplexing the other
protocols is more challenging in terms of source repetition
rate and memory bandwidth, as we discussed in the previous
section, but certainly worth investigating in detail. We
have focused on temporal multiplexing because this seems
particularly promising in practice, however other forms of
multiplexing (spatial, frequency) may be possible as well,
and promise additional benefits in addition to improved
rate, such as greater robustness with respect to storage
time limitations \cite{Collins2007}, cf. section
\ref{Collins}.

%%%%%%%%%%%%%%%%%%%%%%%%%%%%%%%%
\subsection{Robustness}
\label{Robustness}

We have mentioned the importance of long memory times and
of long-distance phase stability (for most of the discussed
protocols) repeatedly in the previous sections. Here we
will briefly review what is known about the effects of
imperfections in these respects on the performance of the
various quantum repeater protocols. We will also discuss
imperfections of other important parameters, in particular
memory and detection efficiencies. Let us note that
multi-photon emission errors are not an imperfection in the
same sense, but inherent to the ensemble-based protocols.
For many protocols they directly determine the achievable
rates by forcing one to work with a certain value of the
emission probability $p$, which is why we studied them
already in the previous sections. Eliminating them would be
possible with different resources, such as ideal
single-photon sources or single-pair sources.

\subsubsection{Storage Time}

We have emphasized that it is essential for the storage
time to be long enough to allow the highest-level
entanglement swapping (or the final post-selection in most
protocols) to be performed. This means that the memory time
has to be comparable to the total entanglement creation
time. \cite{Razavi2008} have recently studied
quantitatively how the performance of quantum repeaters
declines if this is not the case. They find, for example,
that the repeater rate declines as a power of
$\exp({-\sqrt{L/c\tau}})$, where $L$ is the total distance,
$c$ the speed of light and $\tau$ the memory time, in the
regime where $\tau \ll L/c$. As discussed in section
\ref{Collins}, \cite{Collins2007} pointed out that certain
kinds of multiplexing can greatly reduce memory time
requirements, whereas simply running several repeaters in
parallel does not. Developing quantum memories with long
storage times is thus essential for the implementation of
quantum repeaters. In certain solid-state atomic ensembles
storage times over 1 second \cite{Longdell2005} and
coherence times up to 30 seconds \cite{Fraval2005} have
already been demonstrated. This, as well as the
experimental status quo for atomic gases, will be discussed
in more detail in section \ref{Implementations}.

\subsubsection{Phase Stability and Entanglement Purification}

As we have already discussed, phase stability is
particularly important for protocols based on single-photon
detections. Not only the fiber links have to be
interferometrically stable over long time scales, but also
the laser phases. The relevant timescale is given by the
creation of entanglement in an elementary link, not over
the whole distance, but this will typically also be at
least in the ms range. This requires stabilization of the
channel, e.g. through active feedback or through the
implementation of self-compensating Sagnac-type setups, and
distribution of a phase reference. This is an active field
of investigation, which will be reviewed in more detail in
section \ref{Implementations}.F. Any phase error not
eliminated by stabilization will typically be amplified by
a factor of 2 in every entanglement swapping operation,
similarly to the vacuum and multi-photon components
discussed in previous sections.

The difficulty is less severe for two-photon detection
based protocols, where propagation and laser phases only
contribute to an irrelevant global phase. However, if fiber
length fluctuations become too large, they reduce the
overlap between the two photons, which also leads to phase
errors. Active stabilization is thus likely to still be
required, but less precision is sufficient. The level of
stabilization required depends on the coherence length of
photons used in a given implementation, which will
typically be on the scale of meters. Again any remaining
errors are amplified in every entanglement swapping step.
This is why we have limited the number of links in our
comparison above. The experimental status of single-photon
and two-photon visibilities is discussed in more detail in
section \ref{Implementations}, in particular sections V.B.1
and V.F.

There is of course the possibility of using entanglement
purification, as originally discussed by
\cite{Briegel1998}. This introduces a supplementary layer
of complexity, causing a further slowdown, which is why we
haven't included it explicitly in our discussions and
comparisons. We believe that for the most immediate goal of
beating direct transmission it will probably be a better
strategy to minimize all errors and do without
purification. However, purification procedures have now
been developed that can be used in all the discussed
protocols. On the one hand, a protocol for the entanglement
purification of photon pairs with linear optics was
proposed by \cite{Pan2001}. The protocol was adapted to
parametric down-conversion sources by \cite{Simon2002},
leading to an experimental realization \cite{Pan2003}. This
topic is well reviewed in \cite{Pan2008}. On the other
hand, single-photon entanglement plays an essential role in
several of the protocols discussed in this review
\cite{Duan2001,Jiang2007,Simon2007,Sangouard2007}. A
protocol for the purification of single-photon entanglement
with linear optics has recently also been proposed
\cite{Sangouard2008b}. The effects of phase errors and the
inclusion of entanglement purification in repeater
protocols were studied in
\cite{Chen2007,Jiang2007,Zhao2007} with some quantitative
results.

\subsubsection{Memory Efficiency}

\begin{figure}
{\includegraphics[scale=0.32]{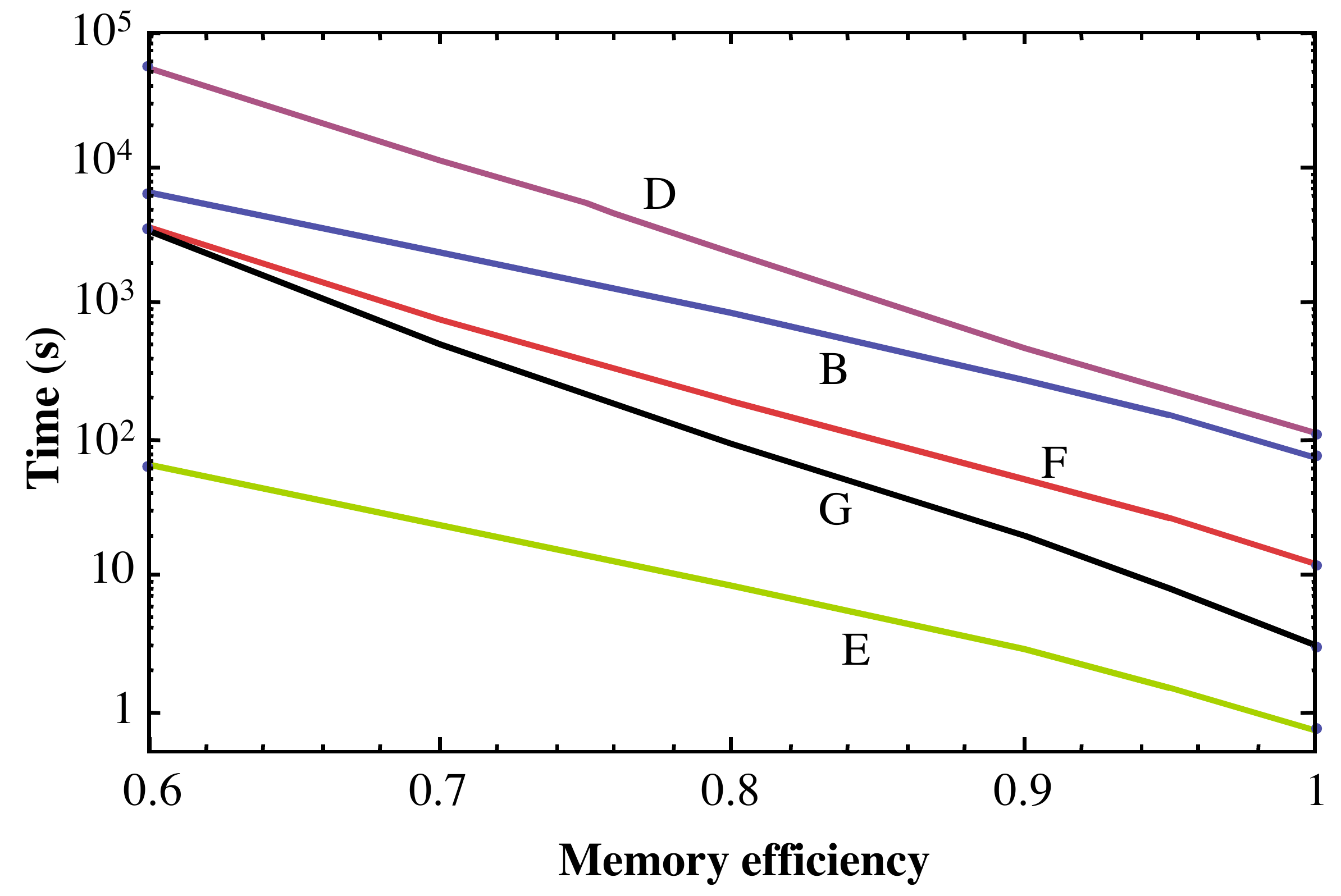}} \caption{(Color
online) Robustness of various protocols with respect to
nonunit memory efficiency. The quantity exhibited is the
average time for the distribution of an entangled pair for
a distance $L=600$km as a function of the memory
efficiency. The letters refer to the same protocols as in
Fig. \ref{comparison}.} \label{memeff}
\end{figure}

Up to now, we have characterized the performance of
protocols by considering a memory efficiency of
$\eta_m=0.9.$ It is important to know how the entanglement
distribution rates vary with the memory efficiency. In Fig.
\ref{memeff}, the average time for the distribution of an
entangled state over 600 km is plotted as a function of the
memory efficiency, with all other parameters as before, in
particular $\eta_d=0.9$. It clearly appears that because
single-photon detection based protocols require less
memories, they are less sensitive to non-unit memory
efficiency than protocols based on two-photon detections.
The main conclusion from Fig. \ref{memeff} however is the
enormous importance of highly efficient memories in order
to achieve reasonable entanglement distribution rates. For
example, a reduction in the memory efficiency from 90 \% to
89 \% leads to an increase in the entanglement distribution
time by 10-14 \%, depending on the protocol. This is
understandable because the memory efficiency intervenes in
every entanglement swapping operation. Intrinsic memory
efficiencies above 80 \% have already been achieved
\cite{Simon2007a}, however overall efficiencies are
typically much lower due to coupling losses. The
experimental status quo will be reviewed in more detail in
section \ref{Implementations}, in particular sections
V.A.1, V.D and V.G.

%%%%%%%%%%%%%%
\subsubsection{Photon Detection Efficiency}

\begin{figure}
{\includegraphics[scale=0.32]{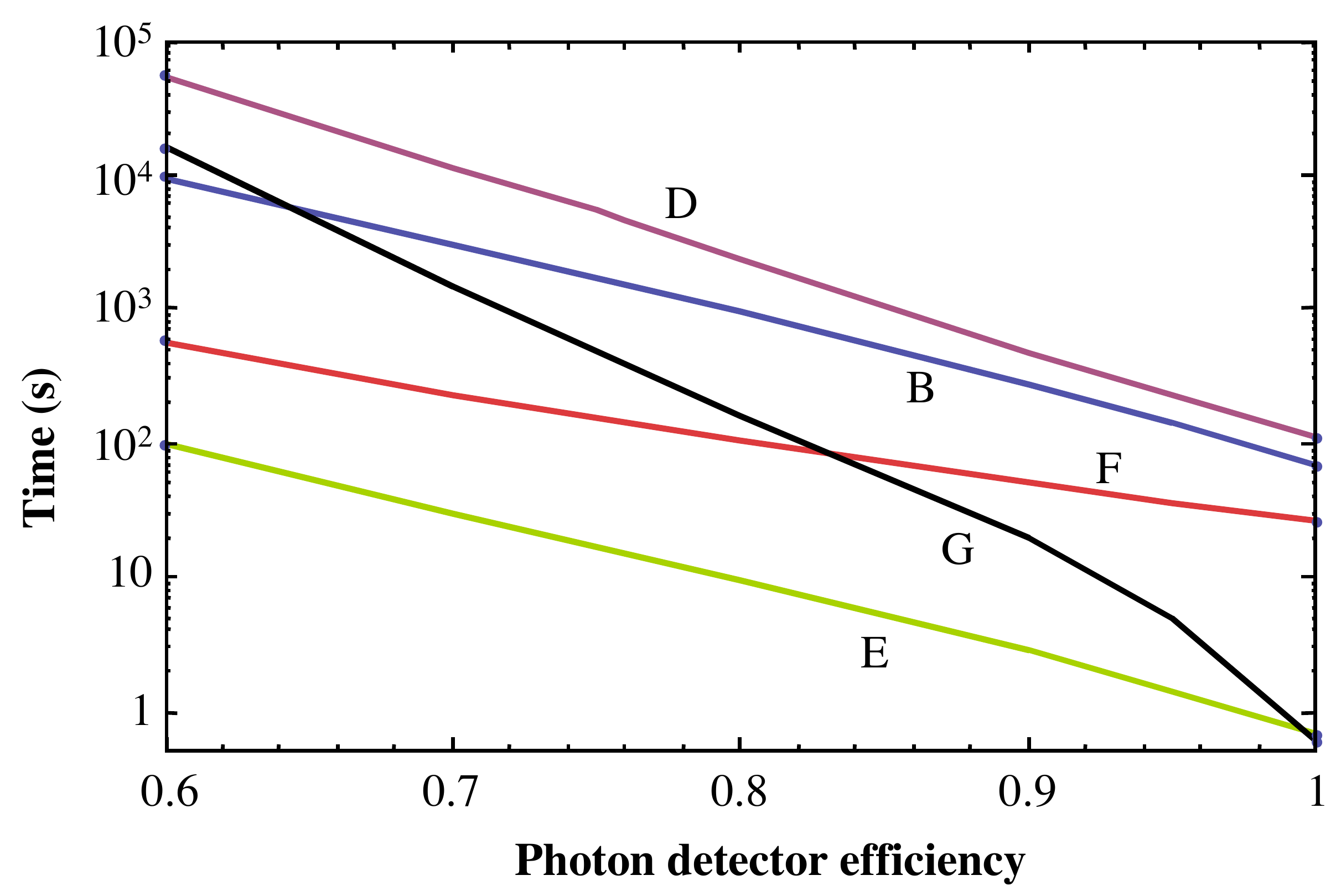}} \caption{(Color
online) Robustness of various protocols with respect to
imperfect photon detector efficiency. The quantity shown is
the average time for the distribution of an entangled pair
for a distance $L=600$km as a function of the photon
detector efficiency. The letters refer to the same
protocols as in Fig. \ref{comparison}.} \label{deteff}
\end{figure}

It is also interesting to know the influence of the photon
detector efficiency on the protocol performance. Hitherto,
we have considered photon detectors capable of resolving
the photon number, and with an efficiency of $\eta_d=0.9$.
Fig. \ref{deteff} shows the average time for the
distribution of one entangled pair for various photon
detector efficiencies, with all other parameters as before,
in particular $\eta_m=0.9$. Since they require less
detectors, the protocols based on single-photon detections
are more robust with respect to photon detector
inefficiency than protocols based on two-photon detections.
Again the main conclusion is that the detection efficiency
is clearly very important, as it too intervenes in every
swapping operation. For example, a reduction in detection
efficiency from 0.90 to 0.89 leads to an increase in the
entanglement distribution time that ranges from 7 \% for
the single-photon source based protocol of
\cite{Sangouard2007} (curve F) to 19 \% for the local
entangled pair based protocol of \cite{Sangouard2008}.

Photon-number resolving detectors with efficiencies as high
as 95 \% have been demonstrated \cite{Lita2008}. We review
the experimental status quo concerning photon detectors in
more detail in section \ref{Implementations}.

%%%%%%%%%%%%%%%%%%%%%%%%%%%%%%%%

\subsubsection{Dark Counts}

Realistic detectors do not only have imperfect efficiency,
but also a certain level of dark counts, which however
depends strongly on the type of detector used. The effect
of dark counts on quantum repeater protocols has been
analyzed for specific protocols, in particular for the
single-photon source based protocol by \cite{Sangouard2007}
and for the DLCZ protocol by \cite{Brask2008}, who studied
the impact of a number of imperfections. It is interesting
to note, as was already remarked in section
\ref{Introduction}, that ``memory-less'' repeaters, also
known as quantum relays, can help alleviate the effects of
dark counts on the transmission of quantum states
\cite{Jacobs2002,Collins2005}.

\subsection{Complexity}
\label{Complexity}

Quantifying and comparing the complexity of different
quantum repeater protocols is not a straightforward task.
One simple way of assessing complexity is counting
elements. For example, outperforming direct transmission
with the protocol of Ref. \cite{Simon2007} will require at
least 2 links, each of which has 4 sources and 4 multi-mode
memories. The cross-over occurs for a distance of 510 km,
for an entanglement distribution time of 2.8 seconds; a
repeater with 4 links is a bit faster, achieving a time of
1.4. seconds. Similar resource counts for the protocols
discussed in the present review are given in
\cite{Sangouard2008}.

However in practice the number of elements is not the only
(nor necessarily the most important) consideration. We have
already emphasized the importance of phase stabilization,
for example, where the required level of precision is
different for single-photon or two-photon detection based
protocols. There are other distinctions where it is less
clear which side is favored. For example, temporal
multi-mode memories would typically be realized in
solid-state atomic ensembles at cryogenic temperatures. On
the other hand, DLCZ-type experiments so far were performed
with atomic gases, requiring optical cooling and trapping.

Our overall conclusion is that outperforming direct
transmission appears possible with repeater architectures
of quite moderate complexity. However, the individual
components have to be excellent. For example, successful
quantum repeaters will probably require storage times of
several seconds, memory and detection efficiencies of 90
percent or more, length-stabilized long-distance fiber
links, and minimal coupling losses between the various
local components, cf. section \ref{Implementations}.

\section{Implementations}
\label{Implementations}

We will now review experiments that are relevant to the different
quantum repeater architectures described above. We first review in
section \ref{implDLCZ} experiments that are directly relevant to
the DLCZ protocol itself. Section \ref{2photons} is devoted to the
experiments relevant to the protocols based on two-photon
entanglement generation and swapping. In section \ref{source}, we
review the various quantum light sources at the single photon
level that are compatible with ensembles based quantum memories
before describing in section \ref{storage} light storage
experiments in atomic ensembles, in particular the storage of
single photons. Section \ref{detector} is devoted to the various
single photon detectors that may be used in a quantum repeater
architecture. Section \ref{channel} briefly describes the quantum
channels that may be used in a quantum repeater, in particular
optical fibers (including phase stabilization) but also free space
links. Finally, we mention in section \ref{loss} an important
practical and technological aspect, the coupling losses. In all
the section \ref{Implementations}, we are mainly concerned with
experiments performed with photon counting in the single
excitation regime. In particular the various experiments
demonstrating the storage and teleportation of quantum continuous
variables of light in atomic ensembles performed with homodyne
detection, e.g.
\cite{Julsgaard2001,Julsgaard2004,Sherson2006,Appel2008,Honda2008,Cviklinski2008}
will not be discussed here. This area of research has been
reviewed recently in \cite{Hammerer2008}.
\subsection{DLCZ Protocol}
\label{implDLCZ} The publication of the article of DLCZ in 2001
triggered an intense experimental effort to realize the basic
elements of this protocol. Over the last few years there have been
a large number of DLCZ type experiments in atomic gases. In this
subsection, we review these experiments. We start in section
\ref{photonpaircreation} by the experimental realization of the
fundamental building block: the generation of strong quantum
correlations between emitted Stokes photons and stored collective
spin excitations, followed by the efficient mapping of the stored
excitation into an anti-Stokes photon. We then describe in section
\ref{DLCZentanglement} experiments demonstrating heralded
entanglement between remote atomic ensembles. Section
\ref{DLCZsegment} describes the experimental realization of an
elementary segment of DLCZ quantum repeater. Finally, section
\ref{DLCZconnection} is devoted to an experiment attempting to
demonstrate entanglement swapping between DLCZ ensembles.

\subsubsection{Creation of Correlated Photon Pairs with a
Programmable Delay} \label{photonpaircreation}

(i) \emph{Quantifying quantum correlations}\\

As we have seen in section \ref{DLCZ-Basic}, the number of
Stokes photons $|n_S\rangle$ emitted during the spontaneous
Raman process is in the ideal case strongly correlated with
the number of collective spin excitations $|n_a\rangle$
stored in the level $|g_2\rangle$. The joint atom-photon
state can be written as (cf Eq.(5)):
\begin{equation}
\left(1-\frac{p}{2}\right)|0_S\rangle
|0_a\rangle+\sqrt{p}|1_S\rangle|1_a\rangle+p
|2_S\rangle|2_a\rangle +O(p^{3/2}). \label{Stokesspinexcitation}
\end{equation}
This state is sometimes referred as a two-mode squeezed state. The
probability $p$ of creating a pair Stokes photon-collective
excitation is directly proportional to the write laser intensity.
In practice, various sources of noise can degrade the quantum
correlations and it is important to experimentally quantify these
correlations. The first step is to convert the atomic collective
excitations into anti-Stokes photons, with a read laser. The
correlations between Stokes photons and stored excitations will
now be mapped into correlations between Stokes and anti-Stokes
fields, which can be measured by photon counting techniques. In
particular, the various probabilities $p_S$,$p_{AS}$ to detect a
Stokes and an anti-Stokes photon, respectively and the joint
probabilities $p_{S,AS}$, $p_{S,S}$, $p_{AS,AS}$ to detect a pair
Stokes-anti-Stokes, two Stokes and two anti-Stokes photons in a
given trial can be easily accessed.

With these measured probabilities, it has been shown that there
exists a well defined border between classical and quantum fields
\cite{Clauser1974,Kuzmich2003}. Specifically, for classical
fields, we have the following Cauchy Schwartz inequality:
\begin{equation}
R=\frac{g^2 (S,AS)}{g(S,S)g(AS,AS)}\leq 1 \label{CS}
\end{equation}
where \begin{equation} g(S,AS)=\frac{p_{S,AS}}{p_Sp_{AS}}
\end{equation}
is the normalized cross-correlation function between Stokes and
anti-Stokes fields, and $g(S,S)=p_{S,S}/p_S^2$ and
$g(AS,AS)=p_{AS,AS}/p_{AS}^2$) are the normalized autocorrelation
function for the Stokes and anti-Stokes fields, respectively.

For a perfect two mode squeezed state as in Eq.
(\ref{Stokesspinexcitation}), the non-conditional Stokes and
anti-Stokes fields exhibit thermal statistics and hence bunching,
with $g_{S,S}$=$g_{AS,AS}$=2. In that case, a measured value of
$g_{S,AS}>2$ is a signature of non classical correlations between
Stokes and anti-Stokes fields.

While a formal proof of non classical correlations requires the
measurement of the Cauchy-Schwartz inequality, the measurement of
a cross correlation function $g_{S,AS}>2$ already gives strong
evidence of non classical behavior. This is because in practice,
the presence of background noise such as leakage of excitation
lasers and dark counts decreases the bunching of the non
conditional fields, such as $g_{S,S}$=$g_{AS,AS}< $2. In addition,
the cross correlation function is very important, since many
parameters crucial for applications, such as the autocorrelation
function of the heralded anti-Stokes photon
\cite{Chou2004,Laurat2006,Matsukevich2006,Chen2006}, the
visibility of two photon interference
\cite{deRiedmatten2006,Chen2007a}, the visibility of the
Hong-Ou-Mandel interference \cite{Felinto2006,Yuan2007} and the
visibility in measurement induced entanglement experiments
\cite{Laurat2007} are directly related to $g_{S,AS}$. For a
perfect two-mode squeezed state, the normalized cross-correlation
function is linked to the probability $p$ of creating a
Stokes-anti-Stokes pair as :
\begin{equation}
g_{S,AS}=1+\frac{1}{p}
\end{equation}
for $p\ll1$. In that case the finite value of $g_{S,AS}$ is due to
multiple pair creation. As mentioned in Section
\ref{DLCZ-Protocol}, multiple pair creation is a major source of
errors in the DLCZ architecture that should be minimized by
working in the regime of high quantum correlations. For an ideal
state, this regime can be in principle accessed by using very low
pump power. This is true of course as long as the background noise
is negligible. A major experimental challenge of this type of
experiment is to preserve the quantum character of the emitted
single photons, which requires excellent filtering of the various
sources of noise, such as leakage of the excitation lasers,
fluorescence and stray light. \\
\begin{figure}[hr!]
{\includegraphics[scale=0.4]{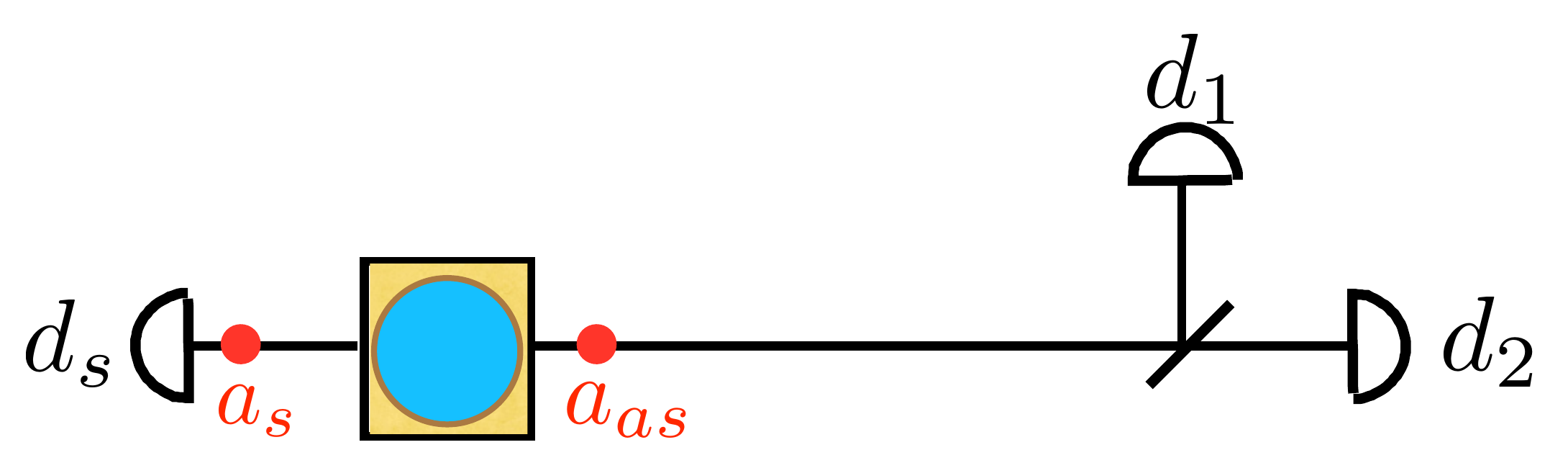}} \caption{Typical
experimental configuration used to measure the single photon
character of the conditional anti-Stokes fields. A detection of
the Stokes field in detector $D_s$ is used as a trigger and the
anti-Stokes field is split at a beam splitter and detected by
detectors $D_1$ and $D_2$.} \label{HBT}
\end{figure}
The strong correlations between Stokes and anti-Stokes photons
enable the generation of heralded single photon states. For
example, the detection of a Stokes photon by detector $D_S$
projects the anti-Stokes field on a single photon state. The
quality of this single photon state can be measured with a
Hanbury-Brown-Twiss setup, i.e. by splitting the anti-Stokes field
at a beam splitter and recording the detection events in the two
output modes with detectors $D_1$ and $D_2$ (see Fig. \ref{HBT}).
As shown by \cite{Grangier1986}, the single photon character of
the conditional anti-Stokes field can be characterized with the
auto-correlation function:
\begin{equation}
\alpha=\frac{p_{(1,2)\mid S}}{p_{(1\mid S)}p_{(2\mid
S)}}=\frac{p_{(S)}p_{(S,1,2)}}{p_{(S,1)}p_{(S,2)}} \label{alpha}
\end{equation}
where the various $p$'s correspond to the probability of a joint
detection event in the corresponding combination of detectors
$D_S$, $D_1$ and $D_2$. For classical fields, a Cauchy-Schwartz
inequality leads to $\alpha \geq 1$. For coherent fields, we have
$\alpha$=1 while $\alpha$=2 for thermal fields. In contrast, for a
perfect conditional single photon field $\alpha$=0. The
measurement of the "anticorrelation parameter" $\alpha$ thus
provides a way to quantify the two photon suppression of the
conditional field with respect to a coherent field.\\

(ii) \emph{Initial experiments} \\

 The first enabling step towards a
practical realization of the DLCZ quantum repeater is the
observation of non classical correlations between the Stokes and
anti-Stokes fields emitted with a controllable delay by one atomic
ensemble. The first experiments have been performed simultaneously
in 2003 in Caltech \cite{Kuzmich2003} and Harvard
\cite{vanderWal2003}. The Caltech experiment used ensembles of
cold Cesium atoms in a magneto-optical trap (MOT) and observed
quantum correlations in the single excitation regime. The write
and read pulses were separated by 400 ns and sent in a collinear
co-propagating configuration through the ensemble. A challenging
aspect of the experiment was to separate the classical pulses from
the weak nonclassical fields, since they were temporally and
spatially overlapped, and their frequencies were only 9 GHz apart.
In the first experiment the filtering had three stages. First, the
Stokes (anti-Stokes) fields was separated from the write (read)
pulse in a polarizing beam splitter (PBS) right after the MOT
chamber. Later, the leakage of the excitation pulses that still
escapes the PBS in the wrong direction was spectrally filtered by
optically pumped vapor cells. Finally, the Stokes (anti-Stokes)
field was distinguished from the read (write) pulse by temporal
gating of the detection.

The non classical character of the fields was demonstrated
using the Cauchy-Schwartz inequality of (Eq.\ref{CS}). A
value of $R=1.84\pm0.06>1$ was measured, thereby
demonstrating that the Stokes and anti-Stokes fields in the
single photon regime were non classically correlated. The
size of the violation of the inequality was limited mostly
by uncorrelated fluorescence from individual atoms in the
atomic sample and by the residual leakage of excitation
pulses.

 The Harvard experiment used a hot vapor of Rubidium
atoms. The first experiment was carried out in the regime
of high excitation number ($10^3-10^4)$
\cite{vanderWal2003}. Strong
 intensity correlation between the Stokes and anti-Stokes fields
 were observed and their quantum nature was demonstrated by an analysis of the fluctuation
spectral density with respect to the shot-noise (or vacuum-state)
limit. In more recent experiments, non classical correlations were
also observed in the single excitation regime with hot vapors
\cite{Eisaman2004,Eisaman2005}.

Since the initial experiments, tremendous progress has been
made on several fronts in various experiments. We shall now
review relevant experiments, based on three important
properties of the DLCZ source: the quality of non-classical
correlations between Stokes and anti-Stokes photons, the
efficiency of atomic to photonic conversion for the
Anti-Stokes field (called retrieval
efficiency) and the storage time of the stored excitation.\\

 (iii) \emph{Quantum Correlations}\\

In the initial experiment \cite{Kuzmich2003} the measured
$g_{S,AS}$ was only slightly above 2. Various improvements in the
experiments have allowed the observation of substantially higher
quantum correlations between Stokes and anti-Stokes field. A first
important step was to set the write laser slightly off resonance
to avoid uncorrelated fluorescence. A further improvement was to
use a four-level scheme of excitation, in which write and read
pulses are 42 nm apart. This allows a fourth filtering stage by
narrow-bandwidth optical filters, and the study of correlations
with temporally overlapped write and read pulses. Experiments in
that regime yielded values of $g_{S,AS}$ of order 10
\cite{Chou2004,Polyakov2004,Felinto2005}. More recently,
\cite{Chen2006} have achieved a value of $g_{S,AS}$ =100 with an
improved version of the setup of \cite{Kuzmich2003}. Another
important step was achieved by using an off axis geometry in which
the Stokes (anti-Stokes) photon is collected with a small angle
with respect to the write (read) beam direction. This
configuration was first used in the classical regime in
\cite{Braje2004}, and in the quantum regime in
\cite{Matsukevich2005}. This geometry allows for a very efficient
spatial filtering of the write and read lasers, which decreases
the background light by a substantial amount. Much higher values
of cross-correlation function have been obtained in this case,
e.g.  $g_{S,AS} \simeq$ 300 in \cite{Matsukevich2006} and
$g_{S,AS}\simeq$ 600 in \cite{Laurat2006}.

Note that the detection of Stokes and anti-Stokes photons with an
angle cannot be used with hot atomic vapors, due to the motion of
atoms (see below). In addition, the detuning of the write pulse in
order to be off resonance must be much larger than in cold gases,
because of the doppler inhomogeneous broadening (which is
approximately 500 MHz). Consequently the write pulse intensity is
also much larger than in cold gases. For these reasons, the
filtering of the write and read pulses is more challenging in hot
vapors than in cold gases. Strong non classical correlations have
nevertheless been observed by using a counterpropagating
configuration for the write and read beams
\cite{Eisaman2005,Walther2007}.\\

(iv) \emph{Retrieval efficiency}\\

\begin{figure}[hr!]
{\includegraphics[scale=0.9]{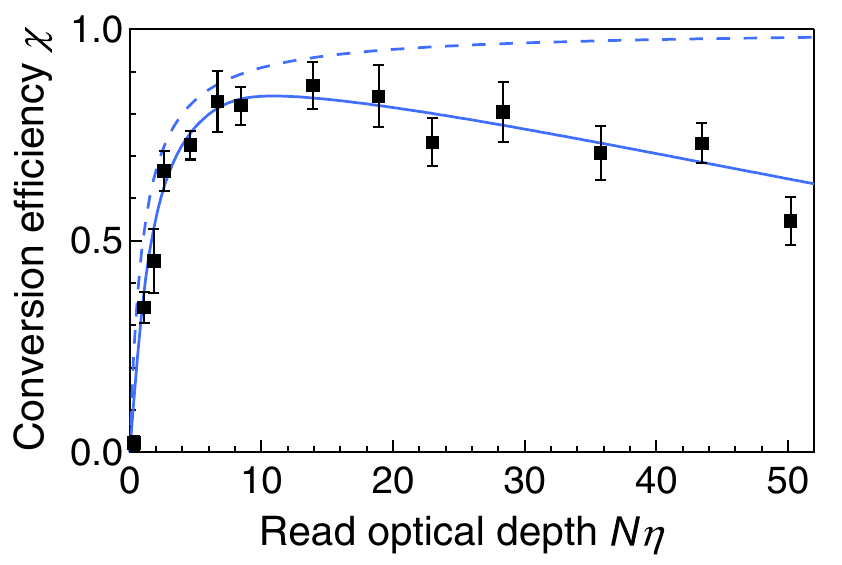}} \caption{Retrieval
efficiency versus read optical depth at a write-read delay of 120
ns with the atomic ensemble inserted into an optical cavity
\cite{Simon2007a}. The dashed line shows the predicted retrieval
efficiency for a three-level system; the solid line is the
prediction from a model including dephasing from additional
excited states.} \label{retrievalOD}
\end{figure}

The single spin excitation stored in the ensemble can in principle
be retrieved with unit efficiency in a well defined
spatio-temporal mode, due to the collective enhancement effect, as
discussed in section \ref{DLCZ-Basic}. The retrieval efficiency is
defined as the probability to have an anti-Stokes photon in a well
defined spatio-temporal mode at the output of the atomic ensemble
conditioned on the successful detection of a Stokes field. In
practice however, several factors can limit the retrieval
efficiency. For example, it depends on the available optical depth
and read beam power. The collective interference can also be
decreased by various dephasing effects due e.g. to spatial
intensity profile mismatch between read beam and stored
excitation, inhomogeneous broadening \cite{Ottaviani2009} or
atomic motion (for atomic motion, see below and section
\ref{DLCZ-Basic}). The effect of atomic motion has been discussed
in the the While early experiments suffered from low retrieval
efficiencies, progress has been made on several fronts, leading to
retrieval efficiencies of 50 $\%$ in free space \cite{Laurat2006}
and of more than 80$\%$ in cavities \cite{Simon2007a}. Fig.
\ref{retrievalOD} shows the measured retrieval efficiency as a
function of the optical depth with the atomic ensemble inserted
inside an optical cavity \cite{Simon2007a}. Note that the
conditional probability to \textit{detect} an anti-Stokes photon
is usually much lower than these values, due to the various
passive losses and to detection inefficiency (see section \ref{loss}).\\

(v) \emph{Storage time}
\\

A crucial parameter for using DLCZ quantum memories in a repeater
scheme is the storage time of the collective spin excitation,
which is limited by the coherence time of the ground state
transition $|g_1\rangle$-$|g_2\rangle$. There are several factors
that affect the storage time at different timescales. The main
factors can be divided in two classes: the inhomogeneous
broadening of the spin transition and the atomic motion. We are
now going to describe these effects in more detail.
\\

(a) Inhomogeneous broadening of the spin transition.\\

In most experiments to date, the spin excitation was stored in a
hyperfine state containing a Zeeman state manifold. In absence of
external magnetic fields, all the Zeeman states with different
$m_F$ are degenerate. In practice however, it is difficult to
suppress  external magnetic fields completely. For experiments
with cold atoms for example, a strong magnetic field gradient is
needed in order to trap the atoms. This leads to  an inhomogeneous
broadening of the spin transition. In that case, the collective
spin excitation undergoes a strong inhomogeneous dephasing, which
effectively suppresses the collective enhancement necessary for
efficient retrieval of the photons. This decoherence mechanism was
studied in detail in \cite{Felinto2005}. The inhomogeneous
broadening due to the the trapping magnetic field gradient leads
to storage times of the order of a few hundreds nanoseconds
\cite{Kuzmich2003,Polyakov2004,Matsukevich2004}. A direct solution
to this problem is to switch off the magnetic field gradient
during the write/read sequence. This solution was tested
experimentally and led to an increase of storage time by two
orders of magnitude (of order of 10$\mu
s$)\cite{Felinto2005,Matsukevich2005,Matsukevich2006,Black2005,Chen2006}.
The storage time was in that case mostly limited by the residual
magnetic field. It is experimentally difficult to further decrease
residual magnetic fields. Another solution to avoid the
inhomogeneous broadening of the spin transition is to use first
order magnetically insensitive hyperfine transitions, known as
'clock transitions', connecting two specific Zeeman states. Such
transitions exists in Cs and Rb atoms. This requires to prepare
all the atoms in a specific Zeeman state (typically with $m_F$=0).
This preparation can be implemented with optical pumping
techniques. A storage time of 1 ms has been recently demonstrated
using this technique in a collinear configuration \cite{Zhao2009}.

While turning off the trapping magnetic field allows a strong
reduction in the inhomogeneous broadening of the spin transition,
it has a major drawback: in that case, the atoms are no more
trapped and are free to fly away, which severely decrease the
available optical depth in the time scale of a few ms. To overcome
this problem, it was suggested to use an optical dipole trap to
maintain a sufficient atom density. This solution was tested
experimentally in \cite{Chuu2008}. The storage time was however
limited to a few tens of $\mu s$ by atomic motion since the
experiment was performed in the configuration with an angle
between the
write (read) and the Stokes (anti-Stokes) fields (see below) \\

(b) Atomic motion\\

\begin{figure}
{\includegraphics[scale=0.9]{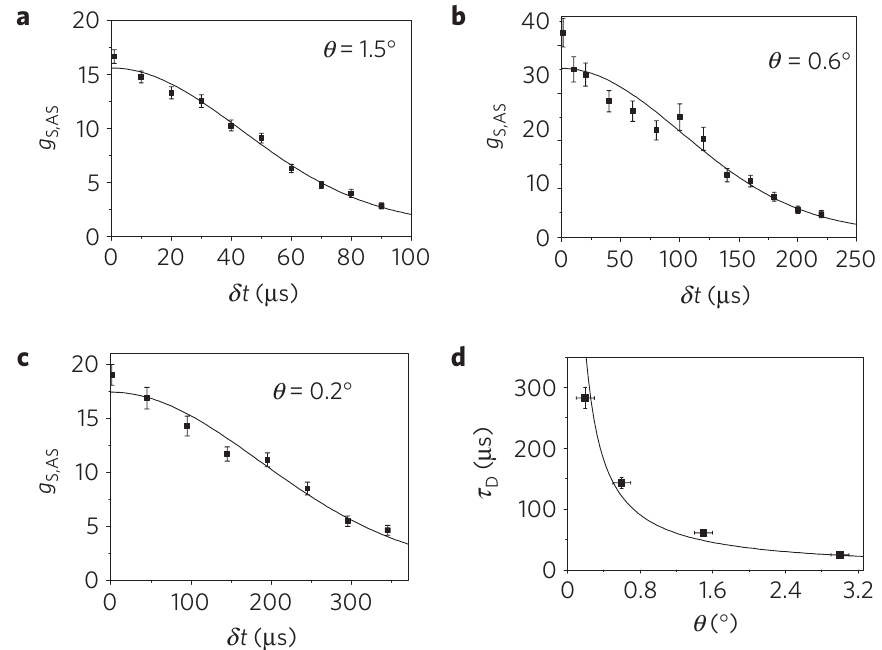}} \caption{(a-c),
The cross-correlation $g_{A,AS}$ versus the storage time $\delta
t$ for different angles $\theta$ between the write beam and the
Stokes field. \cite{Zhao2009}. By reducing the angle, the lifetime
is increased from 25 to 283 $\mu s$, which implies that the
decoherence is mainly caused by the dephasing induced by atomic
random motion. (d) The measured lifetime $\tau_D$ as a function of
detection angle theta, where the horizontal error bars indicate
measurement errors in the angles. The solid line is the
theoretical curve with T=100 $\mu K$. The experimental results are
in good agreement with the theoretical predictions.} \label{angle}
\end{figure}
Another important cause of decoherence is the motion of
atoms. This is obviously a bigger problem for experiment
with hot gases, but is also a strong limitation for cold
ensembles, as we shall see. The motion of atoms can cause
two different problems. The first one is the diffusion of
the atoms out of the excitation region during the storage
of the spin excitation. This is the prime cause of
decoherence for the experiments with hot gases realized to
date. For hot gases, this leads to coherence time of a few
$\mu s$ \cite{Eisaman2004,Eisaman2005}, while for cold
ensembles the diffusion time is of order of 1 ms
\cite{Felinto2005}. This diffusion can be mitigated by
using bigger beams and/or colder atoms. There is however a
much more severe effect of the atomic motion: the
disturbance of the phase of the collective spin excitation.
In section \ref{DLCZ-Basic} we have seen that the motion of
the atoms is not a problem for the phases of the collective
state, as long as a collinear configuration with ${\bf
k}_w={\bf k}_s$, and ${\bf k}_r={\bf k}_{as}$ is used. For
all other configurations, the motion of the atoms will
induce a dephasing that depends on the angle between ${\bf
k}_w$ and ${\bf k}_s$, as was nicely demonstrated
experimentally in \cite{Zhao2009}. The wavelength of the
stored spin wave can be written as :
\begin{equation}
\Lambda=\frac{2\pi}{\Delta k_{SW}}=\frac{2\pi}{|{\bf
k}_w-{\bf k}_s|}\approx\frac{2\pi}{k_w sin\theta}
\end{equation}
where $\theta$ is the angle between the write beam and the
Stokes field. The time scale of the dephasing can be
estimated by calculating the average time to cross $1/2\pi$
of the wavelength of the spin wave, leading to storage
lifetime of $\tau\sim(\Lambda/2\pi v)$ with
$v=\sqrt{k_BT/m}$ the one dimensional speed of the atoms,
where $k_B$ is the Boltzmann constant, T the temperature
and m the mass of the atoms. The reduced wavelength of the
spin wave due to the angle $\theta$ severely limits the
achievable storage time. For example, for a typical
$\theta=3 ^\circ$ and for T=100 $\mu K$, we find $\Lambda =
15\mu m$ and $\tau=25 \mu s$. \cite{Zhao2009} confirmed
this prediction experimentally by  measuring the storage
time as a function of $\theta$ using a clock transition in
a cold Rb ensemble, as shown in Fig. \ref{angle}. For
$\theta$=0$^\circ$ (collinear configuration) they achieved
a storage life time of order 1 ms.

\begin{figure}[hr!]
{\includegraphics[scale=1.3]{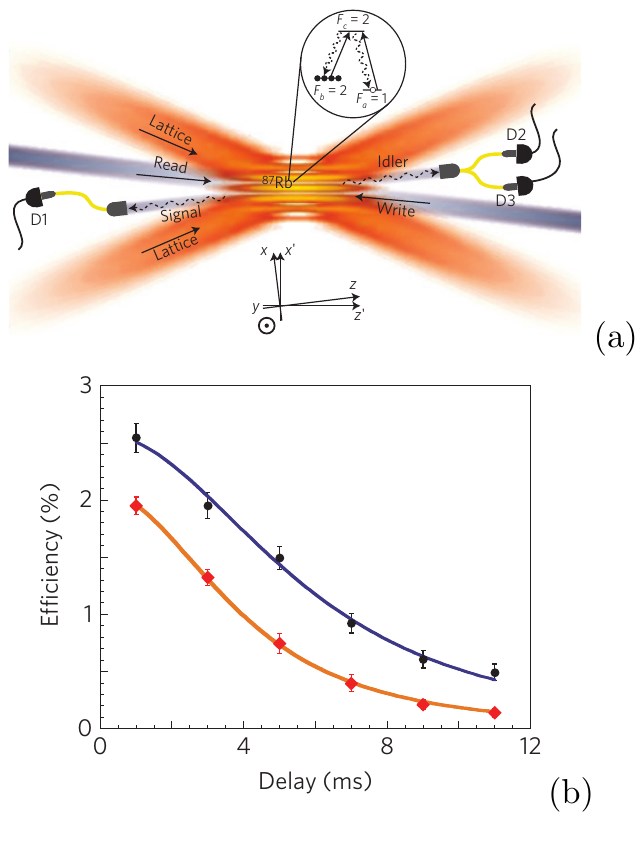}} \caption{(a)
Experiment demonstrating a DLCZ quantum memory with an atomic
ensembles loaded into an 1 D optical lattice, from
\cite{Zhao2009a}. Between $10^5$ and $10^6$ sub-Doppler-cooled
87Rb atoms are loaded into an optical lattice (see the Methods
section), and detection of the signal field, generated by Raman
scattering of the write laser pulse (red-detuned by 20 MHz),
heralds the presence of a write spin wave excitation. A resonant
read/control field converts the surviving atomic excitation into
an idler field after a storage period Ts. The inset shows the
atomic level scheme of 87Rb with levels a and b being the
hyperfine components of the ground $^5S_{1/2}$ level, and level c
being a hyperfine component of the excited $^5P_{1/2}$ level. (b)
Retrieval efficiency (including detection) as a function of
storage time for atoms optically pumped in clock states in the
optical lattice for two different lattice depths $U_0$ (
Diamonds,$U_0$ 80 $\mu K$; circles $U_0$ 40 $\mu K$ circles.) The
intrinsic retrieval efficiency at the output of the ensemble is
roughly four times larger. } \label{DLCZlattice}
\end{figure}

In practice however, there is a great advantage of using a
non-collinear configuration since it enables a very efficient
spatial filtering to suppress the excitation beams in the quantum
channel. In that case, the only way to avoid motion induced
dephasing is to suppress the atomic motion. With atomic gases, one
possibility is to load the atoms into an optical lattice.
\cite{Zhao2009a} have demonstrated a DLCZ quantum memory using a
clock transition in rubidium atoms confined in a one dimensional
optical lattice. They achieved a storage lifetime exceeding 6 ms,
which is currently the longest storage lifetime observed in the
single photon regime (see Fig. \ref{DLCZlattice}).
A light storage experiment with bright coherent pulses based on
Electro-magnetically induced transparency (EIT) has also been
recently demonstrated with Rb atoms confined in a 3 dimensional
optical lattice, leading to a storage lifetime of 240 ms
\cite{Schnorrberger2009}.  Another possibility may be to use
colder atoms, for example a Bose Einstein condensate where
collective coherences in the high excitation number ($>10^4$)
regime have been created and stored recently
\cite{Yoshikawa2007,Yoshikawa2009}. These two techniques are
however complex and technically demanding. Another potential
solution, which may be more practical, is the use of atomic
ensembles in the solid state, implemented with rare-earth ion
doped solids. In such a medium, the atoms behave as a 'frozen gas'
and a storage lifetime exceeding 1 s has been demonstrated, though
not yet in the quantum regime \cite{Longdell2005}.

\subsubsection{Heralded Entanglement between two Atomic Ensembles}
\label{DLCZentanglement}

\begin{figure}
{\includegraphics[scale=1.2]{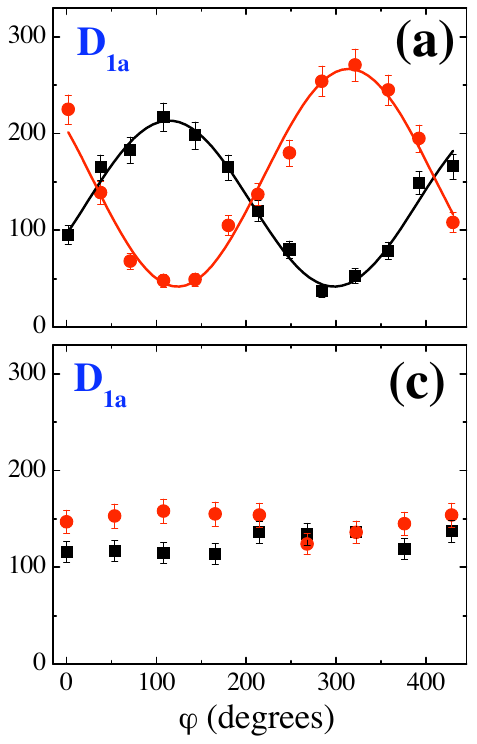}} \caption{Signature
of the coherent superposition of a single excitation delocalized
between two atomic ensembles located 3 m away \cite{Chou2005}.
After the detection heralding entanglement, the stored excitation
are converted into anti-Stokes photons and combined at a beam
splitter, forming a Mach-Zehnder interferometer. Panel (a)
corresponds to the
 detection of the anti-Stokes fields after the beam
splitter as a function of the phase of the interferometer
conditioned on the detection of a Stokes photon (heralding event),
when the Stokes fields are combined with the same polarization.
Panel (c) shows the same measurement when the Stokes fields are
combined with orthogonal polarizations. This highlights the
importance of the indistinguishability of the Stokes photons in
order to generate an entangled state } \label{fringeDLCZ}
\end{figure}

 A crucial step towards the implementation
of a quantum repeater is the demonstration of heralded
entanglement between two spatially separated atomic ensembles in
the single excitation regime. This was first demonstrated in
\cite{Chou2005} with two cold Cs atomic ensembles in two vacuum
chambers separated by 3 meters. Entanglement for excitation stored
in remote ensembles was created by a quantum interference in the
detection of light emitted by the quantum memories. Following the
DLCZ protocol described in section \ref{DLCZ-Protocol}, the two
ensembles are simultaneously and coherently excited by a weak
write beam and the two Stokes fields created by spontaneous Raman
scattering are collected into single mode optical fibers and mixed
at a beam splitter, forming a long Mach-Zehnder interferometer. If
the two Stokes fields are indistinguishable, the information about
the origin of the photon is erased and a detection after the beam
splitter projects the ensembles in the ideal case onto an
entangled state with one de-localized excitation, of the form:
\begin{equation}
|\Psi_{ab}\rangle=\frac{1}{\sqrt{2}}(|1_a\rangle|0_b\rangle+e^{i\theta_{ab}}|0_a\rangle|1_b\rangle)
\label{numberstate}
\end{equation}
where the phase $\theta_{AB}=\phi_B-\phi_A+\xi_B-\xi_A$, with
$\phi_{A,B}$ the phase of the laser at ensemble $A$ and $B$
respectively, and $\xi_{A,B}$ the phase acquired by the Stokes
photons from the ensembles to the beam spitter. In order to
generate a measurable entangled state, it is important that the
phase $\theta_{AB}$ is stable during the duration of the
experiment. In \cite{Chou2005}, it was actively stabilized using
an auxiliary laser. The state of Eq. (\ref{numberstate}) is an
idealized state.  In practice, various sources of noise can  turn
the heralded state into a mixed state. For example, due to the
probabilistic nature of the spontaneous Raman process, there is an
unavoidable finite probability to create higher order terms with
two or more excitations (see section \ref{DLCZ-Basic}). Non
perfect filtering of the excitation light will also alter the
heralded state. In order to prove entanglement experimentally, it
is therefore crucial to demonstrate the single excitation
character of the atomic state as well as the coherent
superposition of the de-localized excitation \cite{Enk2007}.
\cite{Chou2005} have devised a way to prove unambiguously the
entanglement of the heralded atomic state, based on quantum
tomography. They reconstructed the density matrix of the stored
state in a Hilbert space spanned by the state $|0_A0_B\rangle,
|1_A0_B\rangle, |0_A1_B\rangle, |1_A1_B\rangle$, where
$|n_An_B\rangle$ is the state with $n$ excitations in ensemble A
and $n$ excitation in ensemble B. In order to measure the density
matrix, the atomic state is first transferred into a photonic
state and the state of the atoms is inferred from the state of the
electromagnetic fields. The diagonal terms of the density matrix
are measured by direct photon counting, while the coherences are
inferred from an interference measurement with the conditional
anti-Stokes photons (see Fig. \ref{fringeDLCZ}). The density
matrix can then be used to calculate the amount of entanglement
using an entanglement measure, for example the concurrence $C$ ,
where $C=0$ for unentangled states, and $C=1$ for maximally
entangled states \cite{Wootters1998}. Using this technique,
\cite{Chou2005} were able to demonstrate measurement induced
entanglement between the two spatially separated atomic ensembles,
albeit with a low concurrence (of order $C=2\cdot10^{-2}$ at the
output of the ensembles).

\begin{figure}
{\includegraphics[scale=1]{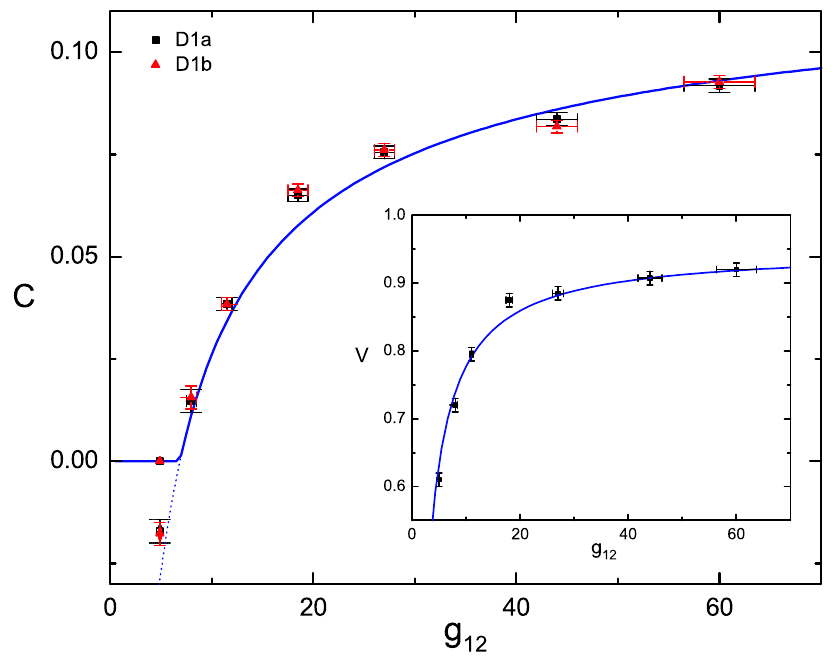}}
\caption{Heralded entanglement between two atomic ensembles
\cite{Laurat2007}. Concurrence C (without correcting for
propagation and detection losses) as a function of the
normalized cross correlation function $g_{S,AS}$ (denoted
here $g_{12}$, for the two possible heralding events
(detection at D1a or D1b after the beam splitter).  Inset:
Average visibility of the interference fringe between the
two field-2 modes.)} \label{concurrence}
\end{figure}

The main reason for the low concurrence in the first experiment
was the limited retrieval efficiency (10$\%$). This was
considerably improved in a more recent experiment with two
ensembles in the same MOT \cite{Laurat2007}. The concurrence was
measured as a function of the cross correlation function (see Fig.
\ref{concurrence}), with a maximum value of $g_{S,AS}=60$. At this
value, a concurrence of $C=0.35\pm0.1$ has been measured at the
output of the ensemble, leading to an inferred $C=0.9\pm 0.3$ for
the atomic state.  The decoherence of the stored entangled state
was also analyzed in this experiment, with entanglement persisting
for at least 20$\mu s$.

\subsubsection{Elementary Segment of DLCZ Quantum Repeater}
\label{DLCZsegment}

 Number state entanglement of the form of Eq.
(\ref{numberstate}) is not practical for performing quantum
communication tasks, such as quantum key distribution. It is
indeed difficult to implement single qubit rotations in the
excitation number basis. The solution proposed by DLCZ to this
problem is the implementation of two chains of entangled number
state ensembles in parallel (see section \ref{DLCZ-Protocol}). In
this way, it is possible to create effective two-excitation
entangled state by post selection when the two chains are combined
at the remote locations. This architecture also considerably
relaxes the constraints for phase stability. If the light fields
for the two chains are combined together and multiplexed in the
same quantum channel, the phase of the quantum channel must be
constant only during the time interval $\Delta t$ between the
successful entanglement generation in the two chains. An advantage
of this scheme compared to two photon schemes (see below) is that
the entanglement by single photon detection can be generated
independently in the two chains, which leads to higher generation
rates for the elementary link.
\begin{figure}[hr!]
{\includegraphics[scale=0.8]{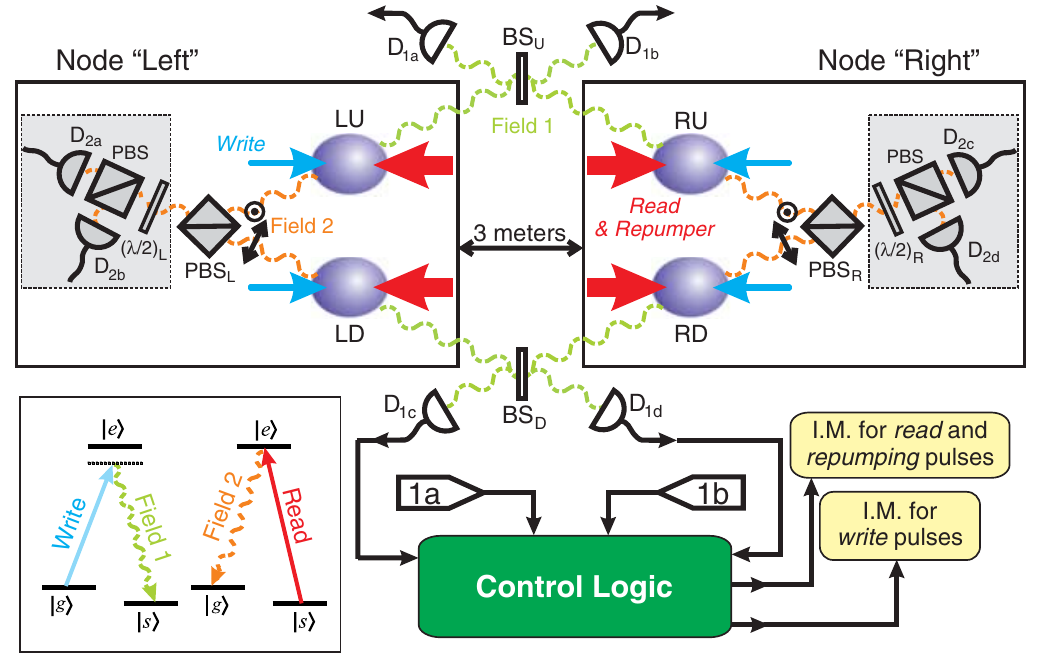}} \caption{Setup for the
elementary link of a DLCZ quantum repeater between two quantum
nodes (L,R) separated by 3 m \cite{Chou2007}. The inset at the
bottom left shows the relevant atomic levels for the
$^6S_{1/2}\rightarrow ^6P_{3/2}$ transition in atomic cesium, as
well as the associated light fields. With this setup, a
photo-detection event at either detector D1a or D1b indicates
entanglement between the collective excitation in LU and RU, and a
photo-detection event at either detector D1c or D1d indicates
entanglement between the collective excitation in LD and RD (20).
Two orthogonal polarizations in one fiber beamsplitter implement
BS$_U$ and BS$_D$, yielding excellent relative path stability. A
heralding detection event triggers the control logic to gate off
the light pulses going to the corresponding ensemble pair (U or D)
by controlling the intensity modulators (I.M.). The atomic state
is thus stored while waiting for the second ensemble pair to be
prepared. After both pairs of ensembles U,D are entangled, the
control logic releases strong read pulses to map the states of the
atoms to photons that are combined with orthogonal polarizations
on the polarizing beamsplitter PBS$_L$ and PBS$_R$. If only
coincidences between the fields at both nodes are registered, the
state is effectively equivalent to a polarization maximally
entangled state.)} \label{DLCZlink}
\end{figure}
The realization of this scheme \cite{Chou2007} was the first
experimental demonstration of an elementary segment of a quantum
repeater. The experiment was realized with 4 cold atomic
ensembles, as shown in Fig. \ref{DLCZlink}. The two quantum nodes
were in different apparatuses, about 3 m apart. Each node was
composed of two ensembles about 1mm apart, implemented by
addressing two different parts of a cold atomic cloud with
orthogonal polarizations. The Stokes light emitted by the
ensembles was recombined and coupled in the same single mode
fiber, with two orthogonal polarization (polarization
multiplexing). The light from the two nodes was then mixed at a
beam splitter, followed by polarization optics to separate the two
polarizations corresponding to the two chains of ensembles. The
entanglement was generated in a heralded fashion, using a
conditional control \cite{Felinto2006} that stopped sending
excitation pulses once a successful detection was obtained for the
corresponding chain. In this way, the entanglement could be
obtained independently in the two chains (up to the limited memory
time of about 10$ \mu s$ in the present experiment).

Once the two chains of ensembles have been successfully entangled,
the stored excitations are retrieved simultaneously in all
ensembles, and the retrieved light is combined at each node. The
desired effective two-photon state entangled is finally
post-selected by conserving only the events where one detection is
present at each node. The effective entanglement was verified by
violating a Bell inequality with the two fields.

Note that the optical phases were not actively stabilized
in this experiment. The passive phase stability of the
quantum channels during the time interval corresponding to
the memory time (10 $\mu s$) was good enough to ensure a
proper phase compensation.

\subsubsection{Entanglement Connection}
\label{DLCZconnection} Entanglement connection is obviously a
crucial step in order to extend the entanglement distance in
quantum repeater architectures. While many experimental
demonstrations of entanglement swapping have been realized with
entangled photons
\cite{Pan1998,Riedmatten2005,Halder2007,Kaltenbaek2009}, only one
attempt has been made so far to demonstrate the swapping of
entanglement with ensemble quantum memories \cite{Laurat2007}. The
experiment was realized in a setup similar to the one of ref
\cite{Chou2007}. Heralded number state entanglement was again
generated independently in two chains of ensembles. One ensemble
of each chain is then read-out simultaneously and the retrieved
light is combined at a beam splitter. A detection after the beam
splitter transfers the entanglement to the remaining ensembles,
which have never interacted. While the authors have been able to
demonstrate the transfer of a substantial amount of quantum
coherence, the demonstration of entanglement was not possible
using the method developed in \cite{Chou2005}. Actually, this
experiment highlights one of the main limitation of the DLCZ
protocol: the quadratic propagation of two photon errors with the
number of links. In order to keep the two-photon error low enough
to demonstrate entanglement, the excitation probability has to be
kept to a very low level. The resulting count rate was not high
enough to be able to determine the two excitation probability, as
required for the quantum tomography, in a reasonable time.

\subsection{Entanglement Creation and Swapping based on Two-Photon Detections}
\label{2photons}

We will now review experiments that are particularly relevant to
the schemes where the entanglement creation and/or connection are
based on two-photon detections.

\subsubsection{Two Photon Quantum Interference from Separate
Ensembles}

Two-photon quantum interference plays an essential role in all of
the protocols discussed in this review. Depending on the
protocols, it is used for entanglement swapping, see sections
III.A, III.B and III.F, and for entanglement generation in the
elementary links, see sections III.B and III.F. It also intervenes
in the protocols that are primarily based on single-photon
detections (see sections II, III.C, III.E) in the final step,
where two-photon entanglement is post-selected. At the heart of
two-photon interference lies the Hong-Ou-Mandel effect
\cite{Hong1987}. Due to the bosonic nature of photons, two
indistinguishable photons mixed at a beam splitter stick together
(photon bunching) and always exit in the same spatial mode. This
is due to a destructive interference between the probability
amplitude of both photons being reflected and both transmitted.
This effect manifests itself by the absence of coincidence
detection between the two output modes of the beam splitter when
the two photons are made indistinguishable, a property known as
Hong-Ou-Mandel (HOM) dip. The observation of a HOM dip is an
efficient way to quantify the degree of indistinguishability of
photons generated by spatially separated atomic ensembles.

In 2006, Felinto and coworkers \cite{Felinto2006} reported the
first observation of a two photon quantum interference with
photons emitted by separate atomic memories. The authors used
heralded single photons generated independently in two cold Cs
ensembles. Write pulses were sent in the two ensembles
simultaneously and Stokes light was collected in an optical fiber
and sent to a single photon detector. In order to be able to
address independently the two clouds, the authors used a
conditional control that stopped sending write pulses in the
corresponding ensemble when a Stokes photon was detected. In this
way, they could generate single spin excitations independently in
the two ensembles. After a time corresponding to the memory time
of the device, the two spin excitations were converted into single
photons, coupled in a single-mode optical fiber and combined at a
beam splitter in order to measure the two photon quantum
interference. A visibility of 70$\%$ was measured for the
Hong-Ou-Mandel dip. From this, value, taking into account the loss
of visibility due to the remaining two-photon contribution, the
 authors inferred an indistinguishability of 90$\%$ between the
 two photons.

 Beyond the two photon interference, this experiment was also the
 first one to show that the use of a quantum memory could increase
 the generation rate of quantum state of light in separated
 sources. The conditional control resulted in a 28-fold increase in
 the probability of obtaining a pair of single photons, relative to the case
 without memory.

 A similar setup with Rb atoms was used in a more recent experiment by \cite{Yuan2007}.
 The Hong-Ou-Mandel dip was measured by varying
the relative delay between the two read pulses. In this
way, the authors could infer
 the coherence time of the photons, 25 ns. A HOM visibility of 80
  $\%$was obtained, also limited by two-photon contributions. The
 authors also measured the HOM dip in the frequency domain, by
 changing the relative detuning between the two read beams. They
 found a similar visibility and a dip width of 35 MHz, in
 accordance with the time measurement.

 In the two experiments mentioned above, the quantum interference
 is realized with conditional anti-Stokes single photons retrieved from the
 stored excitations. However, if the creation of entanglement is realized with a two-photon detection (as in the protocol of \cite{Zhao2007}),
 the interference will take place between the Stokes photons. This
 configuration was tested experimentally by Chaneli\`{e}re and
 coworkers \cite{Chaneliere2007} with two cold Rb ensembles separated by 5.5m.
 The two photon interference was measured by recording the coincidence rate after the beam splitter for
 photons combined with the same and with orthogonal polarization. When the two unconditional Stokes
 photons are combined at the beam splitter, a HOM visibility of 33
 $\%$ is observed. This low visibility reflects the fact that the non-conditional Stokes fields are thermal fields. The probability of
 creating two Stokes photons in the same ensemble is equal
 to the probability of creating one Stokes photon in each
 ensemble. This is similar to what has been observed with two
 separate parametric down conversion sources
 \cite{Riedmatten2003}. However, if only those cases are taken into account
 where the stored excitations are converted into anti-Stokes photons and detected (using a four photon delayed coincidence
procedure), then
 the conditional Stokes fields are single photon fields, and a high
 visibility HOM dip can be achieved, provided that
 the two fields are indistinguishable. The authors observed a visibility of 86 $\pm 3 \%$.

In order to keep a high fidelity in a quantum repeater
architecture, it is essential that the visibility of the
Hong-Ou-mandel interference is very high. The dip
visibility indeed determines the fidelity of the swapping
operations. The errors acquired during each swapping will
then grow linearly with the number of links. The visibility
of the Hong-Ou-Mandel interference for the experiments
described here is still too low in that context. However,
it is mainly limited by the two photon components, which is
already taken into account in the theoretical description
of the protocols in section III. By working in a lower
excitation regime, it should thus be possible to
significantly improve the visibility. Besides the two
photon components, other factors can also decrease the
visibility, such as waveform or polarization
distinguishability. In that context, it is informative to
look at the experiments demonstrating Hong-Ou-Mandel
interferences with independent parametric down conversion
sources. The visibility has steadily improved over the last
few years with the best result so far being 0.96
\cite{Kaltenbaek2009}. Note that the two photon error is
suppressed only in the very low excitation regime. That
regime can be experimentally accessed if the measurement is
done with two photons from the same source. In that case,
visibilities approaching unity (0.994) have been measured
\cite{Pittman2003}.

\subsubsection{Entanglement between a Photon an a Stored
Excitation} \label{entanglementphotonexcitation}

Entanglement between photons and atomic excitations plays
an important role in a number of repeater protocols, cf.
sections III.B and III.F. This entanglement can be realized
in different ways: by encoding one logical qubit in two
ensembles and by using internal spin states. We now
describe these techniques in more detail. \\

(i) \emph{Collective excitations in different spatial modes}\\

 This first technique uses two collective excitations in different ensembles/spatial modes to encode one logical
 atomic qubit. It was first proposed and experimentally realized in \cite{Matsukevich2004}.
 In this experiment two nearby ensembles $A$ and $B$ within the same atomic cloud  are simultaneously
 excited with orthogonally polarized write beams. Similarly to Eq. (\ref{realstate}), the joint state of the atom-photon
 system after the Raman excitation can be written:
 \begin{equation}
 \left(1+\sqrt{\frac{p}{2}}\left (\alpha s_a^\dagger a_H^\dagger e^{i\phi_a}+\beta s_b^\dagger b_V^\dagger
 e^{i\phi_b}\right )+O(p)\right)|0\rangle.
 \end{equation}
 The resulting
 orthogonally polarized Stokes fields in different spatial modes are then
 combined into a single spatial mode at a polarizing beam
 splitter. Neglecting vacuum and higher order terms, the state can
 then be written:
 \begin{equation}
 |\Psi\rangle=\alpha|1_a,V\rangle+\beta e^{i\Phi}|1_b,H\rangle
 \label{spatial}
 \end{equation}
where $1_{a,b}$ represents the terms with one collective
spin excitation in ensemble $A$ and $B$, respectively, and
$|H\rangle$ ($|V\rangle$) is a photon with horizontal
(vertical) polarization. By measuring the Stokes photon in
the polarization basis, Matsukevich and Kuzmich were able
to project the atomic ensembles into a superposition state.
Then they showed that the atomic qubit could be mapped into
a photonic qubit with a fidelity exceeding classical
thresholds, by simultaneously reading out the two
ensembles. In 2007, this technique was used to demonstrate
the quantum teleportation of a polarization qubit carried
by a weak photonic coherent state onto a matter qubit
implemented with two cold Rb ensembles \cite{Chen2008}
\begin{figure}[hr!]
{\includegraphics[scale=0.8]{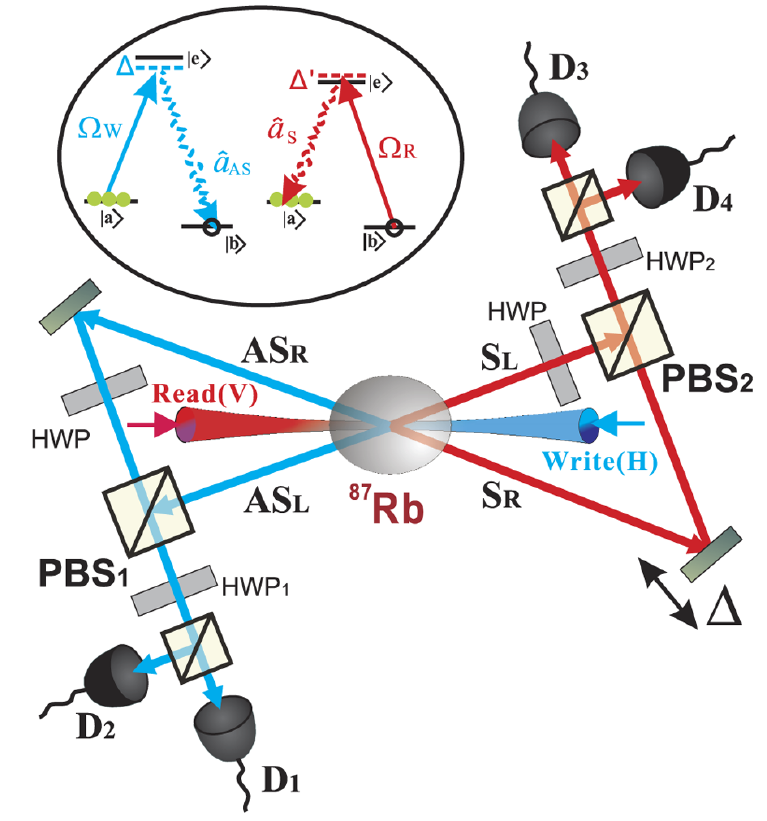}} \caption{Illustration
of the scheme of the experimental setup to generate entanglement
between a photon a stored atomic excitation using the two modes
approach \cite{Chen2007}. A weak horizontal polarized write pulse
illuminates the cold $^{87}$Rb atom cloud. The spontaneous Raman
scattered anti-Stokes field $AS_L$ and $AS_R$ with vertical
polarization are collected at $\pm 3^\circ$ to the propagating
direction of the write beam, defining the spatial mode of the
atomic ensembles L and R, respectively (Note that in this
experiment the storage state has a lower energy than the initial
state. Hence, the field generated by the write beam is called
 anti-Stokes field and the collective field generated by the read
beam is called Stokes field, contrary to the convention used in
this review). The $AS_R$ mode is rotated to be horizontal
polarized, combined with $AS_L$ mode on a polarizing beam splitter
PBS1 and sent to the polarization analyzer. This creates the
entanglement between the polarization of the anti-Stokes field and
the spatial modes of spin excitation in the atomic ensemble.}
\label{twomodes}
\end{figure}

An interesting development has been proposed in 2007
\cite{Chen2007}. Instead of using two separate ensembles,
the authors used a single ensemble, but collected the
Stokes photon in two different spatial modes separated by a
small angle, as shown in Fig (\ref{twomodes}). After
rotating the polarization of one of the mode by
$90^{\circ}$, the two modes are combined at a PBS. The read
beam is then sent in a counter-propagating way. Thanks to
the phase matching, the anti-Stokes photon are emitted in
the opposite direction of their respective Stokes photon.
The entanglement between the two photons was verified by a
violation of a Bell inequality, persisting for a storage
time of 20 $\mu s$.

The advantage of the entanglement with collective
excitation in different spatial modes is that the relative
probability of excitation of the two modes can be tuned,
contrary to the entanglement with internal spin states (see
below). The drawback is that it requires interferometric
stability between the two spatial modes.

The experiments mentioned here can also be seen as an elementary
realization of spatial multiplexing, as described in section
\ref{Collins}.\\

 (ii) \emph{Internal spin states}\\

 The second possibility to generate entanglement between light and stored excitation uses entanglement between the polarization
 of the Stokes photon and the
 internal spin state of the ensemble \cite{Matsukevich2005,deRiedmatten2006}.
This technique is similar to the one used to entangle single atoms
or ions and emitted photons \cite{Duan2003,Simon2003,Feng2003}.
Suppose an ensemble of three-level atoms with two ground state
$|g_1\rangle$ and $|g_2\rangle$, and one excited state
$|e\rangle$.
 The atoms are initially in $|g_1\rangle$. A weak write pulse induces spontaneous Raman scattering via the excited state,
 and the excitation is transferred to $|g_2\rangle$ while
emitting a Stokes photon. The emission of this photon can
follow two different decay path,
 leaving the atoms in different spin states. For example, for a circularly polarized write beam and for
 atoms initially in $|g_1,m_F\rangle$, the spontaneous Raman
 scattering can lead to spin excitation in $|g_2,m_F\rangle$ with
 emission of a Stokes photon with polarization $\sigma^-$, and to a
 spin excitation in $|g_2,m_F+2\rangle$ with a $\sigma^+$ polarized
 Stokes photon. As long as the final states are indistinguishable in all
 other degrees of freedom, the non vacuum part of the joint state of the light is
 given by:
 \begin{equation}
|\psi\rangle=\sqrt{p}\left(\cos\eta_{m_F}|\sigma^+,1_a^+\rangle+\sin\eta_{m_F}|\sigma^-,1_a^-\rangle
\right)+O(p)
 \end{equation}
 The coefficient $\eta_{m_F}$ is given by the Clebsch-Gordan
 coefficient for the relevant transition. In the more general case
 where the initial state is an incoherent mixture of that various $|g,m_F\rangle$, the collective atomic
 states
 are mixed states.
 This scheme was first demonstrated in \cite{Matsukevich2005},
 with a cold Rb ensemble. The entanglement was verified by the
 violation of a Bell-CHSH inequality between the Stokes and
 anti-Stokes photon, after a storage time of 200 ns. More recent
 results with cold Cs atoms \cite{deRiedmatten2006} have led to a violation of Bell
 inequality close to the quantum limit, and to the measurement of
 decoherence of entanglement, with violation of Bell inequality up
 to 20 $\mu s$.

Besides these two techniques, two other experiments have
demonstrated entanglement between collective matter qubits
and photonic qubits. The first one is based on frequency
encoded photonic qubits and dual species atomic ensembles
\cite{Lan2007}. The matter qubit basis consists in single
collective excitations in each of the co-trapped atomic
species (Rb$^{85}$ and Rb$^{87}$). The second experiment
demonstrated entanglement between the orbital angular
momentum of the Stokes photon a the
stored excitation \cite{Inoue2006}.\\

\subsubsection{Elementary Segment of Quantum Repeater}
\cite{Yuan2008} reported the experimental realization of an
elementary segment of quantum repeater, following the protocol of
\cite{Zhao2007}. In that protocol, the entanglement creation is
based on two photon detection. This requires the combination of
light matter entanglement and of two photon quantum interference.
Probabilistic entanglement between stored excitations and emitted
photons is first generated simultaneously in remote atomic
ensembles. The two Stokes photons are then combined at a beam
splitter (or a Polarizing beam splitter) in the middle station for
the Bell State measurement (BSM). A successful BSM projects the
atomic ensembles in an entangled state. This stored atomic
entangled state can be retrieved on demand by simultaneously
reading out the memories. \cite{Yuan2008}, used two cold Rb
ensembles connected by 6 m and 300 m of optical fibers. In each
ensemble, they created entanglement between the polarization of
the Stokes photon and the spatial mode of the stored excitations,
as in Eq. \ref{spatial}, using the technique introduced in
\cite{Chen2007}. The two Stokes photons were then combined at a
polarizing beam splitter, thus analyzing the projection on the
Bell state:
\begin{equation}
|\Phi^+\rangle=\frac{1}{\sqrt{2}}\left(|H,H\rangle+|V,V\rangle\right)
\end{equation}
A successful BSM  projects the atomic ensembles in the state:
\begin{equation}
|\Phi^+\rangle_{1,2}=\frac{1}{\sqrt{2}}\left(|L_1,L_{2}\rangle+|R_1,R_{2}\rangle\right)
\label{Bell_ensemble}
\end{equation}
where $|L_{i}\rangle$ and $|R_{i}\rangle$ corresponds to an
excitation in spatial mode $L$ and $R$ with $i={1,2}$ denoting the
remote quantum nodes. In order to verify the entanglement, atomic
qubits are converted into photonic qubits. In this scheme, the
double excitations in each ensemble induce spurious events in the
BSM that do not result in successful entanglement swapping. Hence,
the probability to successfully project the ensembles in the state
of Eq. (\ref{Bell_ensemble}) conditioned on a BSM is 1/2
\cite{Yuan2008}. The events that lead to a state with two
excitations in one ensemble and none in the other can be
eliminated by post-selection during the entanglement verification
stage. More importantly they can be in principle discarded after
the first entanglement connection, with a properly designed BSM
\cite{Zhao2007}.

In the present experiment, the quality of the post-selected
atomic state was good enough to violate a Bell inequality
with the retrieved photons when the two ensembles were
connected by 6 m of fiber for a storage time of up to $
\sim 4 \mu s$. When the ensembles were connected by 300 m
of fiber, a post-selected fidelity $F=Tr \left
(\rho_{exp}|\Phi^+\rangle_{1,2}\langle\Phi^+| \right
)=0.83\pm0.02$ was measured.
\subsubsection{Deterministic Local Generation of Entanglement}
As mentioned in section \ref{twophotongen}, the local generation
of high fidelity pairs of entangled ensembles is an important
capability for the implementation of robust quantum repeaters
architectures, in particular for the realization of the protocol
of \cite{Chen2007} section III.B. We briefly mention here two
deterministic schemes to create number state entanglement between
two ensembles that may be useful in this context. The first
experiment is based on the adiabatic transfer of one excitation
between two ensembles using a quantum bus in a optical cavity
\cite{Simon2007b}. The second experiment is based on the
absorption of a delocalized one photon state by two atomic
ensembles using Electromagnetically-induced Transparency (EIT)
\cite{Choi2008}.

Note that these techniques, while very effective for
deterministically entangling nearby ensembles, are very
sensitive to loss, and are thus not suitable to generate a
significant amount of entanglement in remote ensembles.

\subsection{Quantum Light Sources compatible with Ensembles based Quantum Memories}
\label{source}

An alternative to DLCZ-like quantum memories, where the photon
source and the memory are implemented within the same atomic
ensembles is to use different systems for the generation and for
the storage of quantum light, cf. section III.C. This
configuration has the distinct advantage that the photon to be
stored and the photon to be transmitted over long distances can
have different wavelengths. The separation of entanglement
creation and storage also allows the use of a class of quantum
memories that are well adapted to the storage of multiple temporal
modes (see section \ref{photonecho}).

In order to be compatible with ensemble based quantum memories,
the single photons must be at the resonance frequency of the atoms
and have a narrow spectrum that match the quantum memory bandwidth
(typically between 10 MHz and 100 MHz). Several ways have been
proposed to create quantum light with such specific properties.
The first technique, described in section \ref{source-ensemble},
is based on the generation of photons pairs and heralded single
photons with atomic ensembles. We also briefly mention two other
promising approaches based on parametric down conversion (section
\ref{narrowPDC}) and on single quantum emitters (section
\ref{single}).

\subsubsection{Photon Pair and Single Photon Sources based on Atomic Ensembles}
\label{source-ensemble}

 A natural way to create single photons or
photon pairs compatible with ensemble based quantum memories is to
use the same atomic
ensembles as photon source.\\

(i) \emph{Photon pair sources} \\

In principle, any DLCZ-like memory such as those described in
section \ref{photonpaircreation}, can be used as a source of
non-classical photon pairs, since the Stokes and anti-Stokes
fields are strongly correlated.
 While the previously described
experiments were performed in the pulsed regime with write and
read beams separated in time, photon pair creation has also been
demonstrated in the continuous regime. In 2005, \cite{Balic2005}
reported a four wave mixing experiment which generated
counterpropagating paired photons with coherence time of about 50
ns and linewidth of about 9MHz using cw write and read lasers in a
cold Rb ensemble. They used a four level system with two hyperfine
ground state. The wavelength of the Stokes and anti-Stokes photons
were 780 nm and 795 nm, respectively. The waveforms of the photon
pairs were shown to be controllable at a rudimentary level by
changing the read beam Rabi frequency. The photons were coupled
into opposing single mode fibers at a rate of 12000 pairs/sec and
violated the Cauchy-Schwartz inequality by $R=400$. Using a
similar setup, \cite{Kolchin2006} generated photon pairs with 5 ns
coherence time with the use of a single driving laser. This is
possible by using a far of resonance (6.8GHz) Raman transition for
the Stokes photon generation. Simultaneously, \cite{Thompson2006}
reported the generation of nearly identical photon pairs of 1.1
MHz bandwidth at a rate of 5 $\times 10^4$ pairs per second from a
cold Cs atomic ensemble inside a low finesse (F=250) optical
cavity. In order to generate degenerate photons, the authors also
used a single driving laser, but two Zeeman ground state level,
instead of two hyperfine ground state level. The two photons were
shown to be non classically correlated ($R=760$). The were also
shown to be indistinguishable to a large degree by performing a
Hong-Ou-Mandel type experiment. The results of \cite{Balic2005}
were recently improved  by the use of a two dimensional magneto
optical trap with optical depth as high as 62 \cite{Du2008}.
Photon pairs with coherence time up 900 ns with a subnatural
linewidth of 0.75 MHz were generated. The authors observed that 74
$\%$ of the Stokes photons were paired, and a strong violation of
the Cauchy-Schwartz inequality ($R=11600$).

\begin{figure}[hr!]
{\includegraphics[scale=1]{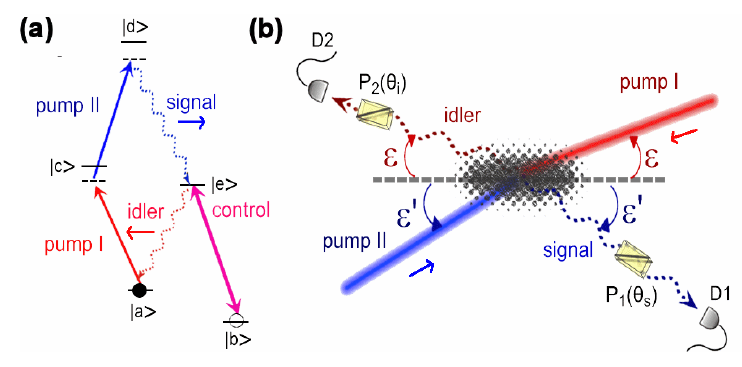}} \caption{Source of
non-degenerate photon pairs using atomic cascade transitions
\cite{Chaneliere2006}. (a) The atomic structure for the proposed
cascade emission scheme involving excitation by pumps I and II.
Pump II and the signal photons lie in the telecommunication
wavelength range when a suitable level of orbital angular momentum
L=0 or L=2 is used as level $|d\rangle$. (b) Schematic of
experimental setup based on ultracold $^{85}$Rb atomic gas.}
\label{cascade}
\end{figure}

The previously described DLCZ-type photon sources do not offer
much flexibility with the photon wavelength (see section
\ref{DLCZ-Basic}). In particular, it is not possible to directly
create a photon at 1.5 $\mu m$, as required for quantum repeater
applications. As mentioned in section \ref{P2M3}, one possibility
is to use wavelength conversion techniques. Another interesting
possibility for creating non-degenerate photon pairs in ensembles
is based on atomic cascade transitions. \cite{Chaneliere2006} have
demonstrated an entangled pair of 1530 nm and 780 nm photons
generated from an atomic cascade transition in a cold Rb ensemble
(see Fig. \ref{cascade}). While the 1530 nm photon can be
transmitted with low loss in optical fibers, the 780 nm is
naturally suited for mapping to an ensemble based Rb memory. They
authors observed polarization entanglement between the two photons
and superradiant temporal profiles for the photon at 780 nm.
\\

(ii) \emph{Single photon sources}\\

\begin{figure}[hr!]
{\includegraphics[scale=0.32]{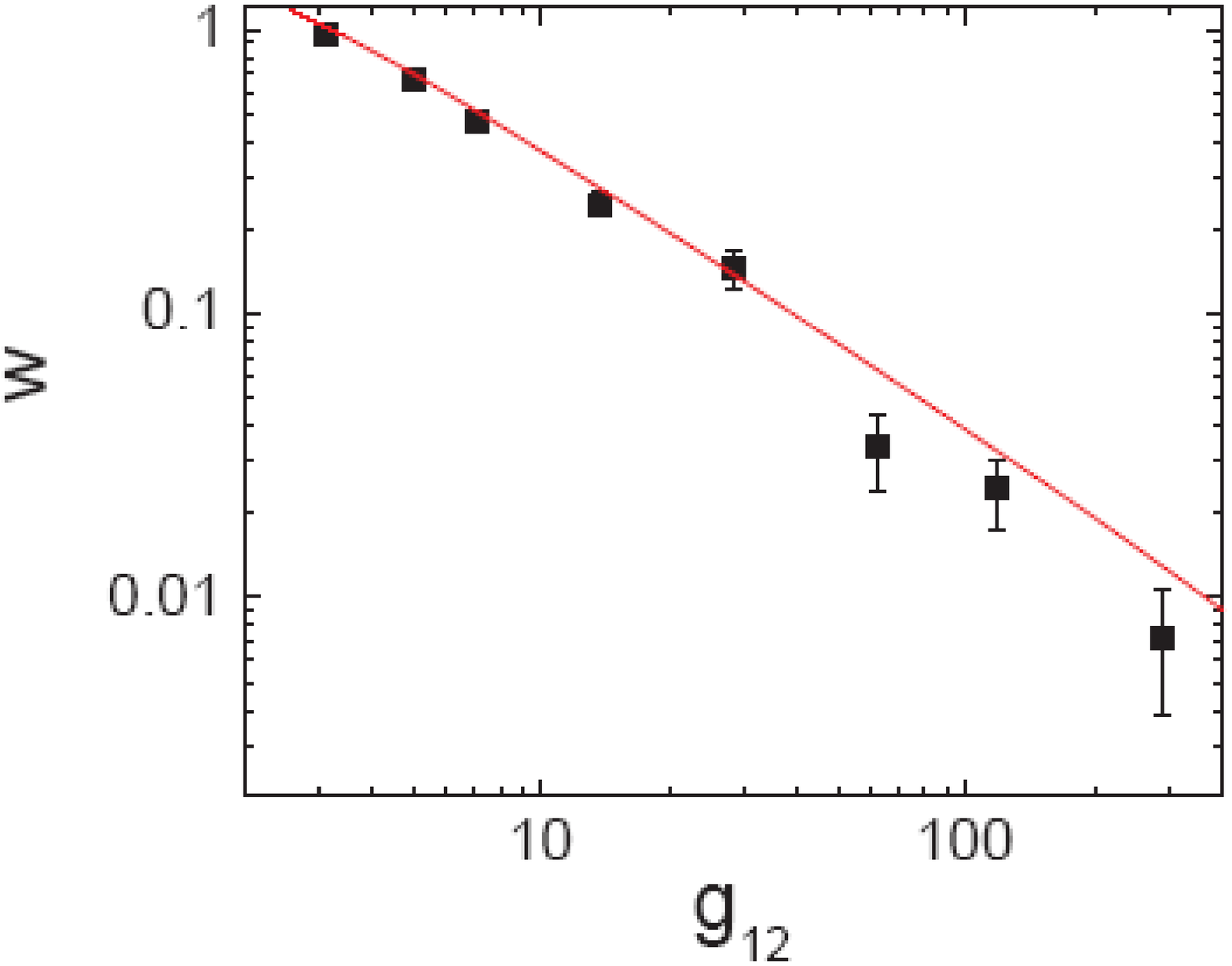}}
\caption{Suppression of the two-photon component of the
anti-Stokes field conditioned on the detection events in
the Stokes field, measured by the autocorrelation parameter
$\alpha$ (denoted here w) as a function of $g_{S,AS}$
(denoted here $g_{12}$), taken from \cite{Laurat2006}. The
lowest value obtained is $0.007\pm0.003$.}
\label{singlephotonensemble}
\end{figure}

The strong correlation between Stokes and anti-Stokes fields
enables the generation of heralded single photon fields with
programmable delay. The detection of the Stokes field heralds the
presence of a stored excitation that can be converted into a
single photon field at a programmable time, as already described
previously. This was first reported by Chou et al \cite{Chou2004}.
The single photon character of the emitted Anti-Stokes field,
conditioned on the detection of a Stokes photon was demonstrated
with a Hanbury-Brown-Twiss setup, leading to an anticorrelation
parameter \cite{Grangier1986} ( see Eq. \ref{alpha}) $\alpha
=0.24\pm0.05$. Improvements in the filtering of the Stokes and
anti-Stokes photons with cold ensembles (see section
\ref{photonpaircreation}) led to improved suppression of the
two-photon component \cite{Chaneliere2005,Chen2006}, with the best
values of $\alpha$ close \cite{Matsukevich2006} or below 0.01
\cite{Laurat2006} (see Fig. \ref{singlephotonensemble}). Another
interesting possibility is to use the heralded source of photons
combined with a measurement feedback protocol for implementing a
deterministic single photon source, as was first proposed and
demonstrated in \cite{Matsukevich2006} and later in
\cite{Chen2006,Zhao2009a}.

For hot gases, the filtering is more challenging and the
two photon components of the heralded anti-Stokes field
reported so far are sensibly higher that with cold
ensembles, with the lowest value being $\alpha=0.1\pm0.1$
\cite{Eisaman2005,Walther2007}.
 \\

\subsubsection{Narrow Band Photon Pair Sources based on Parametric
Down Conversion} \label{narrowPDC} A well known technique to
generate correlated or entangled photon pairs is based on
spontaneous down conversion in non linear crystals
\cite{Burnham1970,Hong1985,Hong1987,Wu1986}. The bandwidth of the
photon generated with this method are however usually of the order
of 10 THz, very far from atomic memory bandwidth, of order 10 to
100 MHz. Two techniques have been proposed to achieve the extreme
reduction in photon bandwidth required to efficiently map the
light produced by spontaneous down conversion in atomic memories.
The first one is based on cavity enhanced down conversion
\cite{Ou1999}. In that case, the non linear crystal is placed
inside an optical cavity and the light is emitted only in the
cavity modes. The system is operated as an optical parametric
oscillator far below threshold. The reduction in spectral
bandwidth is compensated by the enhancement in the efficiency of
the photon pair generation due to the cavity resonance. Without
further filtering, the output of the cavity is spectrally
multimode \cite{Kuklewicz2006,Wang2004,Wolfgramm2008,Scholz2009}.
Ideally, a single cavity mode should be selected, such that the
bandwidth of the parametric light is given by the width of the
cavity mode. This can be done by using etalons or filter cavities
with different free spectral range. \cite{Neergaard-Nielsen2007}
generated single high purity heralded single photons
 of 8 MHz bandwidth at 860 nm using such a system. More recently, the
 generation of
 single mode polarization entangled photons at 780 nm with 10 MHz bandwidth was reported \cite{Bao2008}.
In order to profit from the enhanced generation efficiency,
the cavity must be resonant with the two down conversion
modes. Hence, all the experiments demonstrated so far have
worked in the regime of degenerate  or near-degenerate
photon pairs.

The second technique is based on passive filtering. In that
case, the massive reduction of bandwidth is accompanied by
a corresponding reduction in conversion efficiency. It is
thus extremely challenging to implement such a filtering
with traditional non linear crystals. A possible solution
is the use of highly efficient waveguide sources
\cite{Tanzilli2001} based on periodically poled (PP)
crystals, such as Lithium Niobate (LN) or KTiOPO$_4$ (KTP).
These sources feature a conversion efficiency about four
orders of magnitude larger than conventional bulk crystals.
Using a PPLN waveguide together with fiber Bragg grating
filters, \cite{Halder2008} have generated photon pairs at
telecommunication wavelength with a bandwidth of 1.2 GHz.
With 7 mW of pump power, the source achieved a spectral
radiance of 0.08 photon pairs per coherence time. A photon
pair source adapted for a rubidium atomic memory was also
recently realized with a PPLN waveguide. The 10 THz
bandwidth of the parametric fluorescence was first reduced
to 23 GHz using a holographic grating, and further
decreased to 600 MHz using an etalon before detection
\cite{Akiba2007}. In a more recent experiment
\cite{Akiba2009}, a third filtering stage was introduced
with a moderate finesse optical cavity and the bandwidth of
the photon pairs was reduced to 9 MHz.

%Note that so far no highly non degenerate narrowband source
%compatible with quantum repeater architectures, i.e. with
%one photon adapted to an atomic memory and the other photon
%at telecommunication wavelengths, has been demonstrated
%using parametric down-conversion.

\subsubsection{Single Photon Sources based on Single Quantum Emitters}
\label{single}
 We now briefly mention some single quantum systems that generate single
photons with the required spectral properties to be compatible
with atomic memories. The advantage of using single quantum
emitters is that the photons can be emitted on demand. The first
example is single trapped atoms. Single photons have been emitted
deterministically from single atoms trapped in high finesse
optical cavities \cite{McKeever2004,Hijlkema2007} and in free
space \cite{Darquie2005}. More recently, entangled photons
 have been generated from a single atom trapped in an optical cavity
 \cite{Weber2009}. All these experiments have been performed with
 Cs or Rb atoms emitting at 852 nm and 780 nm, respectively. The emitted photons are therefore naturally
 suited for mapping in a corresponding atomic ensemble,
 but frequency conversion is required for long distance transmission in optical
 fiber. Single photons have also been generated by trapped single ions \cite{Keller2004,Maunz2007,Dubin2007}.
 Some solid state emitters at cryogenic temperature may also be interesting. A first example is single
 quantum dots embedded in microcavities. Quantum dots offer more
 flexibility for the wavelength of the emitted photon, in
 particular, single photons sources at telecommunication wavelength based on quantum dots have
 been demonstrated \cite{Hostein2009,Ward2005,Zinoni2006}. The demonstrated photon
 bandwidths are however still far to large to match the memory
 bandwidths. A promising solid state alternative is based on
 single dye molecules embedded in a solid state matrix. Fourier
 transformed single photons at 590 nm with a bandwidth of 17 MHz have recently been
 generated \cite{Lettow2007}. The use of a different dye molecule may offer some wavelength flexibility.
 Finally a last possible candidate is based on N-V
 center in diamond, where close to Fourier limited single photons
 have recently been observed, with lifetimes of about 10 ns. \cite{Batalov2008}

 %Note that so far the storage of a single photon generated
% by a single quantum system into an ensemble based memory has not
% been demonstrated.

\subsection{Storage of Single Photons in Atomic Ensembles}
\label{storage} We now review the protocols that have been
proposed and demonstrated experimentally for the storage of single
photons in ensembles of atoms. Section \ref{EIT} is devoted to
quantum memories based on Electro-Magnetically-Induced
Transparency (EIT) \cite{Lukin2003,Fleischhauer2005}, and section
\ref{photonecho} describes another more recent class of quantum
memories based on photon echoes \cite{Tittel2008}. A more general
review about light-matter quantum interfaces can be found in
\cite{Hammerer2008}.

%Note that so far, no experiment has demonstrated the
%storage and retrieval of light in a configuration adapted
%for adapted for quantum repeater architectures, i.e.
%involving a pair of photon with at least one of the photon
%at telecommunication wavelength.

\subsubsection{Quantum Memories based on Electromagnetically-Induced Transparency}
\label{EIT}
 A well known technique for mapping quantum state of
light into atomic states is based on Electromagnetically Induced
Transparency (EIT)
\cite{Fleischhauer2000,Fleischhauer2000a,Fleischhauer2002}. It has
been the subject of several reviews
\cite{Lukin2003,Fleischhauer2005} so we only give here the basic
feature of the technique, and review the relevant experiments. EIT
is a quantum interference effect that renders an opaque atomic
medium transparent thanks to a control laser field. This effect is
associated with a strong dispersion in the atomic medium that
leads to strong slowing and compression of the light. It has been
shown that by turning off the control laser when the optical pulse
is completely compressed in the atomic memory, it is possible stop
the light ,i.e.  to map the state of the light onto collective
spin excitations of the atoms
\cite{Fleischhauer2000,Fleischhauer2002}. The atomic state can
then be converted into light again by turning on the control
laser. This scheme has been demonstrated with bright coherent
light both in ultra-cold atoms \cite{Liu2001} with storage time of
about 1 ms and in hot atomic vapor \cite{Phillips2001} with
storage time up to 0.5 ms. More recent experiments in hot atomic
gases have led to storage and retrieval efficiencies close to
50$\%$ \cite{Novikova2007,Phillips2008} for storage times of order
of 100$\mu s$. These experiments implemented the optimal control
strategy introduced in
\cite{Gorshkov2007,Gorshkov2007a,Gorshkov2007b,Gorshkov2007c}.
Very recently, a stopped light experiment has been reported with
ultra-cold atoms loaded in a 3D optical lattice. A storage time of
240 ms was reported \cite{Schnorrberger2009}, albeit with a low
storage and retrieval efficiency (0.3$\%$). Stopped light has also
been observed in rare earth doped solids ( a Pr:Y$_2$SiO$_5$
crystal) \cite{Turukhin2001}. The coherence time of the hyperfine
ground state of Pr:Y$_2$SiO$_5$ without external magnetic fields
is $500 \mu s$. It can be however dramatically increased by using
an appropriate small magnetic field \cite{Fraval2004} and by using
dynamical control of decoherence \cite{Fraval2005}. Using these
techniques, \cite{Longdell2005} have demonstrated a stopped light
experiment (with efficiency of about 1$\%$) in a Pr:Y$_2$SiO$_5$
crystal with storage times exceeding 1 s, which is the longest
light storage time reported do date.

\begin{figure}[hr!]
{\includegraphics[scale=0.9]{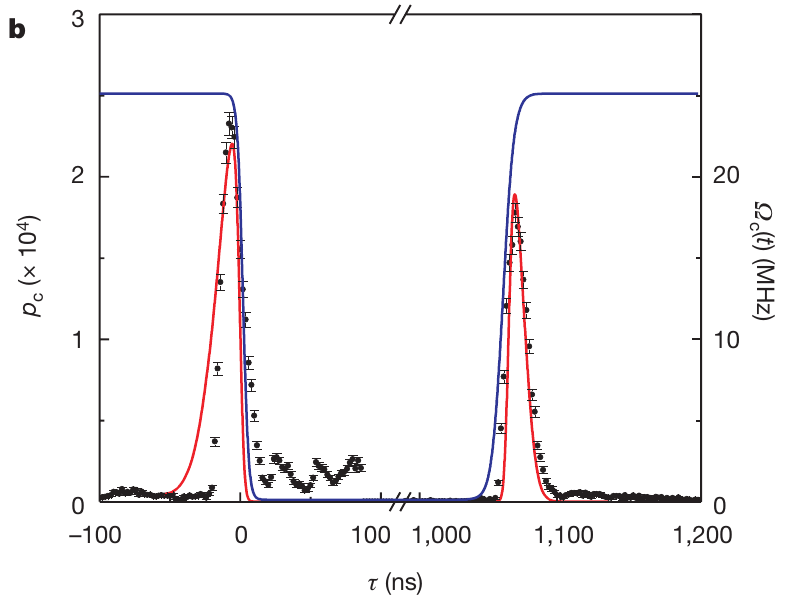}}
\caption{Storage and retrieval of single photon fields
using EIT in a cold Cs ensemble \cite{Choi2008}. The points
around $\tau = 0 \mu s$ represent 'leakage' of the signal
field due to the finite optical depth and length of the
ensemble. The points beyond $\tau = 1 \mu s$ show the
retrieved signal field. The overall storage and retrieval
efficiency is $(17 \pm 1)\%$. The blue solid line is the
estimated Rabi frequency of the control pulse.}
\label{singlephotonstorage}
\end{figure}

All the above mentioned experiments were realized with bright
coherent pulses. The first experiments of storage and retrieval of
single photon fields, which represent an important milestone, were
published in 2005 simultaneously by two groups. \cite{Eisaman2005}
used a hot atomic vapor to generate conditional single photons
fields with the required wavelength and bandwidth properties. The
single anti-Stokes photon pulses were then sent in another distant
hot atomic ensemble where both slow and stopped light was
observed. A storage and retrieval efficiency of 10$\%$ for short
delays and a storage time of 1 $\mu s$ were demonstrated, with the
non classical character of the stored field persisting for 500 ns.
\cite{Chaneliere2005} used two cold Rb atomic ensembles as single
photon source and quantum memory respectively. In this experiment,
the single photon field is implemented with the Stokes field,
conditioned on a subsequent anti-Stokes detection. The single
photons were then directed to another ensemble through 100 m of
optical fiber to be stored. In order to minimize the noise from
the excitation and control beams, the two beams are applied with a
small angle. After a programmable time, the atomic excitation is
converted back into a single photon. The storage and retrieval
efficiency for a storage a time of 500 ns is 6$\%$. The single
photon character of the stored and retrieved field is verified
explicitly by a Hanbury-Brown-Twiss experiment after a storage of
500 ns (with a minimum anticorrelation parameter $\alpha=0.36\pm
0.11$), while non-classical correlations between the retrieved
Stokes photon and the anti-Stokes photon are observed for storage
time exceeding 10 $\mu s$. The storage and retrieval efficiency of
single photons in cold atomic ensembles has been recently
increased to $17\%$ for a storage time of 1 $\mu s$
\cite{Choi2008}, see Fig. \ref{singlephotonstorage}.

\cite{Matsukevich2006a} have shown that the EIT technique also
allows the mapping of a polarization qubit carried by a single
photon. They generated probabilistic entanglement between the
polarization of the Stokes photons and the internal spin state of
the stored excitation, as in \cite{Matsukevich2005}. The
polarization of the Stokes photon was then mapped onto a distant
atomic ensemble, resulting in the probabilistic entanglement of
two remote matter qubits. The entanglement was demonstrated in a
post-selected fashion by converting the two stored qubits in
photonic qubits and by measuring polarization correlations between
the two photons, resulting in a violation of a Bell inequality.

EIT storage of degenerate non classical light generated by
parametric down conversion has also been recently demonstrated
\cite{Akiba2007,Akiba2009}. Correlated photon pairs and heralded
single photons were generated in the PPLN waveguide source
described in section \ref{narrowPDC}, and stored in a cold Rb
ensemble. In the first experiment \cite{Akiba2007}, the parametric
fluorescence light injected into the ensemble was spectrally much
broader (600 MHz) than the EIT transparency window (8.3 MHz) and
the authors demonstrated frequency filtered storage. The small
part of the light that spectrally matched the EIT transparency
window was stopped in the ensemble and thus delayed relative to
the rest of the light, which propagated through the atoms. The
retrieved photons (with 8$\%$ efficiency) could thus be detected
using temporal filtering. The authors demonstrated that the
coherence time of the light retrieved after a storage of 400 ns
was increased to 35 ns, compared to 0.2 ns for the incident
broadband light. Moreover, the conservation of the non-classical
character of the stored and retrieved light was confirmed by
measuring superbunching. In the second experiment
\cite{Akiba2009}, the bandwidth of the incident photons was
reduced to 9 MHz with an moderate finesse optical cavity. The
filtered degenerate parametric fluorescence was stored for 300 ns
in a cold Rb ensemble, before being retrieved with a storage and
retrieval efficiency of $\sim $14 $\%$. The non classical
character of the retrieved light was verified by the experimental
violation a classical inequality for photon counts introduced by
the authors. This confirms that the non classical properties of
the degenerate parametric fluorescence are conserved during the
storage in the ensemble. The storage and retrieval of conditional
single photons states was also demonstrated. The retrieved field
was shown to exhibit anti-bunching (with an autocorrelation
parameter $\alpha=0.52 \pm 0.30$).
%
%We note that quantum continuous variables of light have
%also been stored and retrieved using EIT
%\cite{Appel2008,Honda2008,Cviklinski2008}.

Protocols for storing light using off resonant Raman coupling to
long lived schemes have also been proposed
\cite{Nunn2007,Ham2008}, but not yet demonstrated experimentally.
A detailed study of the theory of photon storage in optically
thick ensembles of three levels atoms in a $\Lambda$ configuration
has been recently published in a series of papers
\cite{Gorshkov2007a,Gorshkov2007b,Gorshkov2007c,Gorshkov2008a}.

To end this section, we discuss the multimode properties of EIT
based storage. It has been shown that the number of modes $N_m$
that can be efficiently stored using EIT scales as
$N_m\sim\sqrt{d}$, where $d$ is the optical depth of the sample
\cite{Nunn2008}. This means that extremely high value of $d$ are
needed in order to store multiple temporal modes. The reason of
this poor scaling is that high efficiency storage of many temporal
modes requires at the same time a very slow group velocity in
order to compress all the pulses in the sample before turning off
the control fields and large transparency windows in order to
store short photons.

\subsubsection{Photon Echo based Quantum Memories}
\label{photonecho}
 We are now going to review other quantum
storage protocols based on photon echo techniques. Contrary to EIT
based protocols which rely on transparency, these protocols rely
on the reversible absorption of a single photon pulse in an
inhomogeneously broadened media. After absorption, the single
photon state is mapped onto a single collective atomic excitation
at the optical transition,
\begin{equation}|1\rangle_A=\sum_i c_i e^{i\delta_i
t}e^{-ikz_i} |g_1\cdot\cdot\cdot e_i\cdot\cdot\cdot g_N
\rangle
\end{equation}
where $z_i$ is the position of atom $i$ and $\delta_i$ is the
detuning of atom $i$ with respect to the central frequency of the
photon.  This collective state rapidly dephases, since each term
acquires a phase $e^{i\delta_i t}$. The goal of the quantum memory
protocols described here is to engineer the atomic system such
that this inhomogeneous dephasing can be reversed. If this
rephasing can be implemented, the light is re-emitted in a well
defined spatio-temporal mode when the atoms are all in phase
again, as a result of a collective interference between all the
emitters.

The rephasing of the dipoles can be triggered by optical pulses,
as it is the case in traditional photon echo techniques. These
techniques, while very successful to store classical light
\cite{Lin1995} and as a tool for high resolution spectroscopy
\cite{Macfarlane2002}, suffer from strong limitations for the
storage of single photons. In particular, it has been shown that
the unavoidable fluorescence due to the atoms excited by the
strong optical rephasing pulse blurs the single photon state and
reduce the fidelity of the storage to an unacceptable level
\cite{Ruggiero2009}.

We here describe two modified photon echo approaches that
allow in principle the storage and retrieval of single
photon fields with unit efficiency and fidelity. The first
one is based on Controlled Reversible Inhomogeneous
Broadening (CRIB), and the second one on Atomic Frequency
Combs (AFC).\\

(i) \emph{Controlled Reversible Inhomogeneous Broadening}\\

The theory of CRIB based quantum memories has been already
reviewed elsewhere \cite{Tittel2008}, and we only give here
short explanation of the principle, before discussing
experimental progress.

The idea of CRIB is to trigger the collective re-emission of light
absorbed by an ensemble of atoms by reversing the detuning of each
emitter at a given time $\tau$ after the absorption, such that
$\delta_i\rightarrow -\delta_i$. In this way, the state of the
atoms evolves as :
\begin{equation}
|1\rangle_A=\sum_i c_i e^{i\delta_i \tau}e^{-i\delta_i
t}e^{-ikz_i} |g_1\cdot\cdot\cdot e_i\cdot\cdot\cdot g_N
\rangle
\end{equation}
and all the atomic dipoles are in phase again when $t=\tau$,
leading to a collective emission a the time $2\tau$ after
absorption.

 The initial proposal exploited the fact that the natural doppler
broadening in a hot gas of atoms can be automatically reversed by
using control pulses with opposite direction \cite{Moiseev2001}.
Achievable storage times are however limited in hot gases, due to
the dephasing induced by atomic motion. Three groups then
described how this protocol could be extended to store single
photons in the optical regime in solid state materials
\cite{Kraus2006,Alexander2006,Nilsson2005}, typically in
rare-earth doped solids. The implementation of CRIB in solids
first requires to isolate a narrow (ideally homogeneously
broadened) absorption peak within a large transparency window.
This can be achieved by spectral hole burning techniques
\cite{Sellars2000,Alexander2006,Rippe2008,Seze2003}. This line
must then be artificially broadened in a controlled way, in order
to spectrally match the photon to be stored. To this end, one can
exploit the fact that some solid state materials have a permanent
dipole moment which gives rise to a linear Stark effect. The
resonance frequency of the atoms can then be controlled with
moderate external electric fields, and a controlled broadening can
be induced by applying an electric field gradient. The photon can
then be absorbed by the broadened peak and stored in the excited
state. After absorption, inhomogeneous dephasing takes place. The
re-emission can then be triggered by changing the polarity of the
electric field, which reverses the detunings and leads to the
rephasing of the dipoles. In that case, the photon is emitted in
the forward direction. It has been shown that in this
configuration and for a broadening applied transversally with
respect to the propagation direction, the storage and retrieval
efficiency is given by \cite{Sangouard2007}:
\begin{equation}
\eta_F(t)=d^2e^{-d}f(t)
\end{equation}
where $d$ is the optical depth of the atoms after
broadening and $f(t)$ is coherence profile in the excited
state, given by the Fourier transform of the initial
absorption peak. The maximal storage and retrieval
efficiency in that configuration is 54$\%$, limited by the
reabsorption of the echo by the optically thick transition.
In addition, the storage time in the excited state is
limited by the finite achievable width of the initial
absorption peak.

In order to overcome these limitations, it has been
proposed to transfer the excitation to an empty long-lived
ground state level, using optical control pulses. The
storage time is now given by the coherence time of the
ground state level, which may be much longer than the
excited state coherence time. Moreover, if the excitation
is brought back to the excited state after storage using a
counterpropagating control pulse, phase matching will
enable the CRIB echo to be emitted backward.
\cite{Kraus2006} have shown that this backward read-out is
equivalent to a time reversal of the Maxwell Bloch
equations, which effectively suppresses re-absorption. In
that case, the storage and retrieval efficiency is given by
\cite{Sangouard2007}:
\begin{equation}
\eta_B(t)=(1-e^{-d})^2g(t)
\end{equation}
with $g(t)=f(t-T_S)$, where $T_S$ is the time spent in the
long lived ground state. We see that $\eta_B$ can reach
100$\%$ for sufficiently high $d$.

Recently it has been shown theoretically that the efficiency of
the storage and retrieval can reach 100 $\%$ even without the
transfer to the ground state, using only a two level system
\cite{Hetet2008}. This requires the application of a longitudinal
broadening, i.e. a broadening where the frequency of the atoms
varies along the propagation direction. Such a broadening can be
obtained by the use of a longitudinal electric field gradient.
This configuration is sometimes referred as 'longitudinal CRIB' or
'Gradient Echo Memory'.

The multimode properties of CRIB have been studied in
\cite{Simon2007,Nunn2008}. It has been shown that, contrary to
EIT, the number of modes $N_m$ that can be stored with high
efficiency is proportional to the initial optical depth, $N_m\sim
d$.

The first proof of principle demonstration of the CRIB scheme with
bright coherent states was realized in a Europium  doped solid
\cite{Alexander2006}. The Europium ions doped in the solid state
matrix have an optical transition at 580 nm, and a level structure
with three hyperfine states in the ground and excited states. The
authors used optical pumping techniques to create a narrow
absorption peak with a width of 25 kHz, within a 3 MHz wide
transparency window. The absorption of the peak was approximately
40 $\%$. This peak was then broadened with a gradient of electric
field implemented with four electrodes in a quadrupole
configuration, thanks to the linear Stark effect. The broadened
spectral feature was excited using 1 $\mu s$ optical pulses and
the polarity of the electric field was reversed after a
programmable time $\tau$. After a further delay $\tau$, two-level
Stark echoes were observed, with a decay time of about 20 $\mu s$.
In an another experiment, the same authors stored and recalled a
train of 4 pulses \cite{Alexander2007}. They also showed that the
phase information of the input pulses was preserved during the
storage. In these experiments, only a very small part of the
incident pulses were re-emitted in the Stark echo (between
$10^{-5}$ and $10^{-6}$). This low efficiency can be partly
explained by the small absorption of the broadened peak (about
$1\%$ ).

In a more recent experiment, the same group demonstrated an
improved storage and retrieval efficiency of 15 $\%$ using a
Praseodymium doped crystal, which features an optical transition
with larger oscillator strength and consequently larger absorption
\cite{Hetet2008}. Using a longer crystal they could achieve an
efficiency as high as $45 \%$. Very recently, a CRIB experiment
has been demonstrated at the single photon level, using an Erbium
doped crystal absorbing at the telecommunication wavelength of
1536 nm \cite{Lauritzen2009}.\\

Finally, an interesting variation has been proposed, where the
reversible inhomogeneous broadening is not on the optical
transition, but on the Raman transition between two ground state
levels \cite{Hetet2008a,Moiseev2008}. The light is mapped on the
atoms by detuned Raman coupling to long lived ground states. A
proof of principle demonstration with bright pulses has been
reported in a rubidium vapor \cite{Hetet2008a}, where the Raman
resonance line was broadened by a magnetic field gradient.\\

(ii) \emph{Atomic Frequency Combs}\\

In order to fully exploit temporal multiplexing in quantum
repeater architectures, the memory should be able to store many
temporal modes with high efficiency. For EIT based quantum
memories, this requires extremely high and currently unrealistic
value of optical depth. The scaling is better for CRIB based
quantum memories but the required optical depth are still very
high (e.g. 3000 for 100 modes with 90$\%$ efficiency)
\cite{Simon2007}. Recently, \cite{Afzelius2009a} proposed a new
scheme, where the number of stored modes does not depend on the
initial optical depth . The scheme is based on ``atomic frequency
combs'' (AFC).

\begin{figure}[hr!]
{\includegraphics[scale=0.9]{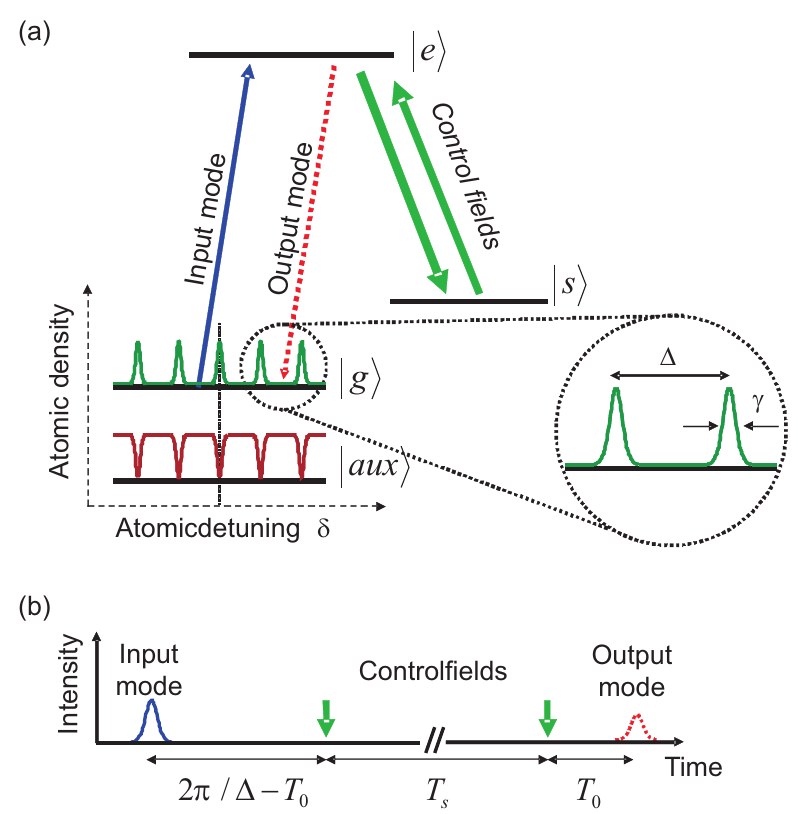}} \caption{The principles of
the atomic frequency comb (AFC) quantum memory
\cite{Afzelius2009a}. (a) An inhomogeneously broadened optical
transition $|g\rangle-|e\rangle$ is shaped into an AFC by
frequency-selective optical pumping to the $|aux\rangle$ level.
The peaks in the AFC have width $\gamma$ (FWHM)
 and are separated
by $\Delta$, where we define the comb finesse as $F =
\Delta/\gamma$ . (b) The input mode is completely absorbed
and coherently excites the AFC modes, which will dephase
and then rephase after a time $2\pi/\Delta$, resulting in a
photon-echo type coherent emission. A pair of control
fields on $|e\rangle-|s\rangle$ allow for long-time storage
as a collective spin wave in $|s\rangle$, and on-demand
read-out after a storage time Ts.} \label{AFC}
\end{figure}
The idea of AFC is to tailor the absorption profile of an
inhomogeneously broadened solid state atomic medium with a series
of periodic and narrow absorbing peaks of width $\gamma$ separated
by $\Delta$ (see Fig. \ref{AFC}). The single photon to be stored
is then collectively absorbed by all the atoms in the comb, and
the state of the light is transferred to collective atomic
excitations at the optical transition. After absorption, the atoms
at different frequencies will dephase, but thanks to the periodic
structure of the absorption profile, a rephasing occurs after a
time $2\pi/\Delta$ which depends on the comb spacing. When the
atoms are all in phase again, the light is re-emitted in the
forward direction as a result of a collective interference between
all the emitters. In order to achieve longer storage times and
on-demand retrieval of the stored photons, the optical collective
excitation can be transferred to a long lived ground state before
the re-emission of the light. This transfer freezes the evolution
of the atomic dipoles, and the excitation is stored as a
collective spin wave for a programmable time. The read out is
achieved by transferring back the excitation to the excited state
where the rephasing of the atomic dipoles takes place. If the two
control fields are applied in a counterpropagating way, the photon
is re-emitted backward. In that case, it has been shown that the
re-absorption of the light can be suppressed thanks to a
collective quantum interference.  In that configuration, the
theoretical storage and retrieval efficiency, assuming that the
decoherence in the long lived ground state is negligible and a
perfect transfer, is given by:
\begin{equation}
\eta_{AFC}\approx\left(1-e^{-\frac{d}{F}}\right
)^2e^{-\frac{7}{F^2}}
\end{equation}
where d is the peak optical depth, $F=\Delta/\gamma$ is the
finesse of the AFC. We see that $\eta_{AFC}$ tends towards
unity for sufficiently large $d$ and $F$.

The number of temporal modes $N_m$ that can be stored in an AFC
quantum memory is proportional to the ratio between the storage
time in the excited state $2\pi/\Delta$ and the duration of the
stored photons, which is inversely proportional to the total AFC
bandwidth $\Gamma=N_p\Delta$, where $N_p$ is the total number of
peaks in the AFC. Hence, we see that $N_m$ is proportional to
$N_p$ and is independent of the optical depth. The total bandwidth
of the AFC is however limited by the ground and excited state
level spacings.

\begin{figure}[hr!]
{\includegraphics[scale=0.9]{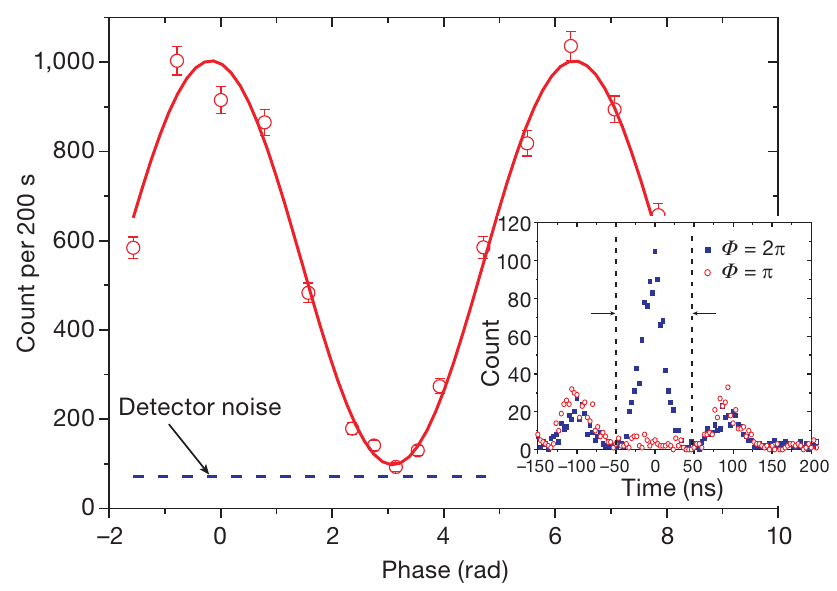}} \caption{Phase
preservation during the storage of a time-bin qubit by an atomic
frequency comb \cite{Riedmatten2008}. Time-bin qubits with
different phases $\Phi$, are stored and analyzed using the
light-matter interface. The analysis is performed by projecting
the time-bin qubit on a fixed superposition basis, which here is
achieved by two partial readouts (see text for details). The inset
shows the histogram of arrival times, where there is constructive
interference for $\Phi=2\pi$ and destructive interference for
$\Phi=\pi$ in the middle time bin. For this particular
interference fringe, a raw visibility of 82$\%$ and a visibility
of 95$\%$ when subtracting detector dark counts are obtained.}
\label{phaseAFC}
\end{figure}

The AFC protocol has been used to realize the first demonstration
of a solid light matter interface at the single photon level
\cite{Riedmatten2008}. The authors demonstrated the coherent and
reversible mapping of weak light fields with less than one photon
per pulse on average onto an ensemble of 10 millions Neodymium
atoms naturally trapped in a solid (a Nd:YVO4 crystal cooled to 3
K). They also showed that the quantum coherence of the incident
weak light field was almost perfectly conserved during the
storage, as demonstrated by performing an interference experiment
with a stored time-bin qubit (see Fig. \ref{phaseAFC}). Finally,
they also demonstrated experimentally that the interface makes it
possible to store light in multiple temporal modes (4 modes). The
storage and retrieval efficiency was low (about 0.5 $\%$) in this
experiment, mainly limited by the unperfect preparation of the
atomic frequency comb and by unperfect optical pumping.
In a more recent experiment with bright pulses,
\cite{Chaneliere2009} demonstrated improved performance in terms
of efficiency (9 $\%$) using a Thulium doped YAG crystal. The
multimode capacity has also been recently increased by almost one
order of magnitude compared to \cite{Riedmatten2008} in a Nd doped
Y$_2$SiO$_5$ crystal. A train of 32 pulses has been stored in the
crystal for 1.5 $\mu s$  \cite{Usmani2009}.

In the previous experiments, only the first part of the protocol
was demonstrated (i.e. the coherent mapping onto collective
excitation at the optical transition and collective re-emission at
a predetermined time). Hence these experiments did not allow for
on demand-read out. Very recently, a proof of principle
demonstration of the full AFC protocol including the transfer to a
long lived ground state has been demonstrated in a Praseodymium
doped Y$_2$SiO$_5$ crystal \cite{Afzelius2009}.

%\subsubsection{Solids}
%Crystals doped with rare-earth ions are promising for the
%realization of quantum memories with long coherence times. They
%also have great potential for multi-mode storage. Experimental
%results so far include the storage of classical light using
%electromagnetically induced transparency \cite{RE-EIT} and
%controlled reversible inhomogeneous broadening
%\cite{Alexander06,Hetet08}, and the demonstration of the phase
%coherence of the light storage in such ensembles \cite{Staudt1},
%including for spatially separated ensembles \cite{Staudt2}.

%%%%%%%%%%%%%%%%%%%%%%%%%%%%%%%%

\subsection{Detectors}
\label{detector}

Highly efficient single photon detectors with photon number
resolution are important for all the quantum repeater
protocols presented in this review. The most common and
most practical single photons detectors are based on
semiconductor avalanche photodiodes (APD). They offer the
advantage of being operated without cryogenic cooling
(using Peltier elements) but their present performance is
insufficient for use in a practical quantum repeater
architecture. Silicon APDs feature detection efficiencies
above 60 $\%$ in the wavelength range from 600 nm to 800 nm
and low dark count rates ($<$ 50 Hz). InGaAs APDs, which
can detect photons at 1550 nm, feature a much worse
efficiency to noise ratio. Moreover, standard APD operation
usually does not allow photon number resolution. It was
recently demonstrated however that using a technique to
measure very weak avalanches at the early stage of their
development, it is possible to obtain photon number
resolution \cite{Kardynal2008}.

\cite{Takeuchi1999} demonstrated a visible light photon
counter with an efficiency of 88$\%$ with an associated
dark count rate of 20 kHz using avalanches across a shallow
impurity conduction band in silicon at cryogenic
temperature. This type of detectors has also been shown to
allow for photon number resolution \cite{Kim1999}.

A new type of detector based on superconducting devices has
recently shown promising performances. Detectors based on
superconducting NbN nanowires \cite{Gol'tsman2001}, usually called
Superconducting Single Photon Detectors (SSPD) feature a very low
dark count rate and an excellent temporal resolution for high
speed counting at cryogenic temperature ($<$ 4K). An efficiency of
57 $\%$ at 1550 nm and of 67$\%$ at 1064 nm has recently been
demonstrated \cite{Rosfjord2006}, by inserting the SSPD into an
optical cavity and by using an anti-reflection coating. Photon
number resolving capability has also been demonstrated recently
\cite{Divochiy2008} in these devices.

The most advanced single photon detector in term of
efficiency and photon number resolution is based on
superconducting transition edge sensors. For example, Ref.
\cite{Lita2008} has recently reported photon-number
resolving detectors with 95$\%$ efficiency at the optimal
wavelength for telecom fibers (1556 nm), with a temporal
response time of about 1 $\mu s$. The drawbacks of these
detectors are the slow response time and the fact that they
have to be operated at very low temperature (100 mK), which
necessitates sophisticated and expensive cooling
techniques. Note that the slow response time is not
necessarily a problem in single mode quantum repeater
architectures, since the repetition rate is given by the
communication time. For architectures based on temporal
multiplexing however, it is important that the detector
response time is fast enough to discriminate between the
successive temporal modes.

Finally, we mention that high efficiency photon number resolving
single photon detection based on atomic ensembles has also been
proposed. These schemes have however not been demonstrated
experimentally \cite{Imamoglu2002,James2002}.

\subsection{Quantum Channels}
\label{channel}

In this section, we describe the quantum channels that can
be used to transmit single photons to remote locations for
the initial entanglement generation. The main focus in this
review is on optical fibers. A detailed review about the
use of optical fibers as quantum channel in quantum
cryptography experiments can be found in \cite{Gisin2002}.
In particular, the effect of birefringence leading to
polarization mode dispersion and of chromatic dispersion
are discussed. The importance of these effects decreases
with the transmitted photon bandwidth and can thus be
considered negligible for the narrow bandwidth photons
required in quantum repeater architectures. Hence, we
concentrate here on two other aspects that are crucial for
quantum repeaters: loss and phase stability.

In a single mode optical fiber, light is guided thanks to
the refractive index profile across the section of the
fiber. To ensure single mode operation, the core of the
fiber is small (diameter of order of a few wavelength).
Over the last 30 years, a considerable effort has been made
in order to reduce the transmission losses (initially
several dB per km). Today, installed commercial fibers
feature an attenuation of 0.35 dB/km at 1310 nm, and of 0.2
dB/km at 1550 nm. The loss around 800 nm is 2 dB/km. Recent
developments have led to the fabrication of ultra-low loss
optical fibers, with attenuation as low as 0.16 dB/km
\cite{Stucki2009}. Note that the loss remains exponential
and that quantum repeater architectures will be useful to
increase the transmission rates even if optical fibers with
lower losses are developed (unless the attenuation can be
reduced dramatically, which seems unlikely in the
foreseeable future).

In quantum repeater architectures where the entanglement
generation is based on singe photon detection, such as the
DLCZ scheme, the phase acquired in long fiber links must
remain constant for times that are typically of order of
seconds. \cite{Minar2008} have studied the phase stability
of installed fiber links for quantum repeater applications.
They have found that the phase of 36 km long Mach Zehnder
interferometer in urban environment remains stable at an
acceptable level (0.1 rad) for a duration of around 100
$\mu s$, which provides information about the time scale
available for active phase stabilization. Note that a phase
noise of 0.1 rad at 1550 nm corresponds to a fiber length
fluctuation of 25 nm, and thus to a timing jitter of 0.12
fs.

The stabilization of phase noise in optical fibers is also
relevant for other applications. There is currently an
active area of research aiming at the transmission of
frequency references over large distances in optical
fibers, in order to synchronize or compare remote
optical-frequency atomic clocks
\cite{Foreman2007,Musha2008,Newbury2007,Coddington2007}. In
this case, the phase noise in the fiber link directly
translates into a spectral broadening of the frequency
reference. To preserve the precision of optical clocks, the
light should be transmitted with subfemtosecond jitter over
long distances through the fiber. Noise cancellation
schemes have been developed, which work well as long as the
phase noise is negligible during the round trip time of the
fiber (see \cite{Foreman2007a} for a recent review).

Another interesting possibility to achieve a phase stable
operation is to excite the two remote memories in a Sagnac
interferometer configuration \cite{Minar2008,Childress2005}. In
this way, the excitation lasers for the two memories and the
emitted photons travel the same path in a counterpropagating
fashion. Hence as long as the phase fluctuations are slower than
the travel time, the phase difference is cancelled automatically.
\cite{Minar2008} have shown that high visibility first order
interference (V$>$ 98$\%$) could be achieved without any active
stabilization for fiber loops as long as 70 km in an urban
environment.

Optical fibers are not the only way of implementing a
quantum channel. The long-distance distribution of photons
through free space is also an active field of
investigation. Single photons and entangled photon pairs
have been transmitted over distances as great as 144 km
\cite{Ursin2007,Fedrizzi2009}. Free-space channels are also
subject to significant loss for long distances. For
example, the total channel loss in \cite{Fedrizzi2009} was
64 dB, where the attenuation was dominated by turbulent
atmospheric effects. An interesting extension of this
approach to long-distance quantum communication is the use
of satellites, in which case only a small part of the
photon path is in the atmosphere and the dominant losses
are due to beam divergence. Realistic links would involve
fast-moving low-orbit satellites. There are several
feasibility studies, both theoretical
\cite{Villoresi2004,Bonato2009} and experimental
\cite{Peng2005,Villoresi2008}. We would like to emphasize
that the quantum repeater principle can be applied to any
kind of lossy channel, including satellite-based
transmission.

\subsection{Coupling Losses}
\label{loss}

In section \ref{Robustness}, we have studied the performances of
various quantum repeater architectures with respect to the quantum
memory and detector efficiencies. In practice, other kinds of loss
need to be taken into account: the passive losses in the optical
elements and the fiber coupling losses. These losses are of
crucial importance for the performance of quantum repeaters.
Passive loss between the memory and the detector (together with
the loss between the photon source and the memory for experiments
with absorptive memories) affects the repeater performances in the
same way as the memory efficiency. One may thus introduce an
effective memory efficiency that take the passive loss into
account.

To illustrate the importance of passive losses, consider the
experiment of \cite{Simon2007a} which demonstrated a DLCZ like
memory with a cold Cs ensemble in an optical cavity. This
experiment has demonstrated the highest intrinsic retrieval
efficiency to date ($\eta_R=84\%)$. However, if one takes into
account the probability to escape the cavity ($T=0.17$), the
transmission efficiency of the interference filter used in the
experiment ($q_1$=0.61) and the fiber coupling efficiency
($q_2=0.65$), the effective retrieval efficiency, given by the
conditional probability to have an anti-Stokes photon in front of
the detector is $\eta_R^{eff}=\eta_RTq_1q_2\simeq0.06$. In free
space experiments the highest $\eta_R^{eff}$ measured in a DLCZ
source so far is of order of 25$\%$ \cite{Laurat2006}.

We emphasize that in order to build a practical quantum
repeater that beats direct transmission, all these passive
losses must be considerably reduced.

\section{Other Approaches towards Quantum Repeaters}
\label{OtherApproaches}

There is a significant number of proposals for realizing
quantum repeaters using ingredients other than atomic
ensembles and linear optics. An exhaustive treatment of
these proposals would be a task for another review paper
comparable to this one. Here we will restrict ourselves to
a very brief overview. One class of proposals keeps atomic
ensembles and linear optical processing as important
ingredients, but supplements them with additional
non-linear elements such as the Kerr effect \cite{He2008}
or light shift induced blockade \cite{Shahriar2007}.

However, most alternative proposals involve individual
quantum systems. One natural system to consider is trapped
ions, since quantum information processing in general is
extremely well developed in this system. There is a number
of proposals for entangling distant ions or atoms, both via
single-photon \cite{Bose1999,Cabrillo1999} and two-photon
detections \cite{Duan2003,Simon2003,Feng2003}, some of
which have already been realized experimentally
\cite{Moehring2007}. Recently quantum repeaters with
trapped ions have been analyzed more quantitatively,
including a discussion of possible multiplexing
\cite{Sangouard2009}. This analysis shows that ion-based
repeaters have the potential to significantly outperform
atomic-ensemble based approaches, notably because all
entanglement swapping operations can be performed with unit
probability.

Another active field of research on quantum repeaters
investigates the potential of single quantum systems in the
solid state. In particular,
\cite{Childress2005,Childress2006} developed a detailed
proposal for a quantum repeater architecture adapted to the
use of NV centers in diamond or of quantum dots as quantum
memories. The basic entanglement generation step in this
proposal is based on single-photon detections. A protocol
based on two-photon detections that uses spins in quantum
dots as quantum memories was proposed by \cite{Simon2007b}.
The ``hybrid quantum repeater'' approach of
\cite{Loock2006,Loock2008} that combines the transmission
of coherent states with the use of individual quantum
systems is also primarily motivated by solid-state systems.

Note that the present review is very focused on the most
immediate goal of outperforming direct transmission, thus
emphasizing simple protocols that are close to current
experimental capabilities. There are also numerous
contributions to the field of quantum repeaters that take a
longer-term and/or more abstract view. Examples include
work on the use of dynamic programming \cite{Jiang2007a},
error correcting codes \cite{Jiang2008}, decoherence-free
subspaces \cite{Dorner2008}, and an analysis of the role of
memory errors \cite{Hartmann2007}.

%%%%%%%%%%%%%%%%%%%%%%%%%%%%%%%%

%%%%%%%%%%%%%%%%%%%%%%%%%%%%%%%%

\section{Conclusions and Outlook}
\label{Conclusions}

Since the seminal DLCZ paper, there has been significant
progress towards the realization of quantum repeaters with
atomic ensembles and linear optics both on the theoretical
and on the experimental front. On the theoretical side,
various improved protocols have been proposed, improving
both the achievable entanglement distribution rate and the
robustness. Quantifying the expected performance of the
various protocols one finds that the prospect of efficient
multiplexing in particular seems to make it realistic to
implement a simple quantum repeater that outperforms the
direct transmission of quantum states in the not too
distant future.

Spurred in part by the original DLCZ proposal, in part by
more recent ideas, experiments are progressing rapidly. On
the one hand, elementary links for certain quantum repeater
protocols have already been demonstrated, though far from
the performance required to be practically useful. On the
other hand, very impressive values have been achieved for
key parameters such as storage time, memory efficiency or
multi-mode capacity, though not yet simultaneously in a
single system. Besides requiring the capacity to generate,
store and swap entanglement, constructing a viable quantum
repeater will also require technological elements such as
stabilized long-distance fiber links, and the virtual
elimination of coupling losses between the various
components.

We suspect that the first quantum repeater that beats
direct transmission will probably be realized using atomic
ensembles, linear optics and photon counting. In the longer
run, this approach may well be overtaken by other systems
with increased capabilities. New capabilities may come from
modifying ensemble-based approaches, for example by
including techniques based on continuous variables, where
good quantum memories have already been demonstrated
\cite{Julsgaard2004}. In a recent experiment,
coherent-state type entanglement was created using photon
counting \cite{Ourjoumtsev2009}, suggesting that hybrid
approaches combining photon counting and homodyne detection
may be promising. Another promising modification may be the
use of ensembles in optical lattices, where long light
storage times have recently been demonstrated for the first
time \cite{Schnorrberger2009}, and where more advanced
information processing may be possible compared to
conventional ensembles. However, it is also quite
conceivable that single quantum systems, as mentioned
briefly in section \ref{OtherApproaches}, will eventually
be more powerful than ensembles. Whichever approach may
turn out to be the most adapted, based on recent progress
we believe that in the long run intercontinental
entanglement will not be out of reach.

\acknowledgements

We thank M. Afzelius, T. Chaneli\`{e}re, Y.-A. Chen, K.S. Choi,
C.W. Chou, T. Coudreau, H. Deng, R. Dubessy, D. Felinto, M.
Halder, S. Hastings-Simon, H.J. Kimble, S. Kr\"{o}ll, J. Laurat,
B. Lauritzen, J.-L. Le Gou\"{e}t, I. Marcikic, J. Min\'{a}\v{r},
C. Ottaviani, J.-W. Pan, S. Polyakov, V. Scarani, W. Sohler, M.
Staudt, S. Tanzilli, W. Tittel, S. Van Enk, I. Usmani, H. Zbinden,
and B. Zhao for fruitful collaborations, as well as H. Briegel, M.
Eisaman, Q. Glorieux, P. Goldner, P. Grangier, S. Guibal, L.
Guidoni, O. Guillot-No\"{e}l, B. He, G. H\'{e}tet, D. Jaksch, A.
Kuzmich, A. Lvovsky, D. Matsukevich, C. Monroe, J. Nunn, A.
Ourjoumtsev, M. Plenio, E. Polzik, M. Razavi, S. Removille, M.
Sellars, A. S\o rensen, K. Surmacz, R. Tualle-Brouri, P. Van
Loock, and I. Walmsley for helpful comments and interesting
discussions. We gratefully acknowledge support by the EU
Integrated Project {\it Qubit Applications}, the Swiss NCCR {\it
Quantum Photonics}, and the ERC Advanced Grant {\it QORE}.

\appendix

\section{Calculating the Entanglement Distribution Time}
\label{calculatetime}

{\it (i) Creation of entanglement for an elementary link}

For a success probability $P_0$ one has an exponential
distribution of waiting times $n$ (in units of
$\frac{L_0}{c}$)
\begin{equation}
p(n)=(1-P_0)^{n-1} P_0, \nonumber
\end{equation}
which gives an expectation value
\begin{equation}
\langle n \rangle=\sum \limits_{n=0}^\infty n
p(n)=\frac{1}{P_0}. \nonumber
\end{equation}

{\it (ii) Waiting for a success in two neighboring
elementary links.}

In order to be able to attempt the first swapping,
entanglement creation has to succeed in two neighboring
elementary links. The corresponding waiting time is the
maximum of two waiting times, each of which follows the
distribution given for the elementary link. We will denote
the distribution for this combined waiting time by
$\tilde{p}(n)$ and its expectation value by $\langle
\tilde{n} \rangle$. One has
\begin{equation}
\tilde{p}(n)=p(n)^2+ 2 p(n) \sum \limits_{k=1}^{n-1} p(k),
\nonumber
\end{equation}
taking into account all cases where for at least one of the
two links we have to wait until $n$ for a success. The new
waiting time is
\begin{equation}
\langle \tilde{n} \rangle=\sum \limits_{n=1}^\infty n
\tilde{p}(n)=2 \sum \limits_{n=0}^\infty n p(n) \left(\sum
\limits_{k=1}^{n-1} p(k)+\frac{1}{2} p(n)\right). \nonumber
\end{equation}
It is easy to see that the expression inside the
parenthesis is bounded by 1. As a consequence, one must
have
\begin{equation}
\langle{\tilde{n}} \rangle \leq 2 \langle n \rangle.
\nonumber
\end{equation}
Obviously one also has
\begin{equation}
\langle n \rangle \leq \langle{\tilde{n}} \rangle,
\nonumber
\end{equation}
since waiting for two independent successes has to take at
least as long as waiting for a single success, so defining
\begin{equation}
f \equiv \frac{\langle{\tilde{n}} \rangle}{\langle n
\rangle} \nonumber
\end{equation}
one certainly has
\begin{equation}
1 \leq f \leq 2. \nonumber
\end{equation}
However, we can also calculate $f$ explicitly. This
requires explicitly calculating $\tilde{p}(n)$ and $\langle
\tilde{n} \rangle$. One finds
\begin{equation}
\tilde{p}(n)=P_0^2
(1-P_0)^{2n-2}+2P_0(1-P_0)^{n-1}\left(1-(1-P_0)^{n-1}\right),
\nonumber
\end{equation}
which gives
\begin{equation}
\langle \tilde{n} \rangle = \frac{3-2 P_0}{(2-P_0) P_0}
\approx \frac{3}{2 P_0}, \nonumber
\end{equation}
where the last relation holds in the limit of small $P_0$.

{\it (iii) First entanglement swapping}

The next step is the first entanglement swapping, which
succeeds with a probability $P_1$. Neglecting the time step
required for the swapping itself, the mean waiting time for
a success will be equal to $\langle \tilde{n} \rangle$ with
probability $P_1$. This is for the case where the swapping
works right away. It will be $2 \langle \tilde{n} \rangle$
with probability $(1-P_1) P_1$ (the second swapping attempt
is successful) etc. Taking into account all possibilities,
the waiting time for a successful swapping, and thus for
the establishment of entanglement over the length of 2
elementary links is
\begin{equation}
\langle n_1 \rangle=\langle \tilde{n} \rangle \sum
\limits_{k=0}^\infty (k+1) (1-P_1)^k P_1=\frac{\langle
\tilde{n} \rangle}{P_1} \approx \frac{3}{2 P_0 P_1}.
\end{equation}

{\it (iv) Higher levels in the repeater protocol.}

We have seen that at the lowest level of the repeater the
average waiting time for having a success in two
neighboring links is essentially exactly $\frac{3}{2}$
times longer than the average waiting time for one link.
The situation is more complicated for higher levels of the
repeater, because the corresponding distribution of waiting
times $p(n)$ for each individual link is no longer a simple
exponential distribution. In fact, its form at each level
of iteration depends not only on $P_0$, but also on the
success probabilities for entanglement swapping at all
lower levels ($P_i$). To our knowledge, nobody has so far
succeeded in obtaining useful analytical results for the
general case. However, numerical evidence
\cite{Jiang2007,Brask2008} suggests that $\frac{3}{2}$ is
still a good approximation for the factor $f$. One
certainly always has $1 \leq f \leq 2$. Let us also note
that the exact value of $f$ for each level has a relatively
small impact on log-scale comparison plots such as those in
this paper.

{\it (v) Second entanglement swapping and general formula}.

Whatever the exact value of $\langle \tilde{n}_1 \rangle$,
the same argument as for the first swapping shows that the
expectation value for the waiting time for a successful
second-level swapping, and thus for the establishment of
entanglement spanning four elementary links is
\begin{equation}
\langle n_2 \rangle=\frac{\langle \tilde{n}_1
\rangle}{P_2},
\end{equation}
where $P_2$ is the success probability for the second-level
swapping. The general formula for the waiting time after
$l$ levels (corresponding to $2^l$ elementary links) is
thus (in the limit of small $P_0$)
\begin{equation}
\langle n_l \rangle= \frac{f_0 f_1 ... f_{l-1}}{P_0 P_1 P_2
... P_l},
\end{equation}
with $f_0=\frac{3}{2}$, and $1 \leq f_i \leq 2$ for all
$f_i$.

The final post-selection step in protocols based on
single-photon detections can be treated in full analogy to
entanglement swapping. Again one needs two lower-level
copies to create one higher-level one, so there is an $f$
factor and a success probability $P_{ps}$.

\section{Multiphoton Errors in the DLCZ Protocol}
\label{errors}

Here we explain in more detail how to calculate the
multi-photon errors in the DLCZ protocol
\cite{Minar2009,Jiang2007}. The starting point is Eq.
(\ref{realstate}), where we now explicitly write the $O(p)$
terms, but set the phases $\phi_a=\phi_b=0$,
\begin{eqnarray}
\left(1+\sqrt{\frac{p}{2}}s_a^{\dagger}a^{\dagger}+
\frac{p}{4}(s_a^{\dagger})^2
(a^{\dagger})^2+O(p^{\frac{3}{2}})\right)\nonumber\\
\left(1+\sqrt{\frac{p}{2}}s_b^{\dagger}b^{\dagger}+
\frac{p}{4}(s_b^{\dagger})^2
(b^{\dagger})^2+O(p^{\frac{3}{2}})\right) |0\rangle.
\end{eqnarray}
The $O(\sqrt{p})$ terms give the results that are desired
in the protocol. The higher orders in $p$ give rise to
errors.

The probability for a detector that is photon number
resolving, but that has non-unit efficiency $\eta$, to
detect a single photon in mode $\tilde{b}$, given that
there are $n$ photons present in that mode, is
\begin{equation}
p_n=n \eta (1-\eta)^{n-1}. \label{detprob}
\end{equation}
In the entanglement generation step $\eta=\eta_d \eta_t$,
with $\eta_t \ll 1$, such that one can approximate
$p_1=\eta$ and $p_2=2 \eta$.

As a consequence, the state of the two modes $s_a$ and
$s_b$ conditional on one click in either $d=\frac{1}{\sqrt
2}(a+b)$ or $\tilde{d}=\frac{1}{\sqrt 2}(a-b)$ (for
simplicity we also set the phases $\xi_a=\xi_b=0$) is
\begin{eqnarray}
\rho_{AB}=|\psi_+\rangle \langle
\psi_+|+\frac{p}{2}\left(|20\rangle \langle 20|+|11\rangle
\langle 11|+|02\rangle \langle 02|
\right)\nonumber\\+\frac{p}{2\sqrt{2}}\left( |20\rangle
\langle 11|+|11\rangle \langle 02|+h.c.\right)+O(p^2)
\end{eqnarray}
where
\begin{equation}
|\psi_+\rangle=\frac{1}{\sqrt{2}}\left(
|01\rangle+|10\rangle \right)
\end{equation}
with $|01\rangle=|0\rangle_A |1\rangle_B$ etc. Here we
assume that a corrective phase shift of $\pi$ between $s_a$
and $s_b$ has been applied in the case of a detection in
$\tilde{d}$.

When these states are used to create entanglement between
non-neighboring stations using entanglement swapping, the
errors are amplified. Suppose we have established states
\begin{equation}
\rho_{AB}=\sum \limits_{k,l,k',l'}
\rho^{AB}_{kl,k'l'}|kl\rangle \langle k'l'|
\end{equation}
and $\rho_{CD}$. Entanglement swapping proceeds by
reconverting the atomic modes $s_b$ and $s_c$ into photonic
modes $b'$ and $c'$, and then combining the modes $b'$ and
$c'$ on a beam splitter such that
$b'=\frac{1}{\sqrt{2}}(\tilde{b}+\tilde{c}),c'=\frac{1}{\sqrt{2}}(\tilde{b}-\tilde{c})$.
Entanglement swapping also works with a single click.

The new state created in the entanglement swapping step is:
\begin{eqnarray}
\rho^{AD}_{kn,k'n'}=\sum \limits_{l,l',m,m'}
\rho^{AB}_{kl,k'l'} \rho^{CD}_{mn,m'n'}
\delta_{l+m,l'+m'}\nonumber\\ \eta (1-\eta)^{l+m-1}
2^{-l-m} \sqrt{l!m!l'!m'!}\nonumber\\
\sum \limits_{r=0}^{l+m} r r! (l+m-r)! f(l,m,r) f(l',m',r),
\end{eqnarray}
where now $\eta=\eta_m \eta_d$, with
\begin{equation}
f(l,m,r)=\sum \limits_{p=0}^l
\frac{(-1)^p}{p!(l-p)!(p+m-r)!(r-p)!}.
\end{equation}
The number of iterations depends on the nesting level of
the repeater (which depends on the distance to be covered).
We denote the state after the highest-level swapping
operations by $\rho_{AZ}$.
The final step of the protocol
is the projection onto one photon on each side (cf.
before), which in the ideal case could be written as
\begin{equation}
\sigma_{AZ}=P_{1_A 1_Z}\rho_{A_1 Z_1} \rho_{A_2 Z_2} P_{1_A
1_Z},
\end{equation}
where $P_{1_A 1_Z}$ is the projector onto the states that
have exactly one photon in location $A$ (taking $A_1$ and
$A_2$ together) and one photon in location $Z$. In
practice, the detectors are again not perfect (cf. above).
We therefore have to consider the probabilities for single
detections following Eq. (\ref{detprob}). The correct
formula for the final density matrix is then
\begin{eqnarray}
\sigma^{AZ}_{klmn,k'l'm'n'}=\rho^{A_1
Z_1}_{km,k'm'}\rho^{A_2 Z_2}_{ln,l'n'}p_{k+l} p_{m+n}
p_{k'+l'} p_{m'+n'}.
\end{eqnarray}
Here $k,l,m,n$ refer to the modes $a'_1,a'_2,z'_1,z'_2$
etc. The fidelity is given by the overlap of this state,
properly renormalized, with the ideal final state.

To second order in $p$, the results are the following.
Given as function of the nesting level $n$, the fidelity
$F(n)$ has the form
\begin{equation}
F(n)=1-A_n p (1-\eta)+B_n p^2 (1-\eta)^2,
\end{equation}
where the coefficients for the lowest values of $n$ are
$A_0=8, A_1=18, A_2=56, A_3=204, A_4=788$, $B_0=37,
B_1=250, B_2=2966, B_3=43206, B_4=669702$.

\begin{figure}
{\includegraphics[width=0.8 \columnwidth]{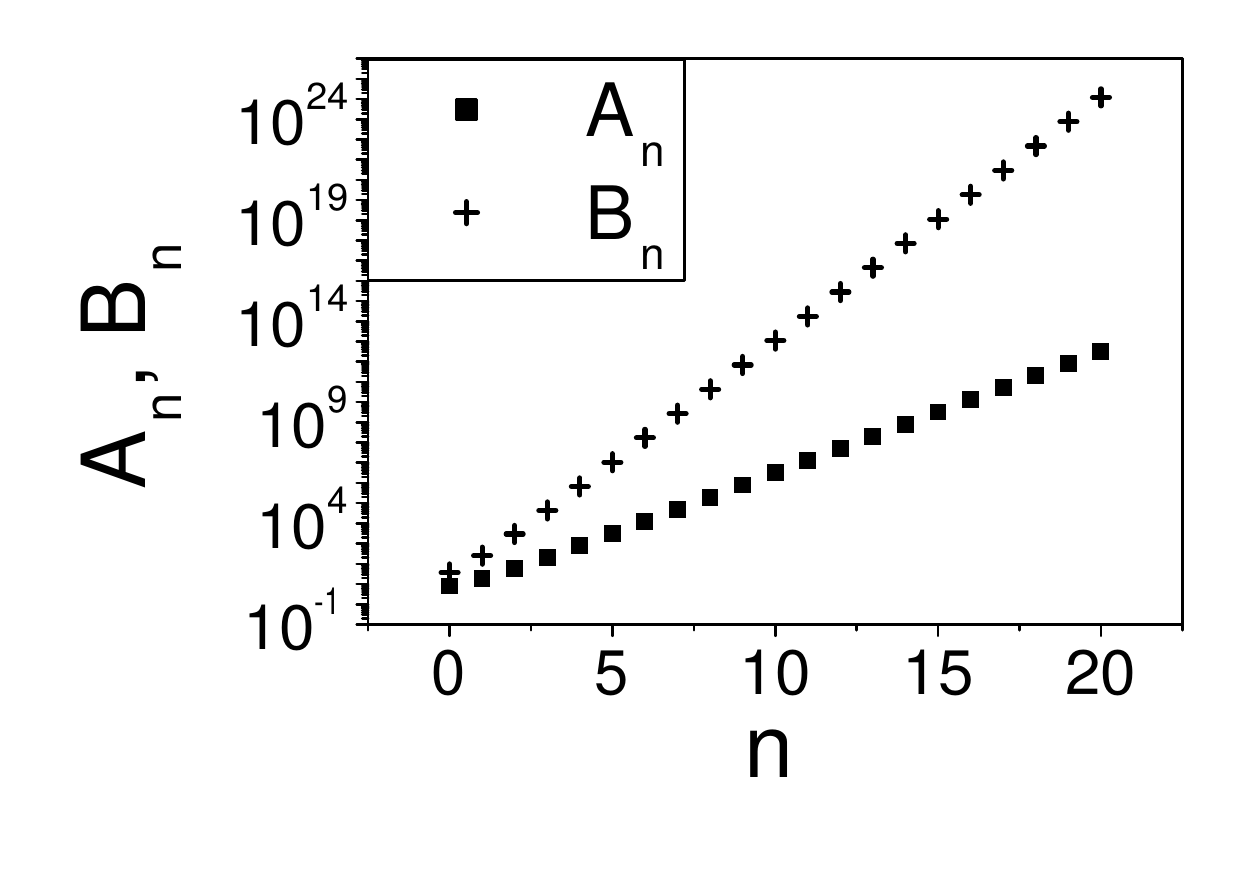}}
\caption{The coefficients $A_n$ and $B_n$ as a function of
$n$. One sees the scaling with $N^2=2^{2n}$ and
$N^4=2^{4n}$ respectively.} \label{Coeff}
\end{figure}

The dependence on $(1-\eta)$ indicates that ideal
photon-number resolving detectors in combination with
perfect memories would allow one to identify all
undesirable multi-photon events, and thus to eliminate the
considered errors. From $n=3$ onwards, $A_n$ scales
approximately like $2^{2n}$ or equivalently like $N^2$,
where $N=2^n$ is the number of links. This scaling becomes
virtually exact for large values of $n$, see Fig.
\ref{Coeff}. Similarly, $B_n$ scales approximately like
$N^4$.

The scaling with $N^2$ of the multi-photon errors is
related to the fact that the size of the vacuum component
in the state scales linearly with $N$. In fact, the errors
in the final post-selected two-photon state that lead to
the given fidelity reduction arise from the combination of
the multi-photon (i.e. two-photon) component for one pair
of ensembles with the vacuum component for the other pair
of ensembles, with both permutations contributing.

\bibliographystyle{apsrmp}
\bibliography{mybibCS}

\end{document}